\documentclass[acmtog,authorversion,screen]{acmart}

\setcopyright{acmlicensed}
\acmJournal{TOG}
\acmYear{2022}
\acmVolume{41}
\acmNumber{3}
\acmArticle{26}
\acmMonth{6}
\acmPrice{15.00}
\acmDOI{10.1145/3504002}

\setcitestyle{nosort,square}

\usepackage{xr}
\makeatletter
\newcommand*{\addFileDependency}[1]{
  \typeout{(#1)}
  \@addtofilelist{#1}
  \IfFileExists{#1}{}{\typeout{No file #1.}}
}
\makeatother

\newcommand*{\myexternaldocument}[1]{
    \externaldocument{#1}
    \addFileDependency{#1.tex}
    \addFileDependency{#1.aux}
}
\usepackage{grffile}  % to be able to use images with periods in their file names, without specifying the file extensions
\usepackage{soul}
\usepackage{microtype}
\usepackage{amsmath}
\usepackage{amsfonts}
\usepackage{booktabs}
\usepackage{multirow}
\usepackage{graphicx}
\usepackage{dsfont}
\usepackage{pifont}
\usepackage{tikz}
\usetikzlibrary{arrows.meta}
\usepackage{wrapfig}
\usepackage{xfrac}
\usepackage{nicefrac}
\usepackage{bbm}
\usepackage{flushend}
\usepackage[capitalize]{cleveref}
\usepackage{enumitem}
\usepackage{tabularx}
\usepackage{ctable}            % nicer tables
\usepackage{mathrsfs}
\usepackage{mathtools}         % for \mathclap, \overbracket
\usepackage[most]{tcolorbox}   % for boxes around equations
\usepackage{MnSymbol}          % for arrow symbols used in algorithm comments
\usepackage[outline]{contour}  % to make outlined text
\usepackage{overpic}      % laying text over images
\usepackage{xspace}
\usepackage{pifont}            % for numbers in circles
\usepackage[nomessages]{fp}    % to perform arithmetic
\usepackage{textcomp}     % for \textnumero symbol
\usepackage{colortbl}     % for \rowcolor

\newcommand{\IGNORE}[1]{}   % use to remove comments
\newcommand{\arXiv}{}

\newcommand{\vertical}{vertical\xspace}
\newcommand{\horizontal}{horizontal\xspace}
\newcommand{\Vertical}{Vertical\xspace}
\newcommand{\Horizontal}{Horizontal\xspace}

\newcommand{\eg}{e.g.,\xspace}
\newcommand{\Eg}{E.g.,\xspace}
\newcommand{\ie}{i.e.,\xspace}
\newcommand{\wrt}{w.r.t.\xspace}

\newcommand{\CwiseProduct}{\odot}
\newcommand{\InnerProd}[1]{\langle #1 \rangle}
\newcommand{\Norm}[1]{\lVert #1 \rVert}
\newcommand{\Abs}[1]{\left| #1 \right|}
\newcommand{\Lnorm}[1]{\mathcal{L}_{#1}}
\newcommand{\Greyscale}[1]{|#1|}

\DeclareMathOperator*{\argmin}{arg\,min}

\newcommand{\PixelCount}{N}
\newcommand{\ImageCount}{M}
\newcommand{\ErrorPixel}{\epsilon}
\newcommand{\ErrorImage}{\pmb{\ErrorPixel}}
\newcommand{\WhiteErrorImage}{\ErrorImage_\mathrm{w}}
\newcommand{\BlueErrorImage}{\ErrorImage_\mathrm{b}}

\newcommand{\EstimatedPixel}{Q}
\newcommand{\EstimatedImage}{\pmb{\EstimatedPixel}}
\newcommand{\EstimatedImageInit}{\EstimatedImage_\mathrm{init}}
\newcommand{\ReferencePixel}{I}
\newcommand{\ReferenceImage}{\pmb{\ReferencePixel}}
\newcommand{\SurrogatePixel}{\ReferencePixel'}
\newcommand{\SurrogatePixelEstimate}{\EstimatedPixel'}
\newcommand{\SurrogateImage}{\pmb{\SurrogatePixel}}

\newcommand{\OutputPixel}{O}
\newcommand{\OutputImage}{\pmb{\OutputPixel}}
\newcommand{\DitherMaskPixel}{B}
\newcommand{\DitherMask}{\pmb{\DitherMaskPixel}}
\newcommand{\DiffusionKernelPixel}{\kappa}
\newcommand{\DiffusionKernel}{\pmb{\DiffusionKernelPixel}}
\newcommand{\KernelPixel}{g}
\newcommand{\Kernel}{\pmb{\KernelPixel}}
\newcommand{\KernelReference}{\pmb{h}}
\newcommand{\DiracKernel}{\pmb{\delta}}
\newcommand{\Energy}{E}
\newcommand{\EnergyS}{\Energy}
\newcommand{\Permutation}{\pi}
\newcommand{\AllPermutations}{\Pi}

\newcommand{\SampleSetPixel}{S}
\newcommand{\SampleSetImage}{\pmb{\SampleSetPixel}}

\newcommand{\SampleSetPixelDomain}{\Omega}
\newcommand{\SampleSetImageDomain}{\pmb{\SampleSetPixelDomain}}

\newcommand{\ErrorDifference}{\pmb{\Delta}}
\newcommand{\EnergyApprox}{\Energy_{\DissimilarityMetric}}

\newcommand{\PixelIndex}{i}
\newcommand{\PixelIndexTwo}{j}
\newcommand{\PixelIndexThree}{k}
\newcommand{\DissimilarityMetric}{d}
\newcommand{\TileRadius}{R}

\newcommand{\ToneMappingOp}{\mathcal{T}}

\newcommand{\WhiteErrorImagePowerSpectrum}{\Abs{{\hat{\ErrorImage}}_\mathrm{w}}^2}

\newcommand{\BlueErrorImagePowerSpectrum}{\Abs{{\hat{\ErrorImage}}_\mathrm{b}}^2}
\newcommand{\ErrorImageSpectrum}{\hat{\ErrorImage}}
\newcommand{\ErrorImagePowerSpectrum}{\Abs{\hat{\ErrorImage}}^2}

\newcommand{\KernelSpectrum}{\hat{\Kernel}}
\newcommand{\KernelPowerSpectrum}{\Abs{\hat{\Kernel}}^2}

\newcommand{\KernelRef}{\pmb{h}}

\newcommand{\ConfidenceBias}{\mathcal{C}}
\newcommand{\CircleNumber}[1]{\FPeval{\result}{clip(191+#1)}$\raisebox{-0.42mm}{\text{\fontsize{11}{8}{\ding{\result}}}}$}

\definecolor{DarkGreen}{rgb}{0.0,0.6,0.0}
\definecolor{TableAltBackColor}{gray}{1}
\tcbset{
    colback=white!99!black,
    colframe=white!75!black,
    boxrule=0.15mm,
    arc=0.3mm,
    boxsep=0pt,left=2pt,right=4pt,top=2pt,bottom=4pt
}

\newcommand{\WrapFigureTemplate}[5]{%
    \setlength{\columnsep}{#2}% horizontal padding
    \begin{wrapfigure}{r}{#1\columnwidth}%
      \begin{center}%
        \vspace{#3}%
        #5
        \vspace{#4}%
      \end{center}%
    \end{wrapfigure}%
    \leavevmode%
}

\newcommand{\WrapFigure}[5]{%
    \WrapFigureTemplate{#2}{#3}{#4}{#5}{
        \includegraphics[width=#2\columnwidth]{#1}%
    }%
}

\usepackage{newfloat}  % for declaring custom environments
\usepackage{savesym}
\usepackage{algorithm}
\usepackage{algorithmicx}
\usepackage[noend]{algpseudocode} 
\savesymbol{Comment}
\algrenewcommand\algorithmicindent{3mm}
\algrenewcommand{\alglinenumber}[1]{\color{black!50}\fontsize{7.5}{6}\selectfont#1\color{black!30}:\phantom{*}}

\DeclareFloatingEnvironment[
    name=Algorithm,
    placement=tbhp,
    within=none,
]{pseudocode}
\crefname{pseudocode}{Alg.}{Algs.}
\Crefname{pseudocode}{Algorithm}{Algorithms}

\newcommand{\undefinecolor}[1]{\expandafter\let\csname\string\color@#1\endcsname\undefined}

\newcommand{\AlgCommentTemplate}[2]{\hfill{\fontsize{7}{6}\selectfont\textcolor{DarkGreen}{\text{#1\;#2}}}}
\newcommand{\AlgComment}[1]{\AlgCommentTemplate{}{#1}}
\newcommand{\AlgCommentLeft}[1]{\AlgCommentTemplate{$\leftarrow$}{#1}}
\newcommand{\AlgCommentDown}[1]{\AlgCommentTemplate{$\rcurvearrowsw$}{#1}}
\newcommand{\AlgCommentUp}[1]{\AlgCommentTemplate{\reflectbox{$\rcurvearrowne$}}{#1}}

\DeclareDocumentCommand{\Outlined}{ O{black} O{white} O{0.55pt} m }{%
    \contourlength{#3}% how thick each copy is
    \contour{#2}{\textcolor{#1}{#4}}%
}

\DeclareDocumentCommand{\TikzLabel}{ O{black} O{white} m m } {%
    \filldraw[white,ultra thick] (#3) circle (0pt) node[anchor=base,rotate=0]{\Outlined[#2][#1]{#4}};
}

\DeclareDocumentCommand{\TikzLabelL}{ O{black} O{white} m m } {%
    \filldraw[white,ultra thick] (#3) circle (0pt) node[anchor=west,rotate=0]{\Outlined[#2][#1]{#4}};
}

\DeclareDocumentCommand{\TikzLabelR}{ O{black} O{white} m m } {%
    \filldraw[white,ultra thick] (#3) circle (0pt) node[anchor=east,rotate=0]{\Outlined[#2][#1]{#4}};
}

\newcommand{\ClipImageRect}[3]{
    \begin{scope}
        \clip (#1) rectangle ++(#2);
        \path[fill overzoom image=#3] (0,0) rectangle (1,1);
    \end{scope}
}

\newcommand{\PlaceImage}[5]{
    \path[fill overzoom image={#5}] (#1,#2) rectangle ++(#3,#4);
}

\DeclareGraphicsExtensions{.ai,.pdf,.jpg,.png}

\newcommand{\Paragraph}[1]{\paragraph{#1}}

\myexternaldocument{supplemental}

\begin{document}

\hypersetup{colorlinks,
    linkcolor=ACMDarkBlue,
    citecolor=ACMDarkBlue,
    urlcolor=ACMDarkBlue,
    filecolor=ACMDarkBlue}

%%%%%%%%%%%%%%%%%%%%%%%%%%%%%%%%%%%%%%%%%%%%%%%%%%%%%%%%%%%%
% Title & authors
%%%%%%%%%%%%%%%%%%%%%%%%%%%%%%%%%%%%%%%%%%%%%%%%%%%%%%%%%%%%

\title{Perceptual error optimization for Monte Carlo rendering}

\author{Vassillen Chizhov}
\orcid{0000-0001-7082-6367}
\affiliation{%
  \institution{MIA Group, Saarland University,}
  \institution{Max-Planck-Institut f{\"u}r Informatik}
  \city{Saarbr{\"u}cken}
  \country{Germany}
}

\author{Iliyan Georgiev}
\orcid{0000-0002-9655-2138}
\affiliation{%
  \institution{Autodesk}
  \country{United Kingdom}
}

\author{Karol Myszkowski}
\orcid{0000-0002-8505-4141}
\affiliation{%
  \institution{Max-Planck-Institut f{\"u}r Informatik}
  \city{Saarbr{\"u}cken}
  \country{Germany}
}

\author{Gurprit Singh}
\orcid{0000-0003-0970-5835}
\affiliation{%
  \institution{Max-Planck-Institut f{\"u}r Informatik}
  \city{Saarbr{\"u}cken}
  \country{Germany}
}

% \renewcommand{\shortauthors}{V.\ Chizhov, et al.}

%%%%%%%%%%%%%%%%%%%%%%%%%%%%%%%%%%%%%%%%%%%%%%%%%%%%%%%%%%%%
% Teaser figure
%%%%%%%%%%%%%%%%%%%%%%%%%%%%%%%%%%%%%%%%%%%%%%%%%%%%%%%%%%%%

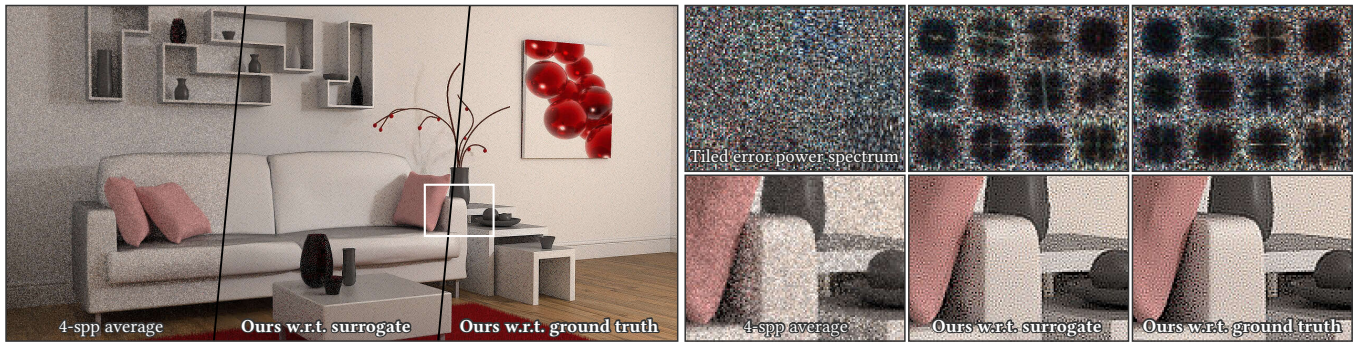
\begin{teaserfigure}
    \centering
    %!TEX root = ../paper.tex

\definecolor{OutlineColor}{gray}{0.2}
\definecolor{TextOutlineColor}{gray}{0.2}
\definecolor{LineColor}{gray}{0}

\footnotesize
\hspace*{-0.82em}
\begin{tabular}{c@{\;}c@{\;}c@{}}
\begin{tikzpicture}[scale=1.05]
    \begin{scope}
        \clip (0,0) -- (2.63,0) -- (3.03,4.25) -- (0,4.25) -- cycle;
        \path[fill overzoom image=figures/teaser/halftoning_aa_spp4_living-room-3_average] (0,0) rectangle (8.5cm,4.25cm);
    \end{scope}
     \begin{scope}
        \clip (2.63,0) -- (5.47,0) -- (5.87,4.25) -- (3.03,4.25) -- cycle;
        \path[fill overzoom image=figures/teaser/halftoning_aa_spp4_living-room-3_VDBS_power_set_DIN] (0,0) rectangle (8.5cm,4.25cm);
    \end{scope}
    \begin{scope}
        \clip (5.47,0) -- (8.5,0) -- (8.5,4.25) -- (5.87,4.25) -- cycle;
        \path[fill overzoom image=figures/teaser/halftoning_aa_spp4_living-room-3_VDBS_power_set_REF] (0,0) rectangle (8.5cm,4.25cm);
    \end{scope}
    \draw[line width=0.25mm, LineColor] (2.63,0) -- (3.03,4.25);
    \draw[line width=0.25mm, LineColor] (5.47,0) -- (5.87,4.25);
    %% Frame around image
    \draw[OutlineColor,line width=0.2mm] (0,0) rectangle (8.5,4.25);
    %% Zoom-in frame
    \draw[white,thick] (5.3,1.33) rectangle ++(0.87,0.65);
    %% Labels
    \filldraw[white,ultra thick] (1.33, 0.12) circle (0pt) node[anchor=base,rotate=0] {\Outlined[white][TextOutlineColor]{4-spp average}};
    \filldraw[white,ultra thick] (4.05, 0.12) circle (0pt) node[anchor=base,rotate=0] {\Outlined[white][TextOutlineColor]{\textbf{Ours \wrt surrogate}}};
    \filldraw[white,ultra thick] (7.0, 0.12) circle (0pt) node[anchor=base,rotate=0] {\Outlined[white][TextOutlineColor]{\textbf{Ours \wrt ground truth}}};
\end{tikzpicture}
&
\begin{tikzpicture}[scale=1.05]
    %% Tiled power spectra zoom-ins
    % \ClipImageRectTwo{0}{2.15}{2.78}{2.10}{0.720}{0.33}{15}{figures/teaser/ps_aa_spp4_living-room-3_average}
    \PlaceImage{0}{2.15}{2.78}{2.10}{figures/teaser/crop_ps_aa_spp4_living-room-3_average}
    \draw[OutlineColor,line width=0.2mm] (0,2.15) rectangle ++(2.78,2.10);
    % \ClipImageRectTwo{2.83}{2.15}{2.78}{2.10}{0.720}{0.33}{15}{figures/teaser/ps_aa_spp4_living-room-3_VDBS_power_set_DIN}
    \PlaceImage{2.83}{2.15}{2.78}{2.10}{figures/teaser/crop_ps_aa_spp4_living-room-3_VDBS_power_set_DIN}
    \draw[OutlineColor,line width=0.2mm] (2.83,2.15) rectangle ++(2.78,2.10);
    % \ClipImageRectTwo{5.66}{2.15}{2.78}{2.10}{0.720}{0.33}{15}{figures/teaser/ps_aa_spp4_living-room-3_VDBS_power_set_REF}
    \PlaceImage{5.66}{2.15}{2.78}{2.10}{figures/teaser/crop_ps_aa_spp4_living-room-3_VDBS_power_set_REF}
    \draw[OutlineColor,line width=0.2mm] (5.66,2.15) rectangle ++(2.78,2.10);
    %% Image zoom-ins
    % \ClipImageRectTwo{0}{0}{2.78}{2.10}{0.720}{0.33}{15}{figures/teaser/halftoning_aa_spp4_living-room-3_average}
    \PlaceImage{0}{0}{2.78}{2.10}{figures/teaser/crop_halftoning_aa_spp4_living-room-3_average}
    \draw[OutlineColor,line width=0.2mm] (0,0) rectangle ++(2.78,2.10);
    % \ClipImageRectTwo{2.83}{0}{2.78}{2.10}{0.720}{0.33}{15}{figures/teaser/halftoning_aa_spp4_living-room-3_VDBS_power_set_DIN}
    \PlaceImage{2.83}{0}{2.78}{2.10}{figures/teaser/crop_halftoning_aa_spp4_living-room-3_VDBS_power_set_DIN}
    \draw[OutlineColor,line width=0.2mm] (2.83,0) rectangle ++(2.78,2.10);
    % \ClipImageRectTwo{5.66}{0}{2.78}{2.10}{0.720}{0.33}{15}{figures/teaser/halftoning_aa_spp4_living-room-3_VDBS_power_set_REF}
    \PlaceImage{5.66}{0}{2.78}{2.10}{figures/teaser/crop_halftoning_aa_spp4_living-room-3_VDBS_power_set_REF}
    \draw[OutlineColor,line width=0.2mm] (5.66,0) rectangle ++(2.78,2.10);
    %% Labels
    \filldraw[white,ultra thick] (1.38, 2.27) circle (0pt) node[anchor=base,rotate=0] {\Outlined[white][TextOutlineColor]{Tiled error power spectrum}};
    \filldraw[white,ultra thick] (1.4, 0.12) circle (0pt) node[anchor=base,rotate=0] {\Outlined[white][TextOutlineColor]{4-spp average}};
    \filldraw[white,ultra thick] (4.2, 0.12) circle (0pt) node[anchor=base,rotate=0] {\Outlined[white][TextOutlineColor]{\textbf{Ours \wrt surrogate}}};
    \filldraw[white,ultra thick] (7.05, 0.12) circle (0pt) node[anchor=base,rotate=0] {\Outlined[white][TextOutlineColor]{\textbf{Ours \wrt ground truth}}};
\end{tikzpicture}
\end{tabular}

\undefinecolor{OutlineColor}
\undefinecolor{TextOutlineColor}
\undefinecolor{LineColor}

    \vspace{-2.5mm}
    \caption{
        We devise a perceptually based model to optimize the error of Monte Carlo renderings. Here we show our \vertical iterative minimization algorithm from \cref{sec:VerticalOptimization}: Given 4 input samples per pixel (spp), it selects a subset of them to produce an image with substantially improved visual fidelity over a simple 4-spp average. The optimization is guided by a surrogate image obtained by regularizing the noisy input; we also show using the ground-truth image~as~a~guide. The power spectrum of the image error, computed on $32\!\times\!32$-pixel tiles, indicates that our method distributes pixel error with locally blue-noise characteristics.
    }
    \label{fig:teaser}
    \vspace{1em}
\end{teaserfigure}

%%%%%%%%%%%%%%%%%%%%%%%%%%%%%%%%%%%%%%%%%%%%%%%%%%%%%%%%%%%%
% Abstract
%%%%%%%%%%%%%%%%%%%%%%%%%%%%%%%%%%%%%%%%%%%%%%%%%%%%%%%%%%%%

\begin{abstract}

Synthesizing realistic images involves computing high-dimensional light-transport integrals. In practice, these integrals are numerically estimated via Monte Carlo integration. The error of this estimation manifests itself as conspicuous aliasing or noise. To ameliorate such artifacts and improve image fidelity, we propose a perception-oriented framework to optimize the  error of Monte Carlo rendering. We leverage models based on human perception from the halftoning literature. The result is an optimization problem whose solution distributes the error as visually pleasing blue noise in image space. To find solutions, we present a set of algorithms that provide varying trade-offs between quality and speed, showing substantial improvements over prior state of the art. We perform evaluations using quantitative and error metrics, and provide extensive supplemental material to demonstrate the perceptual improvements achieved by our methods.

\end{abstract}

%%%%%%%%%%%%%%%%%%%%%%%%%%%%%%%%%%%%%%%%%%%%%%%%%%%%%%%%%%%%
% ACM categories, keywords, etc.
%%%%%%%%%%%%%%%%%%%%%%%%%%%%%%%%%%%%%%%%%%%%%%%%%%%%%%%%%%%%

%% The code below is generated by the tool at http://dl.acm.org/ccs.cfm.
%% Please copy and paste the code instead of the example below.
%%



\maketitle

%%%%%%%%%%%%%%%%%%%%%%%%%%%%%%%%%%%%%%%%%%%%%%%%%%%%%%%%%%%%
\section{Introduction}
%%%%%%%%%%%%%%%%%%%%%%%%%%%%%%%%%%%%%%%%%%%%%%%%%%%%%%%%%%%%

Monte Carlo sampling produces approximation error. In rendering, this error can cause visually displeasing image artifacts, unless control is exerted over the correlation of the individual pixel estimates. A standard approach is to decorrelate these estimates by randomizing the samples independently for every pixel, turning potential structured artifacts into white noise.

In digital halftoning, the error induced by quantizing continuous-tone images has been studied extensively. Such studies have shown that a blue-noise distribution of the quantization error is perceptually optimal \citep{ulichney1987Digital}, achieving substantially higher image fidelity than a white-noise distribution. Recent works have proposed empirical means to transfer these ideas to image synthesis~\citep{georgiev16bluenoise,Heitz2019,Heitz2019b,Ahmed:HierarchicalOrdering}. Instead of randomizing the pixel estimates, these methods introduce \emph{negative} correlation between neighboring pixels, exploiting the local smoothness in images to push the estimation error to the high-frequency spectral range.

We propose a theoretical formulation of perceptual error for image synthesis which unifies prior methods in a common framework and formally justifies the desire for blue-noise error distribution. We extend the comparatively simpler problem of digital halftoning~\citep{Book:Halftoning} where the ground-truth image is given, to the substantially more complex one of rendering where the ground truth is the sought result and thus unavailable. Our formulation bridges the gap between multi-tone halftoning and rendering by interpreting Monte Carlo estimates for a pixel as its admissible `quantization levels'. This insight allows virtually any halftoning method to be adapted to rendering. We demonstrate this for the three main classes of halftoning algorithms: dither-mask halftoning, error diffusion halftoning, and iterative energy minimization halftoning.

Existing methods~\citep{georgiev16bluenoise,Heitz2019,Heitz2019b} can be seen as variants of dither-mask halftoning. They distribute pixel error according to masks that are optimized \wrt a target kernel, typically a Gaussian. The kernel can be interpreted as an approximation to the human visual system's point spread function~\citep{Daly1987, Pappas1999}. We revisit the kernel-based perceptual model from halftoning~\citep{Sullivan1991, Allebach1992, Pappas1999} and adapt it to rendering. The resulting energy can be directly used for optimizing Monte Carlo error distribution without the need for a mask. This formulation help us expose the underlying assumptions of existing methods and quantify their limitations. In summary:
\begin{itemize}[leftmargin=3.4mm]
    \setlength\itemsep{0.5mm}
\item
    We formulate an optimization problem for rendering error by leveraging kernel-based perceptual models from halftoning.
\item    
    Our formulation unifies prior blue-noise error distribution methods and makes all their assumptions explicit, outlining general guidelines for devising new methods in a principled manner.
\item 
    Unlike prior methods, our formulation simultaneously optimizes for both the magnitude and the image distribution of pixel error. 
\item
    We devise four different practical algorithms based on iterative minimization, error diffusion, and dithering from halftoning.
\item
    We demonstrate substantial visual improvements over prior art, while using the same input rendering data.
\end{itemize}

%%%%%%%%%%%%%%%%%%%%%%%%%%%%%%%%%%%%%%%%%%%%%%%%%%%%%%%%%%%%
\section{Related Work}
\label{sec:RelatedWork}
%%%%%%%%%%%%%%%%%%%%%%%%%%%%%%%%%%%%%%%%%%%%%%%%%%%%%%%%%%%%

Our work focuses on reducing and optimizing the distribution of Monte Carlo pixel-estimation error. In this section we review prior work with similar goals in digital halftoning (\cref{sec:DigitalHalftoning}) and image synthesis guided by energy-based  (\cref{sec:QuantitativeErrorAssessmentInRendering}) and perception-based (\cref{sec:PercError}) error metrics. We achieve error reduction through careful sample placement and processing, and discuss related rendering approaches (\cref{sec:SampleErrorDistribution}).

%%%%%%%%%%%%%%%%%%%%%%%%%%%%%%%%%%%%%%%%%%%%%%%%%%%%%%%%%%%%
\subsection{Digital halftoning}
\label{sec:DigitalHalftoning}
%%%%%%%%%%%%%%%%%%%%%%%%%%%%%%%%%%%%%%%%%%%%%%%%%%%%%%%%%%%%

Digital halftoning~\citep{Book:Halftoning} involves creating the illusion of continuous-tone images through the arrangement of binary elements; various algorithms target different display devices. \citet{bayer1973Optimum} developed the widely used dispersed-dot ordered-dither patterns. \citet{allebach1976Color} introduced the use of randomness in clustered-dot ordered dithering. \citet{ulichney1987Digital} introduced \emph{blue-noise} patterns that yield better perceptual quality, and \citet{mitsa1991Digital} mimicked those patterns to produce dither arrays (\ie masks) with high-frequency characteristics. \citet{Sullivan1991} developed a Fourier-domain energy function to obtain visually optimal halftone patterns; the optimality is defined \wrt computational models of the human visual system. \citet{Allebach1992} devised a practical algorithm for blue-noise dithering through a spatial-domain interpretation of \citeauthor{Sullivan1991}'s model. Their approach was later refined by \citet{Pappas1999}. 

The void-and-cluster algorithm \citep{Ulichney1993Void} uses a Gaussian kernel to create dither masks with isotropic blue-noise distribution. This approach has motivated various structure-aware halftoning algorithms in graphics~\citep{ostromoukhov2001Simple,Pang2008Structure,chang2009Structure}. In the present work, we leverage the kernel-based model~\citep{Allebach1992,Pappas1999} in the context of Monte Carlo rendering~\citep{Kajiya86}.

%%%%%%%%%%%%%%%%%%%%%%%%%%%%%%%%%%%%%%%%%%%%%%%%%%%%%%%%%%%%
\subsection{Quantitative error assessment in rendering}
\label{sec:QuantitativeErrorAssessmentInRendering}
%%%%%%%%%%%%%%%%%%%%%%%%%%%%%%%%%%%%%%%%%%%%%%%%%%%%%%%%%%%%

It is convenient to measure the error of a rendered image as a single value; vector norms like the mean squared error (MSE) are most commonly used. However, it is widely acknowledged that such simple metrics do not accurately reflect visual quality as they ignore the perceptually important spatial arrangement of pixels. Various theoretical frameworks have been developed in the spatial~\citep{Niederreiter92Random,Kuipers74Uniform} and Fourier~\citep{singh19analysis} domains to understand the error reported through these metrics. The error spectrum ensemble~\citep{celarek2019Quantifying} measures the frequency-space distribution of the error.

Many denoising methods~\citep{Zwicker2015} employ the aforementioned metrics to obtain noise-free results from noisy renderings. Even if the most advanced denoising techniques driven by such metrics can efficiently steer adaptive sampling \citep{Chaitanya2017,Kuznetsov2018,Kaplanyan2019}, they locally determine the number of samples per pixel, ignoring the aspect of their specific layout in screen space.

Our optimization framework employs a perceptual MSE-based metric that accounts for both the magnitude and the spatial distribution of pixel-estimation error. We argue that the spatial sample layout plays a crucial role in the perception of a rendered image; the most commonly used error metrics do not capture this aspect.

%%%%%%%%%%%%%%%%%%%%%%%%%%%%%%%%%%%%%%%%%%%%%%%%%%%%%%%%%%%%
\subsection{Perceptual error assessment in rendering}
\label{sec:PercError}
%%%%%%%%%%%%%%%%%%%%%%%%%%%%%%%%%%%%%%%%%%%%%%%%%%%%%%%%%%%%

The study of the human visual system (HVS) is still ongoing, and well understood are mostly the early stages of the visual pathways from the eye optics, through the retina, to the visual cortex. This limits the scope of existing HVS computational models used in imaging and graphics. Such models should additionally be computationally efficient and generalize over the simplistic stimuli that have been used in their derivation through psychophysical experiments. 

%%%%%%%%%%%%%%%%%%%%%%%%%%%%%%
\Paragraph{Contrast sensitivity function}
%%%%%%%%%%%%%%%%%%%%%%%%%%%%%%

The contrast sensitivity function (CSF) is one of the core HVS models that fulfills the above conditions and comprehensively characterizes overall optical \citep{Westheimer86,deeley:1991} and neural \citep{SOUZA2011} processes in detecting contrast visibility as a function of spatial frequency. While originally modeled as a band-pass filter \citep{Barten:1999,Daly93}, the CSF's shape changes towards a low-pass filter with retinal eccentricity \citep{Robson1981,Peli1991} and reduced luminance adaptation in scotopic and mesopic levels \citep{Mantiuk2020}. Low-pass characteristics are also inherent for chromatic CSFs \citep{Mullen1985,Mantiuk2020,Bolin98}. In many practical imaging applications, \eg JPEG compression \citep{Rashid2005}, rendering \citep{Ramasubramanian99}, or halftoning \citep{Pappas1999},  the CSF is modeled as a low-pass filter, which also allows for better control of image intensity. By normalizing such a CSF by the maximum contrast-sensitivity value, a unitless function akin to the modulation transfer function (MTF) can be derived \citep{Daly1987,Mannos1974,Mantiuk2005,Sullivan1991,SOUZA2011} that after transforming from the frequency to the spatial domain results in the point spread function (PSF) \citep{Allebach1992,Pappas1999}. Following \citet{Pappas1999}, we approximate such a PSF by a Gaussian filter; the resulting error is practically negligible for a pixel density of 300 dots per inch (dpi) and observer-to-screen distance larger than 60\,cm.

%%%%%%%%%%%%%%%%%%%%%%%%%%%%%%
\Paragraph{Advanced quality metrics}
%%%%%%%%%%%%%%%%%%%%%%%%%%%%%%

More costly, and often less robust, modeling of the HVS beyond the CSF is performed in advanced quality metrics \citep{Lubin95,Daly93,Mantiuk2011}. Such metrics have been adapted to rendering to guide the computation to image regions where the visual error is most strongly perceived \citep{bolin1995Frequency,Bolin98,Ramasubramanian99,ferwerda1996Model,myszkowski1998Visible,volevich2000Using}. An important application is visible noise reduction in path tracing via content-adaptive sample-density control \citep{bolin1995Frequency,Bolin98,Ramasubramanian99}. Our framework enables significant reduction of noise visibility for the same sampling budget.

%%%%%%%%%%%%%%%%%%%%%%%%%%%%%%%%%%%%%%%%%%%%%%%%%%%%%%%%%%%%
\subsection{Blue-noise error distribution in rendering}
\label{sec:SampleErrorDistribution}
%%%%%%%%%%%%%%%%%%%%%%%%%%%%%%%%%%%%%%%%%%%%%%%%%%%%%%%%%%%%

\citet{mitchell1991Spectrally} first observed that high-frequency error distribution is desirable for stochastic rendering. Only recently, \citet{georgiev16bluenoise} adopted techniques from halftoning to correlate pixel samples in screen space and distribute path-tracing error as blue noise, with substantial perceptual quality improvements. \citet{Heitz2019b} built on this idea to develop a progressive quasi-Monte Carlo sampler that further improves quality. \citet{Ahmed:HierarchicalOrdering} proposed a technique to coordinate quasi-Monte Carlo samples in screen space inspired by error diffusion.
 
Motivated by the results of \citet{georgiev16bluenoise}, \citet{Heitz2019} devised a method to directly optimize the distribution of pixel estimates, without operating on individual samples. Their pixel permutation strategy fits the initially white-noise pixel intensities to a prescribed blue-noise mask. This approach scales well with sample count and dimension, though its reliance on prior pixel estimates makes it practical only for animation rendering where it is susceptible to quality degradation.

We propose a perceptual error framework that unifies these two general approaches, exposing the assumptions of existing methods and providing guidelines to alleviate some of their drawbacks.

%%%%%%%%%%%%%%%%%%%%%%%%%%%%%%%%%%%%%%%%%%%%%%%%%%%%%%%%%%%%
\section{Perceptual error model}
\label{sec:Model}
%%%%%%%%%%%%%%%%%%%%%%%%%%%%%%%%%%%%%%%%%%%%%%%%%%%%%%%%%%%%

Our aim is to produce Monte Carlo renderings that, at a fixed sampling rate, are perceptually as close to the ground truth as possible. This goal requires formalizing the perceptual image error along with an optimization problem that minimizes it. In this section, we build a perceptual model upon the extensive studies done in the halftoning literature. We will discuss how to efficiently solve the resulting optimization problem in \cref{sec:Optimization}.

Given a ground-truth image $\ReferenceImage$ and its quantized or stochastic approximation $\EstimatedImage$, we denote the (signed) error image by
\begin{equation}
    \label{eq:ImageErrorHalftoning}
    \ErrorImage = \EstimatedImage - \ReferenceImage.
\end{equation}
To minimize the error, it is convenient to quantify it as a single value. A common approach is to take the $\Lnorm{1}$, $\Lnorm{2}$, or $\Lnorm{\infty}$ norm of the image $\ErrorImage$ interpreted as a vector. Such simple metrics are permutation-invariant, \ie they account for the \emph{magnitudes} of individual pixel errors but not for their \emph{distribution} over the image. This distribution is an important factor for the perceived fidelity, since contrast perception is an inherently spatial characteristic of the HVS (\cref{sec:PercError}). Our model is based on perceptual halftoning metrics that capture both the magnitude and the distribution of error.

%%%%%%%%%%%%%%%%%%%%%%%%%%%%%%
\begin{figure}[t!]
    \centering
    %!TEX root = ../paper.tex

\footnotesize
\hspace*{-1em}
\begin{tabular}{c@{\;}c@{\;}c@{\;}c@{}}
Image & Image spectrum & Kernel spectrum & Product spectrum 
\\[0.2mm]
\includegraphics[width=0.813in,page=1]{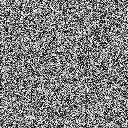}
&
\includegraphics[width=0.813in,page=1]{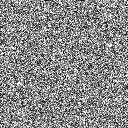}&
\includegraphics[width=0.813in,page=1]{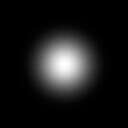}
&
\includegraphics[width=0.813in,page=1]{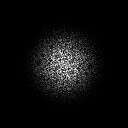}
\\[-0.4mm]
${\WhiteErrorImage}$ & $\WhiteErrorImagePowerSpectrum$ & $\KernelPowerSpectrum$ & $\KernelPowerSpectrum \CwiseProduct \WhiteErrorImagePowerSpectrum$
\\[1mm]
\includegraphics[width=0.813in,page=1]{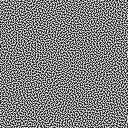}
&
\includegraphics[width=0.813in,page=1]{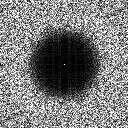}
&
\includegraphics[width=0.813in,page=1]{{figures/product-blue-white-gaussian/power-gaussian-gx0.5-gy0.5-sx0.01-sy0.01-regular-n16384-128x128}}
&
\includegraphics[width=0.813in,page=1]{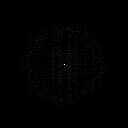}
\\[-0.4mm]
${\BlueErrorImage}$ & $\BlueErrorImagePowerSpectrum$ & $\KernelPowerSpectrum$ & $\KernelPowerSpectrum \CwiseProduct \BlueErrorImagePowerSpectrum$
\end{tabular}%

    \vspace{-2mm}
    \caption{
        Error images $\WhiteErrorImage$ and $\BlueErrorImage$ with respective white-noise, $\WhiteErrorImagePowerSpectrum$, and blue-noise, $\BlueErrorImagePowerSpectrum$, power spectra. For a low-pass kernel $\Kernel$ modeling the PSF of the HVS (here a Gaussian with std.\,dev.\ $\sigma = 1$), the product of its spectrum $\KernelPowerSpectrum$ with $\BlueErrorImagePowerSpectrum$ has lower magnitude than the product with $\WhiteErrorImagePowerSpectrum$. This corresponds to lower perceptual sensitivity to $\BlueErrorImage$, even though $\WhiteErrorImage$ has the same amplitude as it is obtained by randomly permuting the pixels of $\BlueErrorImage$.
    }
    \label{fig:product-blue-white-gaussian}
\end{figure}
%%%%%%%%%%%%%%%%%%%%%%%%%%%%%%

%%%%%%%%%%%%%%%%%%%%%%%%%%%%%%%%%%%%%%%%%%%%%%%%%%%%%%%%%%%%
\subsection{Motivation}
\label{sec:Motivation}
%%%%%%%%%%%%%%%%%%%%%%%%%%%%%%%%%%%%%%%%%%%%%%%%%%%%%%%%%%%%

 Halftoning metrics model the processing done by the HVS as a convolution of the error image $\ErrorImage$ with a kernel~$\Kernel$:
\begin{align}
    \label{eq:KernelErorr}
    \Energy
        \,=\, \Norm{\Kernel * \ErrorImage}^2_2
        \,=\, \Norm{\KernelSpectrum \,\CwiseProduct\, \ErrorImageSpectrum}^2_2
        \,=\, \InnerProd{\KernelPowerSpectrum\!\!, \, \ErrorImagePowerSpectrum}.
\end{align}
The convolution is equivalent to the element-wise product of the corresponding Fourier spectra $\KernelSpectrum$ and $\ErrorImageSpectrum$, whose 2-norm in turn equals the inner product of the power spectra images $\KernelPowerSpectrum$ and $\ErrorImagePowerSpectrum$. \citet{Sullivan1991} optimized the error image $\ErrorImage$ to minimize the error~\eqref{eq:KernelErorr} \wrt a kernel $\Kernel$ that approximates the HVS's modulation transfer function $\Abs{\KernelSpectrum}$ (MTF)~\citep{Daly1987}. \citet{Allebach1992} used a similar model in the spatial domain with a kernel that approximates the PSF\footnote{The MTF is the magnitude of the Fourier transform of the PSF.} of the human eye. That kernel is low-pass, and the optimization naturally yields blue-noise\footnote{The term ``blue noise'' is often used loosely to refer to any isotropic spectrum with minimal low-frequency content and no concentrated energy spikes.} distribution in the error image~\citep{Allebach1992}, as we show later in \cref{fig:Kernels}. The blue-noise distribution can thus be seen as byproduct of the optimization which pushes the spectral components of the error to the frequencies least visible to the human eye (see \cref{fig:product-blue-white-gaussian}).

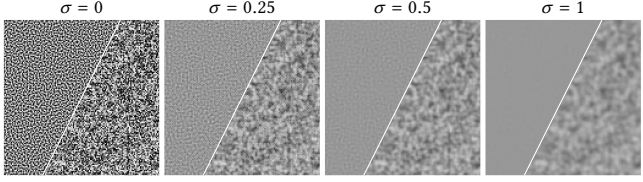
\begin{figure}[t]
    \centering
    %!TEX root = ../paper.tex

\newcommand{\Crops}[1]{%
    \begin{tikzpicture}[x=0.813in,y=0.813in]
        \path[fill overzoom image={figures/noise-viewing-distance/8size_white-#1}] (0.25,0) -- (1,0) -- (1,1) -- (0.75,1) -- cycle;
        \path[fill overzoom image={figures/noise-viewing-distance/8size_blue-#1}]  (0,0) -- (0.25,0) -- (0.75,1) -- (0,1) -- cycle;
        \draw[draw=white,thin] (0.25,0) -- (0.75,1);
    \end{tikzpicture}%
}

\footnotesize
\hspace*{-1em}
\begin{tabular}{c@{\;}c@{\;}c@{\;}c@{}}
    $\sigma=0$ & $\sigma=0.25$ & $\sigma=0.5$ & $\sigma=1$
    \\
    \Crops{0} & \Crops{0.25} & \Crops{0.5} & \Crops{1}
\end{tabular}

\let\Crops\SomeUndefinedMacroBrotha
    \vspace{-2.5mm}
    \caption{
        The appearance of blue noise (left images) converges to a constant image faster than white noise (right images) with increasing observer \mbox{distance}, here emulated via the standard deviation $\sigma$ of a Gaussian kernel. We provide a formal connection between $\sigma$ and the viewing distance in~\cref{sec:Extensions}.
    }
    \label{fig:blue-noise-steady-state}
\end{figure}

To better understand the spatial aspects of contrast sensitivity in the HVS, the MTF is usually modeled over a range of viewing distances~\citep{Daly93}. This is done to account for the fact that with increasing viewer distance, spatial frequencies in the image are projected to higher spatial frequencies onto the retina. These frequencies eventually become invisible, filtered out by the PSF which expands its corresponding kernel in image space. We recreate this experiment to see the impact of distance on the image error. In~\cref{fig:blue-noise-steady-state}, we convolve white- and blue-noise distributions with a Gaussian kernel of increasing standard deviation corresponding to increasing observer-to-screen distance. The high-frequency blue-noise distribution reaches a homogeneous state (where the tone appears constant) faster compared to the all-frequency white noise. This means that high-frequency error becomes indiscernible at closer viewing distances, where the HVS ideally has not yet started filtering out actual image detail which is typically low- to mid-frequency. In~\cref{sec:Extensions} we discuss how the kernel's standard deviation encodes the viewing distance \wrt to the screen resolution.

%%%%%%%%%%%%%%%%%%%%%%%%%%%%%%
%%%%%%%%%%%%%%%%%%%%%%%%%%%%%%

%%%%%%%%%%%%%%%%%%%%%%%%%%%%%%%%%%%%%%%%%%%%%%%%%%%%%%%%%%%%
\subsection{Our model}
\label{sec:RenderingModel}
%%%%%%%%%%%%%%%%%%%%%%%%%%%%%%%%%%%%%%%%%%%%%%%%%%%%%%%%%%%%

In rendering, the value of each pixel $\PixelIndex$ is a light-transport integral. Point-sampling its integrand with a sample set $\SampleSetPixel$ yields a pixel estimate $\EstimatedPixel_\PixelIndex(\SampleSetPixel)$. The signed pixel error is thus a function of the sample set: $\ErrorPixel_\PixelIndex(\SampleSetPixel) = \EstimatedPixel_\PixelIndex(\SampleSetPixel) - \ReferencePixel_\PixelIndex$, where $\ReferencePixel_\PixelIndex$ is the reference (\ie ground-truth) pixel value. The error of the entire image can be~written~as
\begin{equation}
    \label{eq:ImageError}
    \ErrorImage(\SampleSetImage) = \EstimatedImage(\SampleSetImage) - \ReferenceImage,
\end{equation}
where $\SampleSetImage = \{ \SampleSetPixel_1, \ldots, \SampleSetPixel_\PixelCount \}$ is an ``image'' containing the sample set for all $\PixelCount$ pixels. With these definitions, we can express the perceptual error in~\cref{eq:KernelErorr} for the case of Monte Carlo rendering as a function of the sample-set image~$\SampleSetImage$, given a kernel $\Kernel$:
\begin{equation}
    \label{eq:PerceptualError}
    \EnergyS(\SampleSetImage) = \Norm{\Kernel * \ErrorImage(\SampleSetImage)}_2^2.
\end{equation}

Our goal is to minimize the perceptual error~\eqref{eq:PerceptualError}. We formulate this task as an optimization problem:
\begin{tcolorbox}[ams equation,after=,]
    \label{eq:MinimizationProblem}
    \min_{\SampleSetImage \in \, \SampleSetImageDomain} \, \EnergyS(\SampleSetImage)
    \,=\; \min_{\SampleSetImage \in \, \SampleSetImageDomain} \, \Norm{\Kernel * (\EstimatedImage(\SampleSetImage) - \ReferenceImage)}_2^2.
\end{tcolorbox}
The minimizing sample-set image $\SampleSetImage$ yields an image estimate $\EstimatedImage(\SampleSetImage)$ that is closest to the reference $\ReferenceImage$ \wrt the kernel $\Kernel$. The search space $\SampleSetImageDomain$ is the set of all possible locations for every sample of every pixel. The total number of samples in $\SampleSetImage$ is typically bounded by a given target sampling budget. Practical considerations may also restrict the search space $\SampleSetImageDomain$, as we will exemplify in the following section.

Note that the classical MSE metric corresponds to using a zero-width (\ie one-pixel) kernel $\Kernel$ in \cref{eq:PerceptualError}. However, the MSE accounts only for the magnitude of the error $\ErrorImage$, while using wider kernels (such as the PSF) accounts for both magnitude and distribution. Consequently, while the MSE can be minimized by optimizing pixels independently, minimizing the perceptual error requires coordination between pixels. In the following section, we devise strategies for solving this optimization problem.

%%%%%%%%%%%%%%%%%%%%%%%%%%%%%%%%%%%%%%%%%%%%%%%%%%%%%%%%%%%%
\section{Discrete optimization}
\label{sec:Optimization}
%%%%%%%%%%%%%%%%%%%%%%%%%%%%%%%%%%%%%%%%%%%%%%%%%%%%%%%%%%%%

In our optimization problem~\eqref{eq:MinimizationProblem}, the search space for each sample in every pixel is a high-dimensional unit hypercube. Every point in this so-called primary sample space maps to a light-transport path in the scene~\citep{Pharr16Physically}. Optimizing for the sample-set image $\SampleSetImage$ thus entails evaluating the contributions $\EstimatedImage(\SampleSetImage)$ of all corresponding paths. This evaluation is costly, and for any non-trivial scene, $\EstimatedImage$ is a function with complex shape and many discontinuities. This precludes us from studying all (uncountably infinite) sample locations in practice.

To make the problem tractable, we restrict the search in each pixel to a finite number of (pre-defined) sample sets. We devise two variants of the resulting discrete optimization problem, which differ in their definition of the global search space $\SampleSetImageDomain$. In the first variant, each pixel has a separate list of sample sets to choose from (``\vertical'' search space). The setting is similar to that of (multi-tone) halftoning~\citep{Book:Halftoning}, which allows us to import classical optimization techniques from that field, such as iterative minimization, error diffusion, and mask-based dithering. In the second variant, each pixel has one associated sample set, and the search space comprises permutations of these assignments (``\horizontal'' search space). We develop a greedy iterative optimization method for this second variant.

In contrast to halftoning, in our setting the ground-truth image $\ReferenceImage$---required to compute the error image $\ErrorImage$ during optimization---is not readily available. Below we describe our algorithms assuming the ground truth is available; in \cref{sec:Applications} we will discuss how to substitute it with a surrogate to make the algorithms practical.

%%%%%%%%%%%%%%%%%%%%%%%%%%%%%%%%%%%%%%%%%%%%%%%%%%%%%%%%%%%%
\subsection{\Vertical search space}
\label{sec:VerticalOptimization}
%%%%%%%%%%%%%%%%%%%%%%%%%%%%%%%%%%%%%%%%%%%%%%%%%%%%%%%%%%%%

Our first variant considers a ``\vertical'' search space where the sample set for each of the $\PixelCount$ image pixels is one of $\ImageCount$ given sets:\footnote{For notational simplicity, and without loss of generality, we assume that the number of candidate sample sets $\ImageCount$ is the same for all pixels; in practice can vary per pixel.}
\begin{equation}
    \label{eq:VerticalOptimizationSpace}
    \SampleSetImageDomain = \left\{ \SampleSetImage = \{ \SampleSetPixel_1, \ldots, \SampleSetPixel_\PixelCount \} : \SampleSetPixel_\PixelIndex \in \{ \SampleSetPixel_{\PixelIndex,1}, \ldots, \SampleSetPixel_{\PixelIndex,\ImageCount} \} \right\}.
\end{equation}
The objective is to find a sample set $\SampleSetPixel_\PixelIndex$ for every pixel $\PixelIndex$ such that all resulting pixel estimates together minimize the perceptual error~\eqref{eq:PerceptualError}.
\WrapFigureTemplate{0.53}{3.5mm}{-3mm}{-5mm}{
    \begin{overpic}[abs,unit=0.25mm,scale=0.64]{figures/vertical-search-space.ai}
        \put(161,55){\small\Outlined[white][black]{$\OutputPixel_1$}}
        \put(45.5,59){\small\Outlined[white][black]{$\EstimatedPixel_{1,\ImageCount}$}}
        \put(156,23.5){\small\Outlined[white][black]{$\OutputPixel_2$}}
        \put(34,5){\small\Outlined[white][black]{$\EstimatedPixel_{2,1}$}}
        \put(126,18.5){\small\Outlined[white][black]{$\OutputPixel_3$}}
        \put(4,20){\small\Outlined[white][black]{$\EstimatedPixel_{3,2}$}}
    \end{overpic}
}%
This is equivalent to directly optimizing over the $\ImageCount$ possible estimates $\EstimatedPixel_{\PixelIndex,1}, \ldots, \EstimatedPixel_{\PixelIndex,\ImageCount}$ for each pixel, with $\EstimatedPixel_{\PixelIndex,\PixelIndexTwo} = \EstimatedPixel_\PixelIndex (\SampleSetPixel_{\PixelIndex,\PixelIndexTwo})$. These estimates can be obtained by pre-rendering a stack of $\ImageCount$ images $\EstimatedImage_\PixelIndexTwo = \{ \EstimatedPixel_{1,\PixelIndexTwo}, \ldots, \EstimatedPixel_{\PixelCount,\PixelIndexTwo} \}$, for $\PixelIndexTwo = 1..\ImageCount$. The resulting minimization problem reads:
\begin{align}
    \label{eq:VerticalOptimization}
    \min_{\OutputImage \, : \, \OutputPixel_\PixelIndex \in \, \{ \EstimatedPixel_{\PixelIndex,1}, \ldots, \EstimatedPixel_{\PixelIndex,\ImageCount} \}}\Norm{\Kernel * (\OutputImage - \ReferenceImage)}^2_2. 
\end{align}
This problem is almost identical to that of multi-tone halftoning. The difference is that in our setting the ``quantization levels'', \ie the pixel estimates, are distributed non-uniformly and vary per pixel as they are not fixed but are the result of point-sampling a light-transport integral. This similarity allows us to directly apply existing optimization techniques from halftoning. We consider three such methods, which we outline in  \cref{alg:VerticalOptimization} and describe next.

%%%%%%%%%%%%%%%%%%%%%%%%%%%%%%
\Paragraph{Iterative minimization}
%%%%%%%%%%%%%%%%%%%%%%%%%%%%%%

State-of-the-art halftoning methods attack the problem~\eqref{eq:VerticalOptimization} directly via greedy iterative minimization \citep{Allebach1992,Pappas1999}. After initializing every pixel to a random quantization level, we traverse the image in serpentine order (as is standard practice in halftoning) and for each pixel choose the level that minimizes the energy. Several full-image iterations are performed; in our experiments convergence to a local minimum is achieved within 10--20 iterations.

As a further improvement, the optimization can be terminated when no pixels are updated within one full iteration, or when the perceptual-error reduction rate drops below a certain threshold. Traversing the pixels in random order allows terminating at any point but converges slightly slower.

%%%%%%%%%%%%%%%%%%%%%%%%%%%%%%
\Paragraph{Error diffusion}
%%%%%%%%%%%%%%%%%%%%%%%%%%%%%%

A classical halftoning algorithm, error diffusion scans the image pixel by pixel, snapping each reference value to the closest quantization level and distributing the resulting pixel error to yet-unprocessed nearby pixels according to a given kernel $\DiffusionKernel$. We use the empirically derived kernel of \citet{floyd1976adaptive} which has been shown to produce an output that approximately minimizes \cref{eq:VerticalOptimization} \citep{hocevar2008reinstating}. Error diffusion is faster than iterative minimization but yields less optimal solutions.

%%%%%%%%%%%%%%%%%%%%%%%%%%%%%%
\Paragraph{Dithering}
%%%%%%%%%%%%%%%%%%%%%%%%%%%%%%

The fastest halftoning approach quantizes pixel values using thresholds stored in a pre-computed dither mask (or matrix) \citep{spaulding1997methods}. For each pixel, the two quantization levels that tightly envelop the reference value (in terms of brightness) are found, and one of the two is chosen based on the threshold assigned to the pixel by the mask.

Dithering can be understood as performing the perceptual error minimization in two steps. First, an offline optimization encodes the error distribution optimal for the target kernel $\Kernel$ into a mask. Then, for a given image, the error magnitude is minimized by restricting the quantization to the two closest levels per pixel, and the mask-driven choice between them applies the target distribution of error.

%%%%%%%%%%%%%%%%%%%%%%%%%%%%%%
\begin{pseudocode}[t]
    \caption{
        Three algorithms to (approximately) solve the \vertical search space optimization problem~\eqref{eq:VerticalOptimization}. The output is an image $\OutputImage = \{ \OutputPixel_1, \ldots, \OutputPixel_\PixelCount \}$, given a reference image $\ReferenceImage$ and a stack of initial image estimates $\EstimatedImage_1, \ldots, \EstimatedImage_\ImageCount$. Iterative minimization updates pixels repeatedly, for each selecting the estimate that minimizes the perceptual error~\eqref{eq:PerceptualError}. Error diffusion quantizes each pixel to the closest estimate, distributing the error to its neighbors based on a kernel $\DiffusionKernel$. Dithering quantizes each pixel in $\ReferenceImage$ based on thresholds looked up in a dither mask $\DitherMask$ (optimized \wrt the kernel $\Kernel$).
    }
    \vspace{-2.2mm}
    \label{alg:VerticalOptimization}
    \hrule height 0.6pt
	\begin{algorithmic}[1]
        \Function{IterativeMinimization}{$\Kernel$, $\ReferenceImage$, $\EstimatedImage_1$, \ldots, $\EstimatedImage_\ImageCount$, $\OutputImage$, $T$}
		    \State $\OutputImage = \{\EstimatedPixel_{1,\text{rand}},\ldots,\EstimatedPixel_{\PixelCount, \text{rand}}\}$ \AlgCommentLeft{Init each pixel to random estimate}\vspace{0.5mm}
		    \For{$T$ iterations}
		      \For{pixel $\PixelIndex = 1..\PixelCount$} \AlgCommentLeft{\Eg random or serpentine order}
		          \For{estimate $\EstimatedPixel_{\PixelIndex,\PixelIndexTwo} \in \{ \EstimatedPixel_{\PixelIndex, 1}, \ldots, \EstimatedPixel_{\PixelIndex, \ImageCount} \}$}
        	            \If{$\OutputPixel_\PixelIndex == \EstimatedPixel_{\PixelIndex,\PixelIndexTwo}$ reduces $\Norm{\Kernel * (\OutputImage - \ReferenceImage)}_2^2$}
        	                \State $\OutputPixel_\PixelIndex = \EstimatedPixel_{\PixelIndex,\PixelIndexTwo}$ \AlgCommentLeft{Update estimate}
        	            \EndIf
        	        \EndFor    
		        \EndFor
		    \EndFor
        \EndFunction{}
        \vspace{1mm}
		\Function{ErrorDiffusion}{$\DiffusionKernel$, $\ReferenceImage$, $\EstimatedImage_1$, \ldots, $\EstimatedImage_\ImageCount$, $\OutputImage$}
		    \State $\OutputImage= \ReferenceImage$ \AlgCommentLeft{Initialize solution to reference}
	      \For{pixel $\PixelIndex = 1..\PixelCount$} \AlgCommentLeft{\Eg serpentine order}
	          \State $\OutputPixel_{\PixelIndex}^\text{old} = \OutputPixel_{\PixelIndex}$
  	          \State $\OutputPixel_\PixelIndex \in \argmin_{\EstimatedPixel_{\PixelIndex,\PixelIndexTwo}}\Norm{\OutputPixel_{\PixelIndex}^\text{old} - \EstimatedPixel_{\PixelIndex,\PixelIndexTwo}}^2_2$
  	          \State $\ErrorPixel_\PixelIndex = \OutputPixel_{\PixelIndex}^\text{old} -  \OutputPixel_{\PixelIndex}$ \AlgCommentDown{Diffuse error $\ErrorPixel_\PixelIndex$ to yet-unprocessed neighbors}
	          \For{unprocessed pixel $\PixelIndexThree$ within support of $\DiffusionKernel$ around $\PixelIndex$}
	                \State $\OutputPixel_{\PixelIndexThree} \mathrel{{+}{=}} \ErrorPixel_\PixelIndex \cdot \DiffusionKernelPixel_{\PixelIndexThree-i}$
    	        \EndFor    
	        \EndFor
        \EndFunction{}
        \vspace{1mm}
		\Function{Dithering}{$\DitherMask$, $\ReferenceImage$, $\EstimatedImage_1$, \ldots, $\EstimatedImage_\ImageCount$, $\OutputImage$}
            \For{pixel $\PixelIndex = 1..\PixelCount$} \AlgCommentDown{Find tightest interval $[\EstimatedPixel_\PixelIndex^\text{low},\EstimatedPixel_\PixelIndex^\text{high}]$}
                \State $\EstimatedPixel_\PixelIndex^\text{lower} = \arg\max_{\EstimatedPixel_{\PixelIndex,\PixelIndexTwo}\,:\,\Greyscale{ \EstimatedPixel_{\PixelIndex,\PixelIndexTwo}} \,\leq\, \Greyscale{\ReferencePixel_\PixelIndex}}\Greyscale{ \EstimatedPixel_{\PixelIndex,\PixelIndexTwo}}$ \AlgComment{containing $\ReferencePixel_\PixelIndex$}
                \State $\EstimatedPixel_\PixelIndex^\text{upper} = \argmin_{\,\EstimatedPixel_{\PixelIndex,\PixelIndexTwo}\,:\,\Greyscale{ \EstimatedPixel_{\PixelIndex,\PixelIndexTwo}} \,>\,\Greyscale{\ReferencePixel_\PixelIndex}}\Greyscale{ \EstimatedPixel_{\PixelIndex,\PixelIndexTwo}}$
                \vspace{0.7mm}
                \If{$\Greyscale{\ReferencePixel_\PixelIndex} - \Greyscale{\EstimatedPixel_\PixelIndex^\text{lower}} \, < \, \DitherMaskPixel_\PixelIndex \cdot \big(\Greyscale{\EstimatedPixel_\PixelIndex^\text{upper}}-\Greyscale{\EstimatedPixel_\PixelIndex^\text{low}}\big)$}
                   \State $\OutputPixel_\PixelIndex = \EstimatedPixel_\PixelIndex^\text{lower}$
                   \AlgCommentUp{Set $\OutputPixel_\PixelIndex$ to  $\EstimatedPixel_\PixelIndex^\text{lower}$ or $\EstimatedPixel_\PixelIndex^\text{upper}$ using threshold $\DitherMaskPixel_\PixelIndex$}
                \Else
                    \State $\OutputPixel_\PixelIndex = \EstimatedPixel_\PixelIndex^\text{upper}$
                \EndIf
            \EndFor
        \EndFunction{}
    \end{algorithmic}
    \hrule height 0.6pt
\end{pseudocode}
%%%%%%%%%%%%%%%%%%%%%%%%%%%%%%

%%%%%%%%%%%%%%%%%%%%%%%%%%%%%%%%%%%%%%%%%%%%%%%%%%%%%%%%%%%%
\subsection{\Horizontal search space}
\label{sec:HorizontalOptimization}
%%%%%%%%%%%%%%%%%%%%%%%%%%%%%%%%%%%%%%%%%%%%%%%%%%%%%%%%%%%%

We now describe the second, ``\horizontal'' discrete variant of our minimization formulation~\eqref{eq:MinimizationProblem}. It considers a single sample set $\SampleSetPixel_\PixelIndex$ assigned to each of the $\PixelCount$ pixels, all represented together as a sample-set image $\SampleSetImage$. The search space comprises all possible permutations $\AllPermutations(\SampleSetImage)$ of these assignments:
\begin{equation}
    \label{eq:HorizontalOptimizationSpace}
    \SampleSetImageDomain = \AllPermutations(\SampleSetImage),
    \;\text{with}\;\,
    \SampleSetImage = \{ \SampleSetPixel_1, \ldots, \SampleSetPixel_\PixelCount \}.
\end{equation}
The goal is to find a permutation $\Permutation(\SampleSetImage)$ that minimizes the perceptual error~\eqref{eq:PerceptualError}. The optimization problem~\eqref{eq:MinimizationProblem} thus takes the form
\begin{align}
    \label{eq:HorizontalOptimization}
    \min_{\Permutation \in \, \AllPermutations(\SampleSetImage)} \Norm{\Kernel * (\EstimatedImage(\Permutation(\SampleSetImage)) - \ReferenceImage)}^2_2.
\end{align}
We can explore the permutation space $\AllPermutations(\SampleSetImage)$ by swapping the sample-set assignments between pixels. The minimization requires
\WrapFigure{figures/horizontal-search-space.ai}{0.27}{4mm}{-3.7mm}{-8mm}%
updating the image estimate $\EstimatedImage(\Permutation(\SampleSetImage))$ for each permutation $\Permutation(\SampleSetImage)$, \ie after every swap. Such updates are costly as they involve re-sampling both pixels in each of potentially millions of swaps. We need to eliminate these extra rendering invocations during the optimization to make it practical. To that end, we observe that for pixels solving similar light-transport integrals, swapping their sample sets gives a similar result to swapping their estimates. We therefore restrict the search space to permutations that can be generated through swaps between such (similar) pixels. This enables an efficient optimization scheme that directly swaps the pixel estimates of an initial rendering $\EstimatedImage(\SampleSetImage)$.

%%%%%%%%%%%%%%%%%%%%%%%%%%%%%%
\Paragraph{Error decomposition}
%%%%%%%%%%%%%%%%%%%%%%%%%%%%%%

Formally, we express the estimate produced by a sample-set permutation in terms of permuting the pixels of the initial rendering: $\EstimatedImage(\Permutation(\SampleSetImage)) = \Permutation(\EstimatedImage(\SampleSetImage)) + \ErrorDifference(\Permutation)$. The error $\ErrorDifference$ is zero when the swapped pixels solve the same integral. Substituting into \cref{eq:HorizontalOptimization}, we can approximate the perceptual error by (see \cref{app:error_decomposition})
\begin{subequations}
\begin{align}
    \label{eq:HorizontalOptimizationEnergy}
    \!\!\Energy(\Permutation)
    &= 
    \Norm{\Kernel * (\Permutation(\EstimatedImage(\SampleSetImage)) - \ReferenceImage \, + \, \ErrorDifference(\Permutation))}^2_2 \\
    \label{eq:HorizontalOptimizationEnergyApprox}
    &\approx
    \Norm{\Kernel * (\Permutation(\EstimatedImage(\SampleSetImage)) - \ReferenceImage)}^2_2 \, 
    + \Norm{\Kernel}^2_1 \sum_{\PixelIndex}\DissimilarityMetric(\PixelIndex,\Permutation(\PixelIndex))
    = \EnergyApprox(\Permutation), \,\,
\end{align}
\end{subequations}
where we write the error $\Energy(\Permutation)$ as a function of $\Permutation$ only, to emphasize that everything else is fixed during the optimization. In the approximation $\EnergyApprox$, the term $\DissimilarityMetric(\PixelIndex,\Permutation(\PixelIndex))$ measures the dissimilarity between pixel $\PixelIndex$ and the pixel  $\Permutation(\PixelIndex)$ it is relocated to by the permutation. The purpose of this metric is to predict how different we expect the result of re-estimating the pixels after swapping their sample sets to be compared to simply swapping their initial estimates. It can be constructed based on knowledge or assumptions about the image.

%%%%%%%%%%%%%%%%%%%%%%%%%%%%%%
\Paragraph{Local similarity assumption}
%%%%%%%%%%%%%%%%%%%%%%%%%%%%%%

Our implementation uses a simple binary dissimilarity function that returns zero when $\PixelIndex$ and $\Permutation(\PixelIndex)$ are within some distance $r$ and infinity otherwise. We set $r \in [1,3]$; it should ideally be locally adapted to the image smoothness. This allows us to restrict the search space $\AllPermutations(\SampleSetImage)$ only to permutations that swap adjacent pixels where it is more likely that $\ErrorDifference$ is small. More elaborate heuristics could better account for pixel (dis)similarity.

%%%%%%%%%%%%%%%%%%%%%%%%%%%%%%
\Paragraph{Iterative minimization}
%%%%%%%%%%%%%%%%%%%%%%%%%%%%%%

We devise a greedy iterative minimization scheme for this \horizontal formulation, similar to the one in \cref{alg:VerticalOptimization}. Given an initial image estimate $\EstimatedImage(\SampleSetImage)$, produced by randomly assigning a sample set to every pixel, our algorithm goes over all pixels and for each considers swaps within a $(2\TileRadius+1)^2$ neighborhood; we use $\TileRadius=1$. The swap that brings the largest reduction in the perceptual error $\EnergyApprox$ is accepted. \Cref{alg:HorizontalOptimization} provides pseudocode. In our experiments we run $T=10$ full-image iterations. As before, the algorithm could be terminated based on the swap reduction rate or the error reduction rate. We explore additional optimizations in supplemental 
\ifdefined\arXiv
Section 3.
\else 
\cref{sec:supp_iterative_minimization_optimization}.
\fi

The parameter $\TileRadius$ balances between the cost of one iteration and the amount of exploration it can do. Note that this parameter is different from the maximal relocation distance $r$ in the dissimilarity metric, with $\TileRadius \leq r$.

Due to the pixel (dis)similarity assumptions, the optimization can produce some mispredictions, \ie it may swap the estimates of pixels for which swapping the sample sets produces a significantly different result. Thus the image $\Permutation(\EstimatedImage(\SampleSetImage))$ cannot be used directly as a final estimate. We therefore re-render the image using the optimized permutation $\Permutation$ to obtain the final estimate $\EstimatedImage(\Permutation(\SampleSetImage))$.

%%%%%%%%%%%%%%%%%%%%%%%%%%%%%%
\begin{pseudocode}[t]
    \caption{
        Given a convolution kernel $\Kernel$, a reference image $\ReferenceImage$, an initial sample-set assignment $\SampleSetImage$, and an image estimate $\EstimatedImage(\SampleSetImage)$ computed with that assignment, our greedy algorithm iteratively swaps sample-set assignments between neighboring pixels to minimize the perceptual error $\EnergyApprox$~\eqref{eq:HorizontalOptimizationEnergyApprox}, producing a permutation $\Permutation$ of the initial assignment.
    }
    \vspace{-2.2mm}
    \label{alg:HorizontalOptimization}
    \hrule height 0.6pt
	\begin{algorithmic}[1]
		\Function{IterativeMinimization}{$\Kernel$, $\ReferenceImage$, $\SampleSetImage$, $\EstimatedImage(\SampleSetImage)$, $T$, $\TileRadius$, $\Permutation$}
		    \State $\Permutation = \text{identity permutation}$ \AlgCommentLeft{Initialize solution permutation}
		    \For{$T$ iterations}
		        \For{pixel $\PixelIndex = 1..\PixelCount$} \AlgCommentLeft{\Eg random or serpentine order}
		            \State $\Permutation' = \Permutation$ \AlgCommentDown{Find best pixel in neighborhood to swap with}
    		        \For{pixel $\PixelIndexTwo$ in $(2\TileRadius\!+\!1)^2$ neighborhood around $\PixelIndex$}
        	            \If{$\EnergyApprox(\Permutation_{\PixelIndex \,\leftrightarrows\, \PixelIndexTwo}(\SampleSetImage)) < \EnergyApprox(\Permutation'(\SampleSetImage))$} \AlgCommentLeft{\cref{eq:HorizontalOptimizationEnergyApprox}}
        	                \State $\Permutation' = \Permutation_{\PixelIndex \leftrightarrows \PixelIndexTwo}$ \AlgCommentLeft{Accept swap as current best} \vspace{-0.85mm}
        	            \EndIf
    		        \EndFor
    		        \State $\Permutation = \Permutation'$
		        \EndFor
		    \EndFor
        \EndFunction{}
    \end{algorithmic}
    \vspace{0.5mm}
    \hrule height 0.6pt
\end{pseudocode}
%%%%%%%%%%%%%%%%%%%%%%%%%%%%%%

%%%%%%%%%%%%%%%%%%%%%%%%%%%%%%%%%%%%%%%%%%%%%%%%%%%%%%%%%%%%
\subsection{Discussion}
\label{sec:OptimizationDiscussion}
%%%%%%%%%%%%%%%%%%%%%%%%%%%%%%%%%%%%%%%%%%%%%%%%%%%%%%%%%%%%

%%%%%%%%%%%%%%%%%%%%%%%%%%%%%%
\Paragraph{Search space}
%%%%%%%%%%%%%%%%%%%%%%%%%%%%%%

We discretize the search space $\SampleSetImageDomain$ to make the optimization problem~\eqref{eq:MinimizationProblem} tractable. To make it truly practical, it is also necessary to avoid repeated image estimation (\ie $\EstimatedImage(\SampleSetImage)$ evaluation) during the search for the solution $\SampleSetImage$. Our \vertical~\eqref{eq:VerticalOptimization} and \horizontal~\eqref{eq:HorizontalOptimization} optimization variants are formulated specifically with this goal in mind. All methods in \cref{alg:VerticalOptimization,alg:HorizontalOptimization} operate on pre-generated image estimates that constitute the solution search space.

Our \vertical formulation takes a collection of $\ImageCount$ input estimates $\{ \EstimatedPixel_{\PixelIndex,\PixelIndexTwo} = \EstimatedPixel_\PixelIndex(\SampleSetPixel_{\PixelIndex,\PixelIndexTwo}) \}_{\PixelIndexTwo=1}^\ImageCount$ for every pixel $\PixelIndex$, one for each sample set $\SampleSetPixel_{\PixelIndex,\PixelIndexTwo}$. Noting that $\EstimatedPixel_{\PixelIndex,\PixelIndexTwo}$ are MC estimates of the true pixel value, this collection can be cheaply expanded to a size as large as $2^\ImageCount-1$ by taking the average of the estimates in each of its subsets (excluding the empty subset). In practice only a fraction of these subsets can be used, since the size of the power set grows exponentially with $\ImageCount$. It may seem that this approach ends up wastefully throwing away most input estimates. But note that these estimates actively participate in the optimization and provide the space of possible solutions. Carefully selecting a subset per pixel can yield a higher-fidelity result than blindly averaging all available estimates, as we will show repeatedly in \cref{sec:Experiments}.

In contrast, our \horizontal formulation builds a search space given just a single input estimate $\EstimatedPixel_\PixelIndex$ per pixel. We consciously restrict the space to permutations between nearby pixels, so as to leverage local pixel similarity and avoid repeated pixel evaluation during optimization. The disadvantage of this approach is that it requires re-rendering the image after optimization, with uncertain results (due to mispredictions) that can lead to local degradation of image quality. Mispredictions can be reduced by exploiting knowledge about the rendering function $\EstimatedImage(\SampleSetImage)$, \eg through depth, normal, or texture buffers; we explore this in supplemental 
\ifdefined\arXiv
Section 2.
\else
\cref{sec:supp_our_demodulation}.
\fi
Additionally, while methods like iterative minimization (\cref{alg:HorizontalOptimization}) and dithering (\cref{sec:Aposteriori}) can be adapted to this search space, reformulating other halftoning algorithms such as error diffusion is non-trivial.

A hybrid formulation is also conceivable, taking a single input estimate per pixel (like \horizontal methods) and considering a separate (\vertical) search space for each pixel constructed by borrowing estimates from neighboring pixels. Such an approach could benefit from advanced halftoning optimization methods, but could also suffer from mispredictions and require re-rendering. We leave the exploration of this approach to future work.

Finally, it is worth noting that discretization is not the only route to practicality. \Cref{eq:MinimizationProblem} can be optimized on the continuous space $\SampleSetImageDomain$ if some cheap-to-evaluate proxy for the rendering function is available. Such a continuous approximation may be analytical (based on prior knowledge or assumptions) or obtained by reconstructing a point-wise evaluation. However, continuous-space optimization can be difficult in high dimensions (\eg number of light bounces) where non-linearities and non-convexity are exacerbated.

%%%%%%%%%%%%%%%%%%%%%%%%%%%%%%
\Paragraph{Optimization strategy}
%%%%%%%%%%%%%%%%%%%%%%%%%%%%%%

Another important choice is the optimization method. For the \vertical formulation, iterative minimization provides the best flexibility and quality but is the most computationally expensive. Error diffusion and dithering are faster but only approximately solve \cref{eq:VerticalOptimization}.

One difference between classical halftoning and our \vertical setting is that quantization levels are non-uniformly distributed and differ between pixels. This further increases the gap in quality between the image-adaptive iterative minimization and error diffusion (which can correct for these differences) and the non-adaptive dithering, compared to the halftoning setting. The main advantage of dithering is that it involves the kernel $\Kernel$ explicitly, while the error-diffusion kernel $\DiffusionKernel$ cannot be related directly to $\Kernel$.

%%%%%%%%%%%%%%%%%%%%%%%%%%%%%%%%%%%%%%%%%%%%%%%%%%%%%%%%%%%%
\section{Practical application}
\label{sec:Applications}
%%%%%%%%%%%%%%%%%%%%%%%%%%%%%%%%%%%%%%%%%%%%%%%%%%%%%%%%%%%%

We now turn to the practical use of our error optimization framework. In both our discrete formulations from \cref{sec:Optimization}, the search space is determined by a given collection of sample sets $\SampleSetPixel_{\PixelIndex,\PixelIndexTwo}$ for every pixel $\PixelIndex$, with $\PixelIndexTwo = 1...\ImageCount$ (in the \horizontal setting $\ImageCount = 1$). The optimization is then driven by the corresponding estimates $\EstimatedPixel_{\PixelIndex,\PixelIndexTwo}$. We consider two ways to obtain these estimates, leading to different practical trade-offs: (1)~direct evaluation of the samples by rendering a given scene and (2)~using a proxy for the rendering function. We show how prior works correspond to using either approach within our framework, which helps expose their implicit assumptions.

%%%%%%%%%%%%%%%%%%%%%%%%%%%%%%%%%%%%%%%%%%%%%%%%%%%%%%%%%%%%
\subsection{Surrogate for ground truth}
\label{sec:Surrogate}
%%%%%%%%%%%%%%%%%%%%%%%%%%%%%%%%%%%%%%%%%%%%%%%%%%%%%%%%%%%%

The goal of our optimization is to perceptually match an image estimate to the ground truth $\ReferenceImage$ as closely as possible. Unfortunately, the ground truth is unknown in our setting, unlike in halftoning. The best we can do is substitute it with a \emph{surrogate} image $\SurrogateImage$. Such an image can be obtained either from available pixel estimates or by making assumptions about the ground truth. We will discuss specific approaches in the following, but it is already worth noting that all existing error-distribution methods rely on such a surrogate, whether explicitly or implicitly. And since the surrogate guides the optimization, its fidelity directly impacts the fidelity of the output.

%%%%%%%%%%%%%%%%%%%%%%%%%%%%%%%%%%%%%%%%%%%%%%%%%%%%%%%%%%%%
\subsection{A-posteriori optimization}
\label{sec:Aposteriori}
%%%%%%%%%%%%%%%%%%%%%%%%%%%%%%%%%%%%%%%%%%%%%%%%%%%%%%%%%%%%

Given a scene and a viewpoint, initial pixel estimates can be obtained by invoking the renderer with the input samples: $\EstimatedPixel_{\PixelIndex,\PixelIndexTwo} = \EstimatedPixel_\PixelIndex(\SampleSetPixel_{\PixelIndex,\PixelIndexTwo})$. A surrogate can then be constructed from those estimates; in our implementation we denoise the estimate-average image (\cref{sec:ExperimentSetup}). Having the estimates and the surrogate, we can run any of the methods in \cref{alg:VerticalOptimization,alg:HorizontalOptimization}. \Vertical algorithms directly output an image $\OutputImage$; \horizontal optimization yields a sample-set image $\SampleSetImage$ that requires an additional rendering invocation: $\OutputImage = \EstimatedImage(\SampleSetImage)$.

This general approach of utilizing sampled image information was coined \emph{a-posteriori} optimization by \citet{Heitz2019}; they proposed two such methods. Their histogram method operates in a \vertical setting, choosing one of the (sorted) estimates for each pixel based on the respective value in a given blue-noise dither mask. Such sampling corresponds to using an implicit surrogate that is the median estimate for every pixel, which is what the mean of the dither mask maps to. Importantly, any one of the estimates for a pixel can be selected, whereas in classical dithering the choice is between the two quantization levels that tightly envelop the reference value (\cref{sec:VerticalOptimization}) \citep{spaulding1997methods}. Such selection can yield large error, especially for pixels whose corresponding mask values deviate strongly from the mask mean. This produces image fireflies that do not appear if simple estimate averages are taken instead (see \cref{fig:staircase-all-methods}).

The permutation method of \citet{Heitz2019} operates in a \horizontal setting. Given an image estimate, it finds pixel permutations within small tiles that minimize the distance between the estimates and the values of a target blue-noise mask. This matching transfers the mask's distribution to the image signal rather than to its error. The two are equivalent only when the signal within each tile is constant. The implicit surrogate in this method is thus a tile-wise constant image (shown more formally in supplemental 
\ifdefined\arXiv
Section 5).
\else
\cref{sec:supp_aposteriori_approaches}).
\fi
In our framework the use of a surrogate is explicit, which enables full control over the quality of the error distribution.

%%%%%%%%%%%%%%%%%%%%%%%%%%%%%%%%%%%%%%%%%%%%%%%%%%%%%%%%%%%%
\subsection{A-priori optimization}
\label{sec:Apriori}
%%%%%%%%%%%%%%%%%%%%%%%%%%%%%%%%%%%%%%%%%%%%%%%%%%%%%%%%%%%%

Optimizing perceptual error is possible even in the absence of information about a specific image. In our framework, the goal of such an \emph{a-priori} approach (as coined by \citet{Heitz2019}) is to compute a sample-set image $\SampleSetImage$ by using surrogates for both the ground-truth image $\ReferenceImage$ and the rendering function $\EstimatedImage(\SampleSetImage)$, constructed based on smoothness assumptions. The samples $\SampleSetImage$ can then produce a high-fidelity estimate of any image that meets those assumptions.

Lacking prior knowledge, one could postulate that every pixel $\PixelIndex$ has the same rendering function: $\EstimatedPixel_\PixelIndex(\cdot) \! = \! \EstimatedPixel(\cdot)$; the image surrogate $\SurrogateImage$ is thus constant. While in practice this assumption (approximately) holds only locally, the optimization kernel $\Kernel$ is also expected to have compact support. The shape of $\EstimatedPixel$ can be assumed to be (piecewise) smooth and approximable by a cheap analytical function~$\SurrogatePixelEstimate$. 

With the above surrogates in place, we can run our algorithms to optimize a sample-set image $\SampleSetImage$. The constant-image assumption makes \horizontal algorithms well-suited for this setting as it makes the swapping-error term $\ErrorDifference$ in \cref{eq:HorizontalOptimizationEnergy} vanish, simplifying the perceptual error to $\Energy(\Permutation(\SampleSetImage)) = \Norm{\Kernel * \Permutation(\ErrorImage(\SampleSetImage))}_2^2$. This enables the optimization to consider swaps between \emph{any} two pixels in the error image $\ErrorImage(\SampleSetImage)$. That image can be quickly rendered in advance, by invoking the render-function surrogate $\SurrogatePixelEstimate$ with the input sample-set image.

\citet{georgiev16bluenoise} take a similar approach, with swapping based on simulated annealing. Their empirically motivated optimization energy uses an explicit (Gaussian) kernel, but instead of computing an error image through a rendering surrogate, it postulates that the distance between two sample sets is representative of the difference between their corresponding pixel errors. Such a smoothness assumption holds for bi-Lipschitz-continuous functions. Their energy can thus be understood to compactly encode properties of a class of rendering functions.

\citet{Heitz2019b} adopt the approach of \citet{georgiev16bluenoise}, but their energy function replaces the distance between sample sets by the difference between their corresponding pixel errors. The errors are computed using an explicit render-function surrogate. They optimize for a large number of simple surrogates simultaneously, and leverage a compact representation of Sobol sequences to also support progressive sampling. We relate these two prior works to ours more formally in supplemental 
\ifdefined\arXiv
Section 6, 
\else
\cref{sec:supp_apriori_approaches}, 
\fi
also showing how our perceptual error formulation can be incorporated into the method of \citet{Heitz2019b}.

The approach of \citet{Ahmed:HierarchicalOrdering} performs on-the-fly scrambling of a Sobol sequence applied to the entire image. Image pixels are visited in Morton Z-order modified to breaks its regularity. The resulting sampler diffuses Monte Carlo error over hierarchically nested blocks of pixels giving a perceptually pleasing error distribution. However, the algorithmic nature of this approach introduces more implicit assumptions than prior works, making it difficult to analyze.

Our theoretical formulation and optimization methods enable the construction of a-priori sampling masks in a principled way. For \horizontal optimization, we recommend using our iterative algorithm (\cref{alg:HorizontalOptimization}) which can bring significant performance improvement over simulated annealing; such speed-up was reported by \citet{Allebach1992} for dither-mask construction. \Vertical optimization is an interesting alternative, where for each pixel one of several sample sets would be chosen; this would allow for varying the sample count per pixel. Note that the ranking-key optimization for progressive sampling of \citet{Heitz2019b} is a form of \vertical optimization.

%%%%%%%%%%%%%%%%%%%%%%%%%%%%%%
\begin{figure*}[t!]
    \centering
    %!TEX root = ../paper.tex

\newcommand{\PSFTemplateFour}[4]{
    \begin{scope}
        \clip (0,0) -- (0.5,0) -- (0.487,0.5) -- (0,0.5) -- cycle;
        \path[fill overzoom image=figures/psf-comparisons/#1] (0,0) rectangle (1,1);
    \end{scope}
    \begin{scope}
        \clip (0.487,0) -- (1,0) -- (1,0.5) -- (0.487,0.5) -- cycle;
        \path[fill overzoom image=figures/psf-comparisons/#2] (0,0) rectangle (1,1);
    \end{scope}
    \begin{scope}
        \clip (0,0.5) -- (0.487,0.5) -- (0.487,1) -- (0,1) -- cycle;
        \path[fill overzoom image=figures/psf-comparisons/#3] (0,0) rectangle (1,1);
    \end{scope}
    \begin{scope}
        \clip (0.487,0.5) -- (1,0.5) -- (1,1) -- (0.487,1) -- cycle;
        \path[fill overzoom image=figures/psf-comparisons/#4] (0,0) rectangle (1,1);
    \end{scope}
    \draw[draw=white,thin] (0,0.5) -- (1,0.5);
    \draw[draw=white,thin] (0.487,0) -- (0.487,1);
}

\newcommand{\PSFTemplateThree}[4]{
    \begin{scope}
        \clip (0,0) -- (0.485,0) -- (0.485,#4) -- (0,#4) -- cycle;
        \path[fill overzoom image=figures/psf-comparisons/#1] (0,0) rectangle (1,1);
    \end{scope}
    \begin{scope}
        \clip (0.485,0) -- (1,0) -- (1,#4) -- (0.485,#4) -- cycle;
        \path[fill overzoom image=figures/psf-comparisons/#2] (0,0) rectangle (1,1);
    \end{scope}
    \begin{scope}
        \clip (0,#4) -- (1,#4) -- (1,1) -- (0,1) -- cycle;
        \path[fill overzoom image=figures/psf-comparisons/#3] (0,0) rectangle (1,1);
    \end{scope}
    \draw[draw=white,thin] (0,#4) -- (1,#4);
    \draw[draw=white,thin] (0.485,0) -- (0.485,#4);
}

\newcommand{\PSFTemplateTwo}[2]{
    \begin{scope}
        \clip (0,0) -- (0.485,0) -- (0.485,1) -- (0,1) -- cycle;
        \path[fill overzoom image=figures/psf-comparisons/#1] (0,0) rectangle (1,1);
    \end{scope}
    \begin{scope}
        \clip (0.485,0) -- (1,0) -- (1,1) -- (0.485,1) -- cycle;
        \path[fill overzoom image=figures/psf-comparisons/#2] (0,0) rectangle (1,1);
    \end{scope}
    \draw[draw=white,thin] (0.485,0) -- (0.485,1);
}

\footnotesize
\hspace*{-2.5mm}
\definecolor{contour}{RGB}{81, 82, 59} % {69, 61, 45}
\begin{tabular}{c@{\;}c@{\;}c@{\;}c@{}}
    \begin{tikzpicture}[scale=4.43]
        \PSFTemplateThree{halftoning_aa_spp4_teapot_VDBS_Nasaanen_conv}{halftoning_aa_spp4_teapot_VDBS_AlphaStable_conv}{halftoning_aa_spp4_teapot_VDBS5_conv}{0.5}
        \filldraw[white,ultra thick] (0.245, 0.03) circle (0pt) node[anchor=base,rotate=0] {\fontsize{6.7}{5}\selectfont\Outlined[white][contour]{[N\"as\"anen 1984]}};
        % \filldraw[white,ultra thick] (0.245, 0.03) circle (0pt) node[anchor=base,rotate=0] {\fontsize{6.7}{5}\selectfont\Outlined[white][contour]{\cite{NasanenVisibility}}};
        \filldraw[white,ultra thick] (0.744, 0.03) circle (0pt) node[anchor=base,rotate=0] {\fontsize{6.7}{5}\selectfont\Outlined[white][contour]{\![Gonz\'alez et al.\ 2006]\!}};
        % \filldraw[white,ultra thick] (0.744, 0.03) circle (0pt) node[anchor=base,rotate=0] {\fontsize{6.7}{5}\selectfont\Outlined[white][contour]{\!\cite{Gonzlez2006AlphaSH}\!}};
        \filldraw[white,ultra thick] (0.485, 0.53) circle (0pt) node[anchor=base,rotate=0] {\Outlined[white][contour]{\textbf{Our kernel}}};
    \end{tikzpicture}%
    &
    \begin{tikzpicture}[scale=4.43]
        \PSFTemplateTwo{halftoning_aa_spp4_teapot_VDBS3_conv}{halftoning_aa_spp4_teapot_VDBS3}
        \filldraw[white,ultra thick] (0.245, 0.03) circle (0pt) node[anchor=base,rotate=0] {\Outlined[white][contour]{$\KernelReference = \Kernel$}};
        \filldraw[white,ultra thick] (0.745, 0.03) circle (0pt) node[anchor=base,rotate=0] {\Outlined[white][contour]{$\KernelReference = \DiracKernel$}};
    \end{tikzpicture}%
    &
    \begin{tikzpicture}[scale=4.43]
        \ClipImageRect{0,0}{0.5,1}{figures/psf-comparisons/tonemap_halftoning_avt4_spp4_teapot_VDBS_REF}
        \ClipImageRect{0.5,0}{0.5,1}{figures/psf-comparisons/tonemap_halftoning_avt3_spp4_teapot_VDBS_REF}
        \ClipImageRect{0,0.632}{1,1-0.632}{figures/psf-comparisons/tonemap_aux_reference_avt3_spp4_teapot}
        \draw[draw=white,thin] (0,0.632) -- (1,0.632);
        \draw[draw=white,thin] (0.5,0) -- (0.5,0.632);
        \TikzLabel[contour]{0.245, 0.03}{Linear error}
        \TikzLabel[contour]{0.745, 0.03}{Tone-mapped error}
        \TikzLabel[contour]{0.485, 0.66}{Ground truth}
    \end{tikzpicture}%
    &
    \begin{tikzpicture}[scale=4.43]
        \ClipImageRect{0,0}{0.5,1}{figures/psf-comparisons/color_halftoning_avt_spp4_teapot_VDBS_REF_GREY}
        \ClipImageRect{0.5,0}{0.5,1}{figures/psf-comparisons/color_halftoning_avt_spp4_teapot_VDBS_REF}
        \ClipImageRect{0,0.632}{1,1-0.632}{figures/psf-comparisons/color_aux_reference_avt_spp4_teapot}
        \draw[draw=white,thin] (0,0.632) -- (1,0.632);
        \draw[draw=white,thin] (0.5,0) -- (0.5,0.632);
        \TikzLabel[contour]{0.245, 0.03}{Grayscale error}
        \TikzLabel[contour]{0.745, 0.03}{Color error}
        \TikzLabel[contour]{0.485, 0.66}{Ground truth}
    \end{tikzpicture}%
    \\
    (a) Kernel comparison & (b) Kernel sharpening effect & (c) Tone mapping (ACES) & (d) Color handling
\end{tabular}%
\undefinecolor{contour}

\let\PSFTemplateTwo\undefined
\let\PSFTemplateThree\undefined
\let\PSFTemplateFour\undefined
    \vspace{-2.5mm}
    \caption{
        (a)~Our binomial Gaussian approximation $\Kernel$ ($3\!\times\!3$\,pixels,\,$\sigma\!=\!\sqrt{2/\pi}$) performs on par with state-of-the-art halftoning kernels.
        (b)~Setting the reference-image kernel $\KernelReference$ in \cref{eq:extension_ref_conv} to a zero-width $\DiracKernel$ kernel sharpens the output.
        (c)~Incorporating tone mapping via \cref{eq:extensions_tonemapping}.
        (d)~Incorporating color via \cref{eq:extension_color}.
    }
    \label{fig:Extensions}
\end{figure*}
%%%%%%%%%%%%%%%%%%%%%%%%%%%%%%

%%%%%%%%%%%%%%%%%%%%%%%%%%%%%%
\begin{figure*}[t!]
    \centering
    \vspace{-1mm}
    \input{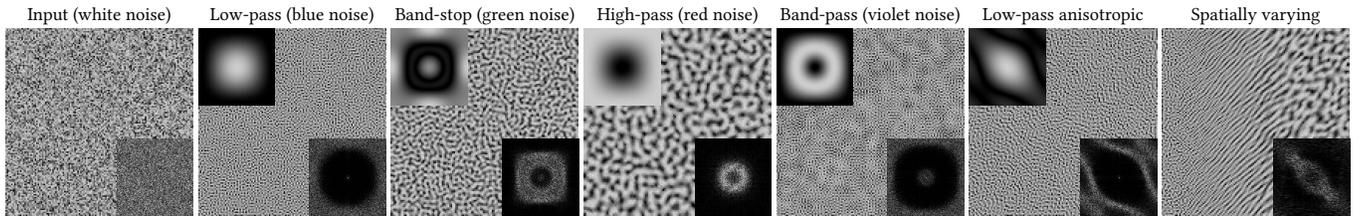}
    \vspace{-2.7mm}
    \caption{
        Our formulation~\eqref{eq:MinimizationProblem} allows optimizing the error distribution of an image \wrt arbitrary kernels. Here we adapt our \horizontal iterative minimization~(\cref{alg:HorizontalOptimization}) to directly swap the pixels of a white-noise input image. Insets show the power spectra of the target kernels (top left) and the optimized images (bottom right).
    }
    \label{fig:Kernels}
    \vspace{-1mm}
\end{figure*}
%%%%%%%%%%%%%%%%%%%%%%%%%%%%%%

%%%%%%%%%%%%%%%%%%%%%%%%%%%%%%%%%%%%%%%%%%%%%%%%%%%%%%%%%%%%
\subsection{Discussion}
\label{sec:ApplicationsDiscussion}
%%%%%%%%%%%%%%%%%%%%%%%%%%%%%%%%%%%%%%%%%%%%%%%%%%%%%%%%%%%%

Our formulation expresses a-priori and a-posteriori optimization under a common framework that unifies existing methods. These two approaches come with different trade-offs. A-posteriori optimization utilizes sampled image information, and in a \vertical setting requires no assumptions except for what is necessary for surrogate construction. It thus has potential to achieve high output fidelity, especially on scenes with complex lighting as it is oblivious to the shape and dimensionality of the rendering function, as first demonstrated by \citet{Heitz2019}. However, it requires pre-sampling (also post-sampling in the \horizontal setting), and the optimization is sensitive to the surrogate quality.

Making aggressive assumptions allows a-priori optimization to be performed offline once and the produced samples $\SampleSetImage$ to be subsequently used to render any image. This approach resembles classical sample stratification where the goal is also to optimize sample distributions \wrt some smoothness expectations. In fact, our a-priori formulation subsumes the per-pixel stratification problem, since the perceptual error is minimized when the error image $\ErrorImage(\SampleSetImage)$ has both low magnitude and visually pleasing distribution. Two main factors limit the potential of a-priori optimization, especially on scenes with non-uniform multi-bounce lighting. One is the general difficulty of optimizing sample distributions in high-dimensional spaces. The other is that in such scenes the complex shape of the rendering function, both in screen and sample space, can easily break smoothness assumptions and cause high (perceptual) error.

To test the capabilities of our formulation, in the following we focus on the a-posteriori approach. In the supplemental document we explore a-priori optimization based on our framework. The two approaches can also be combined, \eg by seeding a-posteriori optimization with a-priori-optimized samples whose good initial guess can improve the quality and convergence speed.

%%%%%%%%%%%%%%%%%%%%%%%%%%%%%%%%%%%%%%%%%%%%%%%%%%%%%%%%%%%%
\section{Extensions}
\label{sec:Extensions}
%%%%%%%%%%%%%%%%%%%%%%%%%%%%%%%%%%%%%%%%%%%%%%%%%%%%%%%%%%%%

Our perceptual error formulation~\eqref{eq:PerceptualError} approximates the effect of the HVS PSF through kernel convolution. In this section we analyze the relationship between kernel and viewing distance, as well as the impact of the kernel shape on the error distribution. We also present extensions that account for the HVS non-linearities in handling tone and color.

%%%%%%%%%%%%%%%%%%%%%%%%%%%%%%
\Paragraph{Kernels and viewing distance}
%%%%%%%%%%%%%%%%%%%%%%%%%%%%%%

As discussed in \cref{sec:Motivation}, the PSF is usually modelled over a range of viewing distances. The effect of the PSF depends on the frequencies of the viewed signal and the distance from which it is viewed. \citet{Pappas1999} have found that the Gaussian is a good approximation to the PSF in the context of halftoning. They derived its standard deviation $\sigma$ in terms of the minimum viewing distance for a given screen resolution:
\begin{equation}
    \label{eq:KernelViewingDistance}
    \sigma = \frac{0.00954}{\tau},
    \quad\text{where}\quad
    \tau = \frac{180}{\pi}2\arctan\left(\frac{1}{2RD}\right).
\end{equation}
Here $\tau$ is the visual angle between the centers of two neighboring pixels (in degrees) for screen resolution $R$ (in $1/$inches) and viewing distance $D$ (in inches). The minimum viewing distance for a given standard deviation and resolution can be contained via the inverse formula: $D = {\left(2R\tan\left(\frac{\pi}{180}\frac{0.00954}{2\sigma}\right)\right)}^{-1}\!$. Larger $\sigma$ values correspond to larger observer distances; we demonstrate the effect of that in~\cref{fig:blue-noise-steady-state} where the images become increasingly blurrier. In~\cref{fig:Extensions}a, we compare that Gaussian kernel to two well-established PSF models from the halftoning literature~\citep{NasanenVisibility,Gonzlez2006AlphaSH}. We have found the differences between all three to be negligible; we use the cheaper to evaluate Gaussian in all our experiments.

%%%%%%%%%%%%%%%%%%%%%%%%%%%%%%
\Paragraph{Decoupling the viewing distances}
%%%%%%%%%%%%%%%%%%%%%%%%%%%%%%

Being based on the perceptual models of the HVS~\citep{Sullivan1991,Allebach1992}, our formulation~\eqref{eq:PerceptualError} assumes that the estimate $\EstimatedImage$ and the reference $\ReferenceImage$ are viewed from the same (range of) distance(s). The two distances can be decoupled by applying different kernels to $\EstimatedImage$ and $\ReferenceImage$:
\begin{align}
    \label{eq:extension_ref_conv}
    \Energy = \Norm{\Kernel * \EstimatedImage - \KernelReference * \ReferenceImage}_2^2.
\end{align}
Minimizing this error makes $\EstimatedImage$ appear from some distance $D_{\Kernel}$ similar to $\ReferenceImage$ seen from a different distance $D_{\KernelReference}$. The special case of using a Kronecker delta kernel $\KernelReference = \DiracKernel$, \ie with the reference $\ReferenceImage$ seen from up close, yields $\Energy = \Norm{\Kernel * \EstimatedImage - \ReferenceImage}_2^2$. This has been shown to have an edge enhancing effect~\citep{anastassiou1989error,Pappas1999} which we show in \cref{fig:Extensions}b. We use $\KernelReference = \DiracKernel$ in all our experiments.

%%%%%%%%%%%%%%%%%%%%%%%%%%%%%%
\Paragraph{Tone mapping}
%%%%%%%%%%%%%%%%%%%%%%%%%%%%%%

Considering that the optimized image will be viewed on media with limited dynamic range (\eg screen or paper), we can incorporate a tone-mapping operator $\ToneMappingOp$ into the perceptual error~\eqref{eq:PerceptualError}:
\begin{equation}
    \label{eq:extensions_tonemapping}
    \Energy
        =\, \Norm{\Kernel * \ErrorImage_\ToneMappingOp}_2^2
        \,=\, \Norm{\Kernel * (\ToneMappingOp(\EstimatedImage) - \ToneMappingOp(\ReferenceImage))}_2^2.
\end{equation}
Doing this also bounds the per-pixel error $\ErrorImage_\ToneMappingOp = \ToneMappingOp(\EstimatedImage) - \ToneMappingOp(\ReferenceImage)$, suppressing outliers and making the optimization more robust in scenes with high dynamic range. We illustrate this improvement in \cref{fig:Extensions}c, where an ACES~\citep{Arrighetti:2017} tone-mapping operator is applied to the optimized image. Optimizing \wrt the original perceptual error~\eqref{eq:PerceptualError} yields a noisy and overly dark image compared to the tone-mapped ground truth. Accounting for tone mapping in the optimization through \cref{eq:extensions_tonemapping} yields a more faithful result.

%%%%%%%%%%%%%%%%%%%%%%%%%%%%%%
\Paragraph{Color handling}
%%%%%%%%%%%%%%%%%%%%%%%%%%%%%%

While the HVS reacts more strongly to luminance than color, ignoring chromaticity entirely (\eg by computing the error image $\ErrorImage$ from per-pixel luminances) can have a negative effect on the distribution of color noise in the image. To that end, we can penalize the perceptual error of each color channel $c \in C$ separately:
\begin{equation}
    \label{eq:extension_color}
    \Energy \,=\, \sum_{\mathclap{c \, \in \, C}} \lambda_c\Norm{\Kernel_c * (\EstimatedImage_c - \ReferenceImage_c)}_2^2\,,
\end{equation}
where $\lambda_c$ is a per-channel weight. In our experiments, we use an RGB space $C = \{\mathrm{r}, \mathrm{g}, \mathrm{b}\}$, set $\lambda_c = 1$, and use the same kernel $\Kernel_c = \Kernel$ for every channel. \Cref{fig:Extensions}d shows the improvement in color noise over using greyscale perceptual error. Depending on the color space, the per-channel kernels may differ (\eg YCbCr)~\citep{Sullivan1991}.

As an alternative, one could decouple the channels from the input estimates and optimize each channel separately, assembling the results into a color image. In a \vertical setting, this decoupling extends the optimization search space size from $\ImageCount$ to $\ImageCount^{|C|}$.

%%%%%%%%%%%%%%%%%%%%%%%%%%%%%%
\Paragraph{Kernel shape impact}
%%%%%%%%%%%%%%%%%%%%%%%%%%%%%%

To test the robustness of our framework, we analyze kernels with spectral characteristics other than isotropic blue-noise in \cref{fig:Kernels}. We run our iterative pixel-swapping algorithm (\cref{alg:HorizontalOptimization}) to optimize the shape of a white-noise input, which produces a spectral distribution inverse to that of the target kernel. The rightmost image in the figure shows the result of using a spatially varying kernel that is a convex combination between a low-pass Gaussian and a high-pass anisotropic kernel, with the interpolation parameter varying horizontally across the image. Our algorithm can adapt the noise shape well.

%%%%%%%%%%%%%%%%%%%%%%%%%%%%%%%%%%%%%%%%%%%%%%%%%%%%%%%%%%%%
\section{Results}
\label{sec:Experiments}
%%%%%%%%%%%%%%%%%%%%%%%%%%%%%%%%%%%%%%%%%%%%%%%%%%%%%%%%%%%%

We now present empirical validation of our error optimization framework in the a-posteriori setting described in \cref{sec:Aposteriori}. We render static images and animations of several scenes, comparing our algorithms to those of~\citet{Heitz2019}.

%%%%%%%%%%%%%%%%%%%%%%%%%%%%%%%%%%%%%%%%%%%%%%%%%%%%%%%%%%%%
\subsection{Setup}
\label{sec:ExperimentSetup}
%%%%%%%%%%%%%%%%%%%%%%%%%%%%%%%%%%%%%%%%%%%%%%%%%%%%%%%%%%%%

%%%%%%%%%%%%%%%%%%%%%%%%%%%%%%
\Paragraph{Perceptual error model}
\label{para:setup_perceptual}
%%%%%%%%%%%%%%%%%%%%%%%%%%%%%%

We build a perceptual model by combining all extensions from \cref{sec:Extensions}. Our estimate-image kernel $\Kernel$ is a  binomial approximation of a Gaussian~\citep{Lindeberg}. For performance reasons and to allow smaller viewing distances we use a $3\!\times\!3$-pixel kernel with standard deviation $\sigma = \sqrt{2/\pi}$ (see \cref{fig:Extensions}a). Plugging this $\sigma$ value into the inverse of \cref{eq:KernelViewingDistance}, the corresponding minimum viewing distance is $D = 4792 / R$ inches for a screen resolution of $R$\,dpi (\eg 16 inches at 300\,dpi). We recommend viewing from a larger distance, to reduce the effect of our $3\!\times\!3$ kernel discretization. We use a Dirac reference-image kernel: $\KernelReference = \DiracKernel$, and incorporate a simple tone-mapping operator $\ToneMappingOp$ that clamps pixel values to $[0,1]$. The final error model reads:
\begin{equation}
    \label{eq:energy_experiments}
    \Energy
    \,=\, \sum_{\mathclap{c \, \in \{\mathrm{r}, \mathrm{g}, \mathrm{b}\}}} \,\Norm{\Kernel * \ToneMappingOp(\EstimatedImage_c) - \DiracKernel * \ToneMappingOp(\SurrogateImage_{\!\!c})}_2^2,
    \vspace{-0.5mm}
\end{equation}
where $\SurrogateImage$ is the surrogate image whose construction we describe below. For dithering we convert RGB colors to luminance, which reduces the number of components in the error~\eqref{eq:energy_experiments} to one.

%%%%%%%%%%%%%%%%%%%%%%%%%%%%%%
\Paragraph{Methods}
%%%%%%%%%%%%%%%%%%%%%%%%%%%%%%

We compare our four methods from \cref{alg:VerticalOptimization,alg:HorizontalOptimization} to the histogram and permutation of \citet{Heitz2019}. For our \vertical and \horizontal iterative minimizations we set the maximum iteration count to 100 and 10 respectively. For error diffusion we use the kernel of \citet{floyd1976adaptive} and for dithering we use a void-and-cluster mask \citep{Ulichney1993Void}. For our \horizontal iterative minimization we use a search radius $R=1$ and allow pixels to travel within a disk of radius $r=1$ from their original location in the dissimilarity metric. For the permutation method of \citet{Heitz2019} we obtained best results with tile size $8\!\times\!8$. (Our $r=1$ approximately corresponds to their tile size $3\!\times\!3$.)

%%%%%%%%%%%%%%%%%%%%%%%%%%%%%%
\Paragraph{Rendering}
%%%%%%%%%%%%%%%%%%%%%%%%%%%%%%

All scenes were rendered with PBRT~\citep{Pharr16Physically} using unidirectional or bidirectional path tracing. None of the methods depend on the sampling dimensionality, though we set the maximum path depth to 5 for all scenes to maintain reasonable rendering times. The ground-truth images have been generated using a Sobol sampler with at least 1024 samples per pixel (spp); for all test renders we use a random sampler. To facilitate numerical-error comparisons between the different methods, we trace the primary rays through the pixel centers.

%%%%%%%%%%%%%%%%%%%%%%%%%%%%%%
\Paragraph{Surrogate construction}
%%%%%%%%%%%%%%%%%%%%%%%%%%%%%%

To build a surrogate image for our methods, we filter the per-pixel averaged input estimates using Intel Open Image Denoise~\citep{inteldenoiser} which also leverages surface-normal and albedo buffers, taking about 0.5\,sec for a $512 \times 512$ image. Recall that the methods of \citet{Heitz2019} utilize implicit surrogates.

%%%%%%%%%%%%%%%%%%%%%%%%%%%%%%
\Paragraph{Image-quality metrics}
%%%%%%%%%%%%%%%%%%%%%%%%%%%%%%

We evaluate the quality of some of our results using the HDR-VDP-2 perceptual metric~\citep{Mantiuk2011}, with parameters matching our binomial kernel. We compute error-detection probability maps which indicate the likelihood for a human observer to notice a difference from the ground truth.

Additionally, we analyze the local blue-noise quality of the error image $\ErrorImage = \ToneMappingOp(\EstimatedImage) - \ToneMappingOp(\ReferenceImage)$. We split the image into tiles of $32\!\times\!32$ pixels and compute the Fourier power spectrum of each tile. For visualization purposes, we apply a standard logarithmic transform $c \ln(1 + |\hat{\ErrorPixel}|)$ to every resulting pixel value $\hat{\ErrorPixel}$ and compute the normalization factor $c$ per tile so that the maximum final RGB value within the tile is $(1,1,1)$. Note that the error image $\ErrorImage$ is computed \wrt the ground truth $\ReferenceImage$ and not the surrogate, which quantifies the blue-noise distribution objectively. The supplemental material contains images of the tiled power spectra for all experiments.

We compare images quantitatively via traditional MSE as well as a metric derived from our perceptual error formulation. Our \emph{perceptual MSE} (pMSE) evaluates the error~\eqref{eq:energy_experiments} of an estimate image \wrt the ground truth, normalized by the number of pixels $\PixelCount$ and channels~$C$: $\text{pMSE} = \frac{\Energy}{\PixelCount \cdot C}$. It generalizes the MSE with a perceptual, \ie non-delta, kernel $\Kernel$. \Cref{tab:Errors} summarizes the results.

%%%%%%%%%%%%%%%%%%%%%%%%%%%%%%
\begin{figure*}[t!]
    \centering
    \input{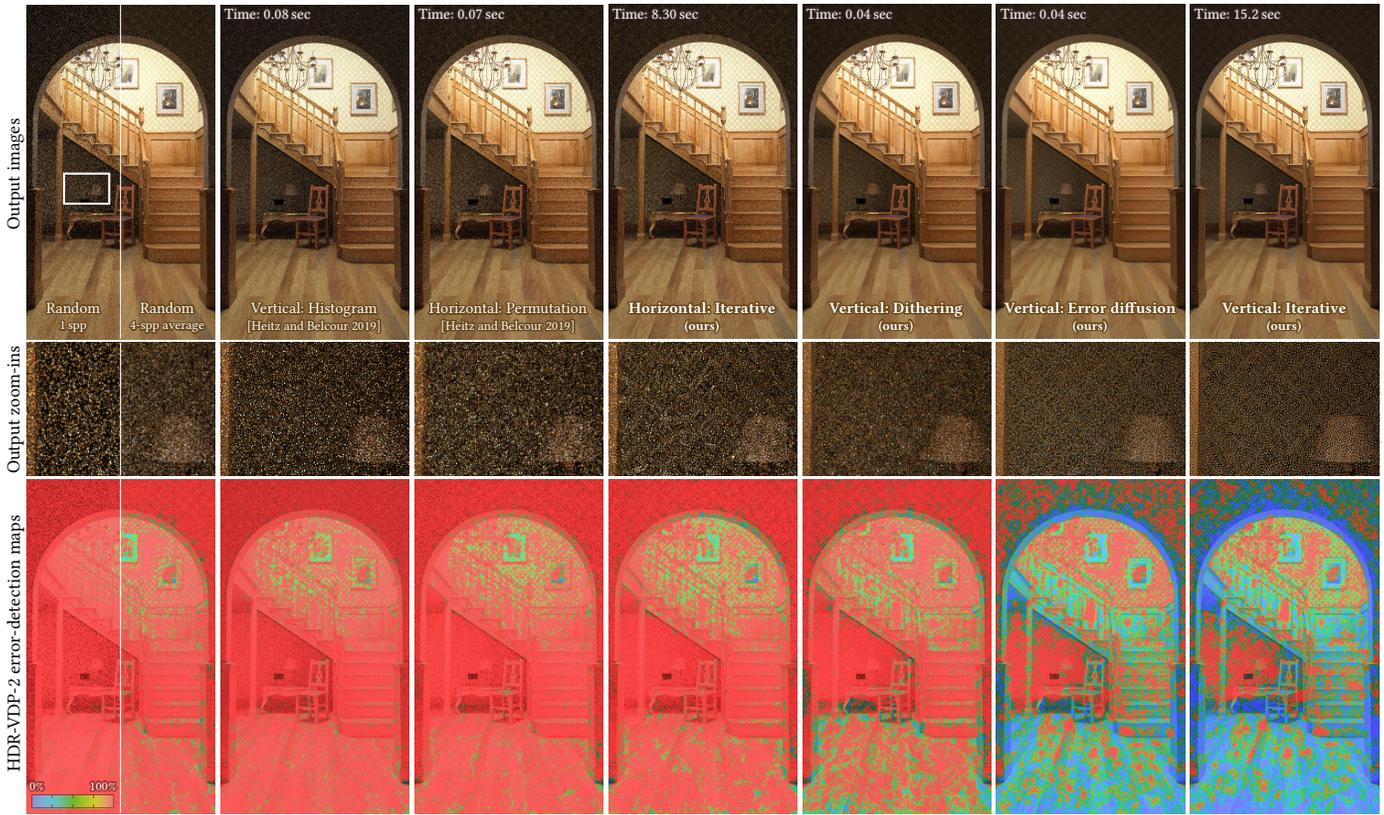}
    \vspace{-3.3mm}
    \caption{ 
        Comparison of our algorithms against the permutation and histogram methods of \citet{Heitz2019} with equal total sampling cost of 4\,spp. Bottom row shows HDR-VPD-2 error-detection maps (blue is better, \ie lower detection probability). The baseline 1-spp and 4-spp images exhibit large perceptual error, while our \vertical iterative minimization achieves highest fidelity. Error diffusion produces similar quality. Dithering is as fast but shows smaller improvement over the baselines, yet significantly outperforms the similar histogram method. Our \horizontal iterative optimization does better than the permutation method. Our methods also reduce MSE compared to the 4-spp baseline, even though they do not focus solely on per-pixel error (see \Cref{tab:Errors}).
    }
    \label{fig:staircase-all-methods}
\end{figure*}
%%%%%%%%%%%%%%%%%%%%%%%%%%%%%%

%%%%%%%%%%%%%%%%%%%%%%%%%%%%%%
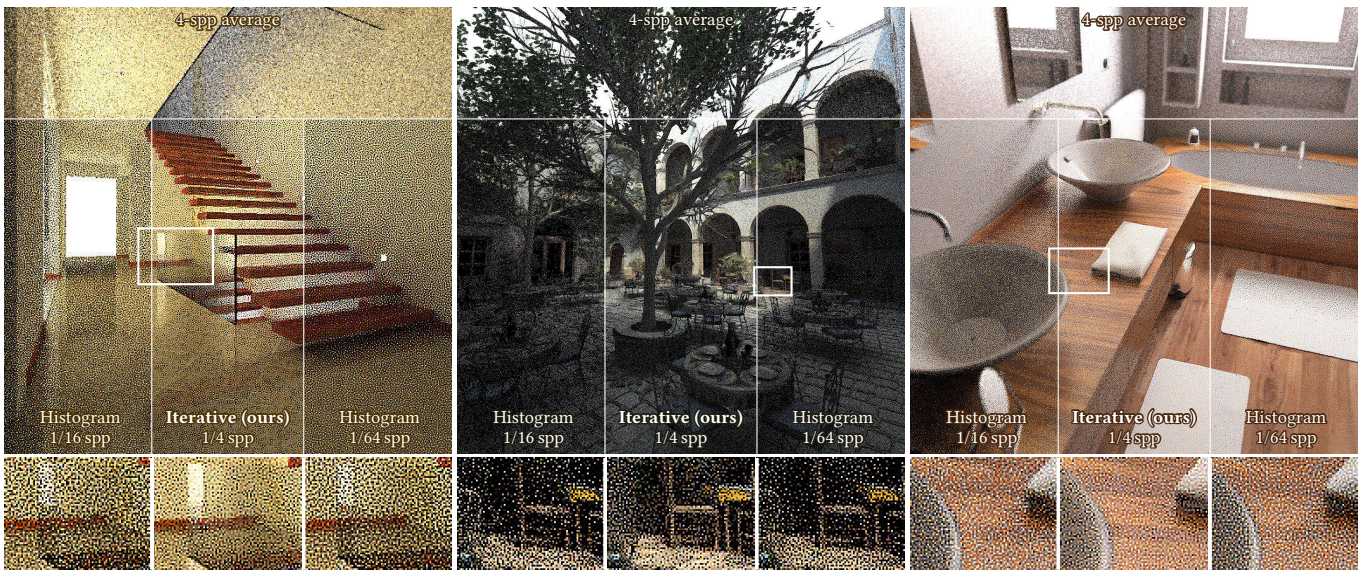
\begin{figure*}[t!]
    \centering
    %!TEX root = ../paper.tex

\newcommand{\ImageTemplate}[5]{
    \ClipImageRect{0,0}{0.33,0.75}{#2}
    \ClipImageRect{0.33,0}{0.34,0.75}{#3}
    \ClipImageRect{0.67,0}{0.33,0.75}{#4}
    \ClipImageRect{0,0.75}{1,0.25}{#5}
    \draw[draw=white,thin] (0.33,0) -- (0.33,0.75);
    \draw[draw=white,thin] (0.67,0) -- (0.67,0.75);
    \draw[draw=white,thin] (0,0.75) -- (1,0.75);
    \begin{scope}
        \TikzLabel[#1]{0.17, 0.07}{Histogram}
        \TikzLabel[#1]{0.17, 0.025}{1/16\,spp}
        \TikzLabel[#1]{0.5, 0.07}{\textbf{Iterative (ours)}}
        \TikzLabel[#1]{0.5, 0.025}{1/4\,spp}
        \TikzLabel[#1]{0.84, 0.07}{Histogram}
        \TikzLabel[#1]{0.84, 0.025}{1/64\,spp}
        \TikzLabel[#1]{0.5, 0.96}{4-spp average}
    \end{scope}
}

\newcommand{\ZoomInTemplate}[3]{
    \PlaceImage{0}{0}{0.33}{0.25}{#1}
    \PlaceImage{0.335}{0}{0.33}{0.25}{#2}
    \PlaceImage{0.67}{0}{0.33}{0.25}{#3}
    \draw[draw=white,line width=0.45mm] (0.33, 0) -- (0.33, 0.25);
    \draw[draw=white,line width=0.45mm] (0.67, 0) -- (0.67, 0.25);
}

\definecolor{contour_stairs}{RGB}{75, 50, 2}
\definecolor{contour_sanmiguel}{RGB}{30, 28, 15}
\definecolor{contour_bathroom}{RGB}{72, 39, 14}

\fontsize{7}{8}\selectfont

\hspace*{-0.85em}
\begin{tabular}{c@{\;}c@{\;}c@{}}
\begin{tikzpicture}[scale=5.94]
    \ImageTemplate{contour_stairs}
        {figures/vertical-vs-histogram/halftoning_aa_spp16_staircase2_VHH}
        {figures/vertical-vs-histogram/halftoning_aa_spp1_staircase2_VDBS}
        {figures/vertical-vs-histogram/halftoning_aa_spp64_staircase2_VHH}
        {figures/vertical-vs-histogram/halftoning_aa_spp1_staircase2_average}
    \draw[white,thick] (0.3,0.38) rectangle ++(0.167,0.125);
\end{tikzpicture}
&
\begin{tikzpicture}[scale=5.94]
    \ImageTemplate{contour_sanmiguel}
        {figures/vertical-vs-histogram/halftoning_aa_spp16_sanmiguel_cam14_VHH}
        {figures/vertical-vs-histogram/halftoning_aa_spp1_sanmiguel_cam14_VDBS}
        {figures/vertical-vs-histogram/halftoning_aa_spp64_sanmiguel_cam14_VHH}
        {figures/vertical-vs-histogram/halftoning_aa_spp1_sanmiguel_cam14_average}
    \draw[white,thick] (0.664,0.355) rectangle ++(0.0833,0.063);
\end{tikzpicture}
&
\begin{tikzpicture}[scale=5.94]
    \ImageTemplate{contour_bathroom}
        {figures/vertical-vs-histogram/halftoning_aa_spp16_bathroom_VHH}
        {figures/vertical-vs-histogram/halftoning_aa_spp1_bathroom_VDBS}
        {figures/vertical-vs-histogram/halftoning_aa_spp64_bathroom_VHH}
        {figures/vertical-vs-histogram/halftoning_aa_spp1_bathroom_average}
    \draw[white,thick] (0.31,0.36) rectangle ++(0.133,0.1);
\end{tikzpicture}
\\[-0.6mm]
\begin{tikzpicture}[scale=5.94]
    \ZoomInTemplate
        {figures/vertical-vs-histogram/halftoning_aa_spp16_staircase2_VHH_cropped}
        {figures/vertical-vs-histogram/halftoning_aa_spp1_staircase2_VDBS_cropped}
        {figures/vertical-vs-histogram/halftoning_aa_spp64_staircase2_VHH_cropped}
\end{tikzpicture}
&
\begin{tikzpicture}[scale=5.94]
    \ZoomInTemplate
        {figures/vertical-vs-histogram/halftoning_aa_spp16_sanmiguel_cam14_VHH_cropped}
        {figures/vertical-vs-histogram/halftoning_aa_spp1_sanmiguel_cam14_VDBS_cropped}
        {figures/vertical-vs-histogram/halftoning_aa_spp64_sanmiguel_cam14_VHH_cropped}
\end{tikzpicture}
&
\begin{tikzpicture}[scale=5.94]
    \ZoomInTemplate
        {figures/vertical-vs-histogram/halftoning_aa_spp16_bathroom_VHH_cropped}
        {figures/vertical-vs-histogram/halftoning_aa_spp1_bathroom_VDBS_cropped}
        {figures/vertical-vs-histogram/halftoning_aa_spp64_bathroom_VHH_cropped}
\end{tikzpicture}
\end{tabular}

\let\ImageTemplate\undefined
\let\ZoomInTemplate\undefined
\undefinecolor{contour_stairs}
\undefinecolor{contour_sanmiguel}
\undefinecolor{contour_bathroom}
    \vspace{-3.3mm}
    \caption{
        With a search space of only 4\,spp, our \vertical iterative minimization outperforms histogram sampling \citep{Heitz2019} with 16$\times$ more input samples. Please zoom in to fully appreciate the differences; the full-size images are included in the supplemental material.
    }
    \label{fig:vertical-vs-histogram-64spp}
    \vspace{-2em}
\end{figure*}
%%%%%%%%%%%%%%%%%%%%%%%%%%%%%%

%%%%%%%%%%%%%%%%%%%%%%%%%%%%%%
\begin{figure*}[t!]
    \centering
    %!TEX root = ../paper.tex

\newcommand{\ImageTemplate}[3]{
    \ClipImageRect{0,0}{0.5,1}{#2}
    \ClipImageRect{0.5,0}{0.5,1}{#3}
    \draw[draw=white,thin] (0.5,0) -- (0.5,1);
    \begin{scope}
        \TikzLabel[#1]{0.25,0.025}{Permutation}
        \TikzLabel[#1]{0.75,0.025}{\textbf{Iterative (ours)}}
    \end{scope}
}

\newcommand{\ZoomInTemplate}[2]{%
    \begin{tikzpicture}[scale=5.92]
        \PlaceImage{0.0}{0.0}{0.5}{0.2}{#1}
        \PlaceImage{0.5}{0.0}{0.5}{0.2}{#2}
        \draw[draw=white,line width=0.45mm] (0.5, 0) -- (0.5, 0.2);
    \end{tikzpicture}%
}

\definecolor{contour_staircase}{RGB}{75, 50, 2}
\definecolor{contour_livingroom}{RGB}{56, 43, 39}
\definecolor{contour_bathroom}{RGB}{72, 39, 14}
\definecolor{arrow_color}{RGB}{100,100,100}

\fontsize{7}{8}\selectfont
\hspace*{-1.8em}
\begin{tabular}{c@{\;}c@{\;}c@{\;}c@{}}
    \rotatebox{90}{\footnotesize \hspace{23mm} Frame 16}
    &
    \begin{tikzpicture}[scale=5.92]
        \ImageTemplate{contour_staircase}
            {figures/horizontal-vs-permutation/halftoning_am_spp1_staircase2_frame16_HHP_8_RE}
            {figures/horizontal-vs-permutation/halftoning_am_spp1_staircase2_frame16_HDBS_ms_RE}
        \draw[white,thick] (0.26,0.38) rectangle ++(0.25,0.1);
    \end{tikzpicture}
    &
    \begin{tikzpicture}[scale=5.92]
        \ImageTemplate{contour_livingroom}
            {figures/horizontal-vs-permutation/halftoning_am_spp1_living-room-2_frame16_HHP_8_RE}
            {figures/horizontal-vs-permutation/halftoning_am_spp1_living-room-2_frame16_HDBS_ms_RE}
        \draw[white,thick] (0.10,0.49) rectangle ++(0.2,0.08);
    \end{tikzpicture}
    &
    \begin{tikzpicture}[scale=5.92]
        \ImageTemplate{contour_bathroom}
            {figures/horizontal-vs-permutation/halftoning_am_spp1_bathroom_frame16_HHP_8_RE}
            {figures/horizontal-vs-permutation/halftoning_am_spp1_bathroom_frame16_HDBS_ms_RE}
        \draw[white,thick] (0.28,0.36) rectangle ++(0.2,0.08);
    \end{tikzpicture}
    \\[-0.7mm]
    \rotatebox{90}{\footnotesize \hspace{1.5mm} Frame 1}
    &
    \ZoomInTemplate
        {figures/horizontal-vs-permutation/cropped_halftoning_am_spp1_staircase2_frame1_HHP_8_RE}
        {figures/horizontal-vs-permutation/cropped_halftoning_am_spp1_staircase2_frame1_HDBS_ms_RE}
    &
    \ZoomInTemplate
        {figures/horizontal-vs-permutation/cropped_halftoning_am_spp1_living-room-2_frame1_HHP_8_RE}
        {figures/horizontal-vs-permutation/cropped_halftoning_am_spp1_living-room-2_frame1_HDBS_ms_RE}
    &
    \ZoomInTemplate
        {figures/horizontal-vs-permutation/cropped_halftoning_am_spp1_bathroom_frame1_HHP_8_RE}
        {figures/horizontal-vs-permutation/cropped_halftoning_am_spp1_bathroom_frame1_HDBS_ms_RE}
    \\[-0.85mm]
    \rotatebox{90}{\footnotesize \hspace{1.3mm} Frame 16}
    &
    \ZoomInTemplate
        {figures/horizontal-vs-permutation/cropped_halftoning_am_spp1_staircase2_frame16_HHP_8_RE}
        {figures/horizontal-vs-permutation/cropped_halftoning_am_spp1_staircase2_frame16_HDBS_ms_RE}
    &
    \ZoomInTemplate
        {figures/horizontal-vs-permutation/cropped_halftoning_am_spp1_living-room-2_frame16_HHP_8_RE}
        {figures/horizontal-vs-permutation/cropped_halftoning_am_spp1_living-room-2_frame16_HDBS_ms_RE}
    &
    \ZoomInTemplate
        {figures/horizontal-vs-permutation/cropped_halftoning_am_spp1_bathroom_frame16_HHP_8_RE}
        {figures/horizontal-vs-permutation/cropped_halftoning_am_spp1_bathroom_frame16_HDBS_ms_RE}
    \\[-0.85mm]
    \rotatebox{90}{\fontsize{6.4}{7}\selectfont \hspace{0.1mm} Prediction}
    \begin{tikzpicture}[overlay]
        \draw[arrow_color, arrows={-Triangle[angle=60:4pt,arrow_color,fill=arrow_color]}, line width=0.3mm]
        (-0.11, 1.05) -- (-0.11, 1.37);
    \end{tikzpicture}
    &
    \ZoomInTemplate
        {figures/horizontal-vs-permutation/cropped_halftoning_am_spp4_staircase2_frame16_HHP_8}
        {figures/horizontal-vs-permutation/cropped_halftoning_am_spp4_staircase2_frame16_HDBS_ms}
    &
    \ZoomInTemplate
        {figures/horizontal-vs-permutation/cropped_halftoning_am_spp4_living-room-2_frame16_HHP_8}
        {figures/horizontal-vs-permutation/cropped_halftoning_am_spp4_living-room-2_frame16_HDBS_ms}
    &
    \ZoomInTemplate
        {figures/horizontal-vs-permutation/cropped_halftoning_am_spp4_bathroom_frame16_HHP_8}
        {figures/horizontal-vs-permutation/cropped_halftoning_am_spp4_bathroom_frame16_HDBS_ms}
\end{tabular}

\undefinecolor{contour_staircase}
\undefinecolor{contour_livingroom}
\undefinecolor{contour_bathroom}
\undefinecolor{arrow_color}
\let\ImageTemplate\undefined
\let\ZoomInTemplate\undefined

    \vspace{-2.5mm}
    \caption{
        Comparison of our \horizontal iterative minimization against the permutation method of \citet{Heitz2019} (with retargeting) on 16-frame sequences of static scenes rendered at 4\,spp. Our method does a better job at improving the error distribution frame-to-frame.
    }
    \label{fig:horizontal-methods-second}
\end{figure*}
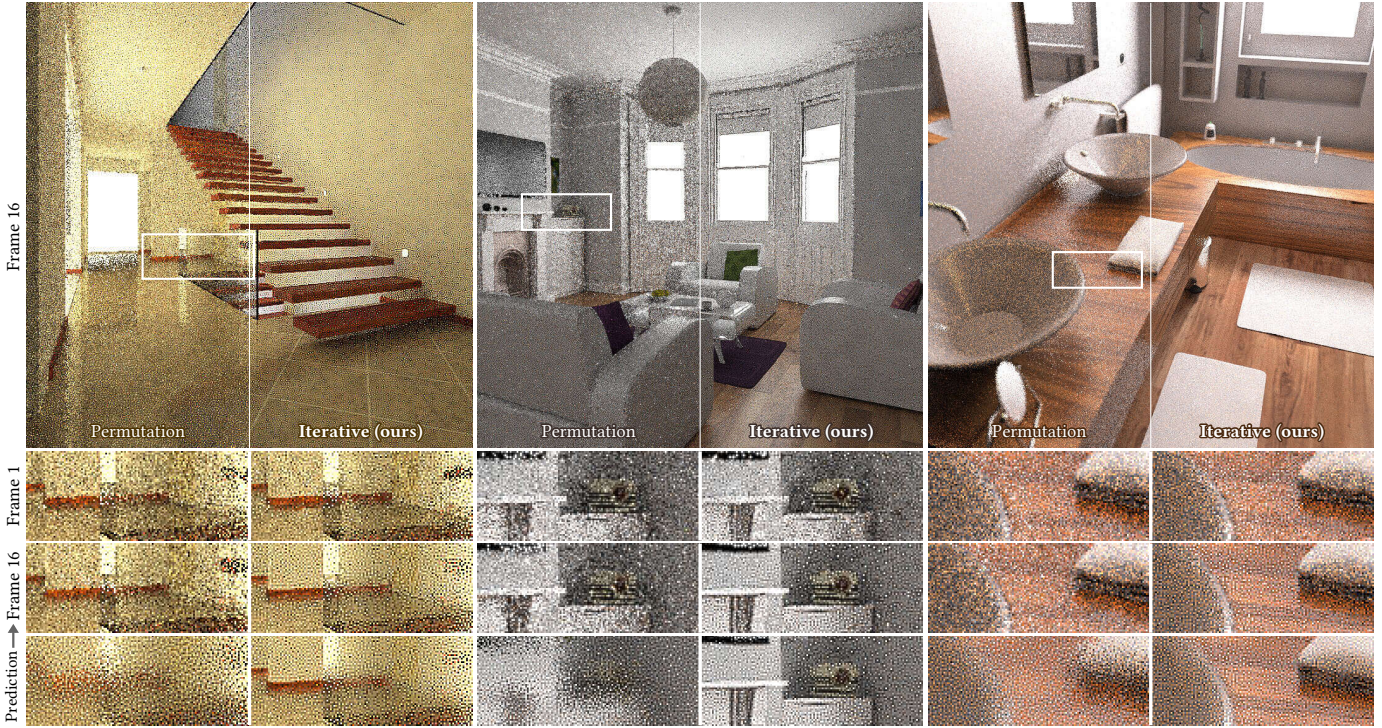
%%%%%%%%%%%%%%%%%%%%%%%%%%%%%%

%%%%%%%%%%%%%%%%%%%%%%%%%%%%%%%%%%%%%%%%%%%%%%%%%%%%%%%%%%%%
\subsection{Rendering comparisons}
\label{sec:rendering_comparisons}
%%%%%%%%%%%%%%%%%%%%%%%%%%%%%%%%%%%%%%%%%%%%%%%%%%%%%%%%%%%%

%%%%%%%%%%%%%%%%%%%%%%%%%%%%%%
\Paragraph{All methods}
%%%%%%%%%%%%%%%%%%%%%%%%%%%%%%

\Cref{fig:staircase-all-methods} shows an equal-sample comparison of all methods. \Vertical methods select one of the 4 input samples per pixel; \horizontal methods are fed a 2-spp average for every pixel, and another 2\,spp are used to render the final image after optimization. Our methods consistently perform best visually, with the \vertical iterative minimization achieving the lowest perceptual error, as corroborated by the HDR-VDP-2 detection maps. Error diffusion is not far behind in quality and is the fastest of all methods along with dithering. The latter is similar to \citeauthor{Heitz2019}'s histogram method but yields a notably better result thanks to using a superior surrogate and performing the thresholding as in the classical halftoning setting (see \cref{sec:Aposteriori}). \Horizontal methods perform worse due to noisier input data (half spp) and worse surrogates derived from it, and also mispredictions (which necessitate re-rendering). Ours still uses a better surrogate than \citeauthor{Heitz2019}'s permutation and is also able to better fit to it. Notice the low fidelity of the 4-spp average image compared to our \vertical methods', even though the latter retain only one of the four input samples for every pixel.

%%%%%%%%%%%%%%%%%%%%%%%%%%%%%%
\Paragraph{\Vertical methods}
%%%%%%%%%%%%%%%%%%%%%%%%%%%%%%

In \cref{fig:vertical-vs-histogram-64spp} we compare our \vertical iterative minimization to the histogram sampling of \citet{Heitz2019}. Both select one of several input samples (\ie estimates) for each pixel. Our method produces a notably better result even when given 16$\times$ fewer samples to choose from. The perceptual error of histogram sampling does not vanish with increasing sample count. It dithers pixel \emph{intensity} rather than pixel error, thus more samples help improve the intensity distribution but not the error magnitude.

\Cref{fig:teaser} shows our most capable method: \vertical iterative minimization with search space extended to the power set of the input samples (with size $2^4 - 1 = 15$ for 4 input spp; see \cref{sec:OptimizationDiscussion}). We compare surrogate-driven optimization against the best-case result---optimization \wrt the ground truth. Both achieve high fidelity, with little difference between them and with pronounced local blue-noise error distribution corroborated by the tiled power spectra.

%%%%%%%%%%%%%%%%%%%%%%%%%%%%%%
\Paragraph{\Horizontal methods \& animation}
%%%%%%%%%%%%%%%%%%%%%%%%%%%%%%

For rendering static images, \horizontal methods are at a disadvantage compared to \vertical ones due to the required post-optimization re-rendering. As \citet{Heitz2019} note, in an animation setting this sampling overhead can be mitigated by reusing the result of one frame as the initial estimate for the next. In \cref{fig:horizontal-methods-second} we compare their permutation method to our \horizontal iterative minimization. For theirs we shift a void-and-cluster mask in screen space per frame and apply retargeting, and for ours we traverse the image pixels in different random order. We intentionally keep the scenes static to test the methods' best-case abilities to improve the error distribution over frames.

Starting from a random initial estimate, our method can benefit from a progressively improving surrogate that helps fine-tune the error distribution via localized pixel swaps. The permutation method operates in greyscale within static non-overlapping tiles. This prevents it from making significant progress after the first frame. While mispredictions cause local deviations from blue noise in both results, these are stronger in the permutation method's. This is evident when comparing the corresponding prediction images---the results of optimization right before re-rendering. The permutation's retargeting pass breaks the blocky image structure caused by tile-based optimization but increases the number of mispredictions.

The supplemental video shows animations with all methods, where \vertical ones are fed a new random estimate per frame. Even without accumulating information over time, these consistently beat the two \horizontal methods. The latter suffer from mispredictions under fast motion and perform similarly to one another, though ours remains superior in the presence of temporal smoothness. Mispredictions could be eliminated by optimizing frames independently and splitting the sampling budget into optimization and re-rendering halves (as in \cref{fig:staircase-all-methods}), though at the cost of reduced sampling quality.

%%%%%%%%%%%%%%%%%%%%%%%%%%%%%%
\Paragraph{Additional comparisons}
%%%%%%%%%%%%%%%%%%%%%%%%%%%%%%

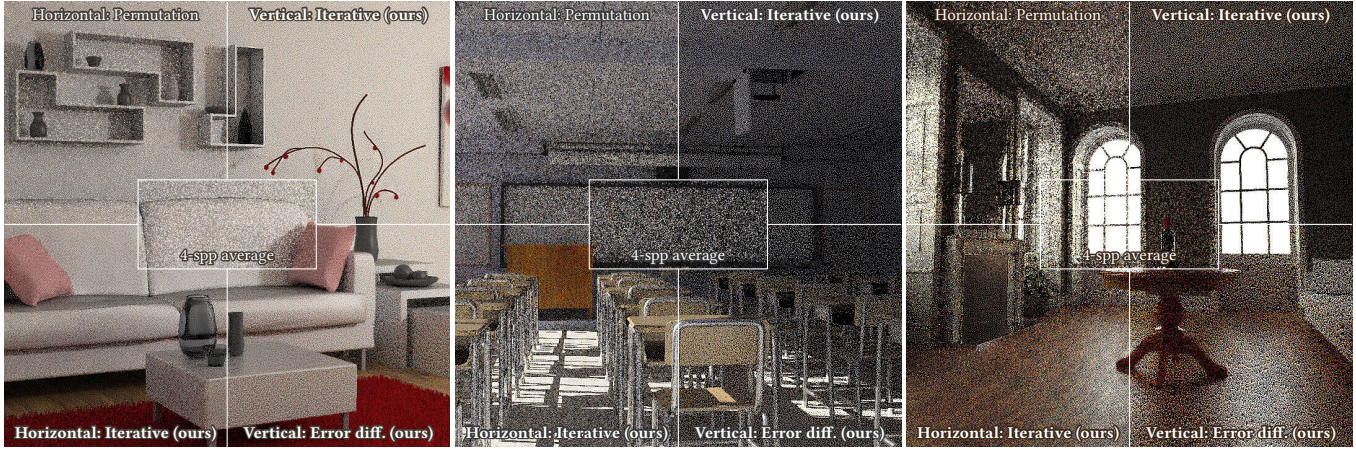
\begin{figure*}[t!]
    \centering
    %!TEX root = ../paper.tex

\newcommand{\VerticalPermHeitzTemplate}[6]{
    \begin{scope}
        \clip (0,0) rectangle ++(0.5,0.5);
        \path[fill overzoom image=figures/horizontal-vertical-vs-permutation/#2] (0,0) rectangle (1,1);
    \end{scope}
    \begin{scope}
        \clip (0.5,0) rectangle ++(0.5,0.5);
        \path[fill overzoom image=figures/horizontal-vertical-vs-permutation/#3] (0,0) rectangle (1,1);
    \end{scope}
    \begin{scope}
        \clip (0.5,0.5) rectangle ++(0.5,0.5);
        \path[fill overzoom image=figures/horizontal-vertical-vs-permutation/#4] (0,0) rectangle (1,1);
    \end{scope}
    \begin{scope}
        \clip (0,0.5) rectangle ++(0.5,0.5);
        \path[fill overzoom image=figures/horizontal-vertical-vs-permutation/#5] (0,0) rectangle (1,1);
    \end{scope}
    \draw[draw=white,thin] (0.5,0) -- (0.5,1);
    \draw[draw=white,thin] (0,0.5) -- (1,0.5);
    \begin{scope}
        \clip (0.3,0.4) rectangle ++(0.4,0.2);
        \path[fill overzoom image=figures/horizontal-vertical-vs-permutation/#6] (0,0) rectangle (1,1);
    \end{scope}
    \draw[white,thin] (0.3,0.4) rectangle ++(0.4,0.2);
    \begin{scope}
        \TikzLabel[#1]{0.25,0.021}{\textbf{\Horizontal: Iterative (ours)}}
        \TikzLabel[#1]{0.75,0.021}{\textbf{\Vertical: Error diff.\ (ours)}}
        \TikzLabel[#1]{0.75,0.96}{\textbf{\Vertical: Iterative (ours)}}
        \TikzLabel[#1]{0.25,0.96}{\Horizontal: Permutation}
        \TikzLabel[#1]{0.5,0.42}{4-spp average}
    \end{scope}
}

\newcommand{\WhiteNoiseVHTemplate}[1]{
    \begin{scope}
        \clip (0,0.3) -- (1,0.3) -- (1,0.55) -- (0,0.55) -- cycle;
        \path[fill overzoom image=figures/horizontal-vertical-vs-permutation/#1] (0,0) rectangle (1,1);
    \end{scope}
}

\definecolor{contour_livingroom}{RGB}{68, 60, 51}
\definecolor{contour_classroom}{RGB}{55,55,55}
\definecolor{contour_livingroom2}{RGB}{60, 40, 27}

\fontsize{6.3}{8}\selectfont
\hspace*{-0.92em}
\begin{tabular}{c@{\;}c@{\;}c@{}}
\begin{tikzpicture}[scale=5.92]
    \VerticalPermHeitzTemplate{contour_livingroom}
        {halftoning_am_spp1_living-room-3_frame16_HDBS_RE}
        {halftoning_aa_spp1_living-room-3_VFS}
        {halftoning_aa_spp1_living-room-3_VDBS}
        {halftoning_am_spp1_living-room-3_frame16_HHP_8_RE}
        {halftoning_aa_spp1_living-room-3_average}
\end{tikzpicture}
&
\begin{tikzpicture}[scale=5.92]
    \VerticalPermHeitzTemplate{contour_classroom}
        {halftoning_am_spp1_classroom_frame16_HDBS_RE}
        {halftoning_aa_spp1_classroom_VFS}
        {halftoning_aa_spp1_classroom_VDBS}
        {halftoning_am_spp1_classroom_frame16_HHP_8_RE}
        {halftoning_aa_spp1_classroom_average}
\end{tikzpicture}
&
\begin{tikzpicture}[scale=5.92]
    \VerticalPermHeitzTemplate{contour_livingroom2}
        {halftoning_am_spp1_living-room_frame16_HDBS_RE}
        {halftoning_aa_spp1_living-room_VFS}
        {halftoning_aa_spp1_living-room_VDBS}
        {halftoning_am_spp1_living-room_frame16_HHP_8_RE}
        {halftoning_aa_spp1_living-room_average}
\end{tikzpicture}
\end{tabular}

\undefinecolor{contour_livingroom}
\undefinecolor{contour_classroom}
\undefinecolor{contour_livingroom2}
\let\VerticalPermHeitzTemplate\undefined
\let\WhiteNoiseVHTemplate\undefined
    \vspace{-2.5mm}
    \caption{
        Comparison of our methods against the permutation approach of \citet{Heitz2019} at 4\,spp; for the \horizontal methods we show the result of the 16\textsuperscript{th} frame of static-scene rendering. Our two iterative minimization algorithms yield the best quality, while error diffusion is fastest (see \cref{tab:Errors,tab:Timings}).
    }
    \label{fig:vertical-vs-permutation-heitz}
\end{figure*}

\Cref{fig:vertical-vs-permutation-heitz} shows additional results from our \horizontal and \vertical minimization and error diffusion. We compare these to the permutation method of \citet{Heitz2019} which we found to perform better than their histogram approach on static scenes at equal sampling rates. For the \horizontal methods we show the results after 16 iterations. Our methods again yield lower error, subjectively and numerically (see~\cref{tab:Errors,tab:Timings}).

%%%%%%%%%%%%%%%%%%%%%%%%%%%%%%
%%%%%%%%%%%%%%%%%%%%%%%%%%%%%%

%%%%%%%%%%%%%%%%%%%%%%%%%%%%%%
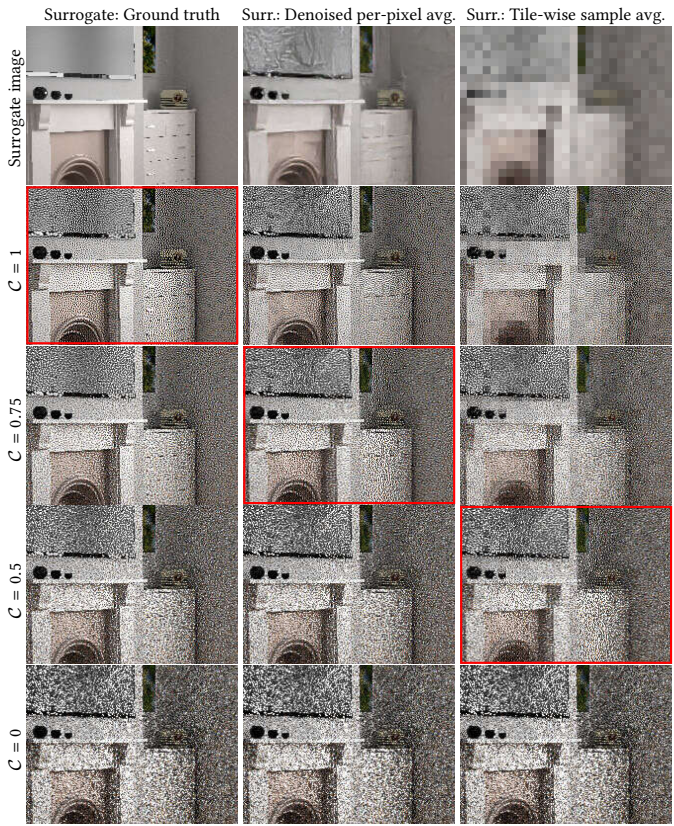
\begin{figure}[t!]
    \centering
    %!TEX root = ../paper.tex

\newcommand{\ClipImageRectT}[8]{
    \begin{scope}
        \clip (#1,#2) rectangle ++(#3,#4);
        \path[fill zoom image=figures/bias-analysis/#8] ({#1-(#5)*#7},{#2-(#6)*#7}) rectangle ++(#7,#7);
    \end{scope}
}

\newcommand{\ImageTemplate}[1]{
    \ClipImageRectT{0}{0.2}{1}{0.75}{0.22}{0.45}{6}{#1}
}

%%%%%%%%%%%%%%%%%%%%%%%%%%%%%%%%%%%%

\definecolor{contour}{RGB}{50,50,50}
% \footnotesize
\fontsize{6.7}{7}\selectfont
\hspace*{-1.85em}
\begin{tabular}{c@{\;}c@{\;}c@{\;}c@{}}
{} & Surrogate: Ground truth & \!\!\! Surr.: Denoised per-pixel avg. \!\!\! & Surr.: Tile-wise sample avg.
\\[0.2mm]
\rotatebox{90}{\hspace{2.3mm}Surrogate image}\!
&
\begin{tikzpicture}[scale=2.8]
    \ImageTemplate{surr_bs_spp4_living-room-2_REF}
\end{tikzpicture}
&
\begin{tikzpicture}[scale=2.8]
    \ImageTemplate{den_aa_spp1_living-room-2_average}
\end{tikzpicture}
&
\begin{tikzpicture}[scale=2.8]
    \ImageTemplate{surr_bs_spp4_living-room-2_HP8}
\end{tikzpicture}
\\[-0.5mm]
\rotatebox{90}{\hspace{7mm}$\ConfidenceBias=1$}
&
\begin{tikzpicture}[scale=2.8]
    \ImageTemplate{halftoning_bs_spp4_living-room-2_VDBSC0_REF}
    \draw[red,thick] (0.004,0.205) rectangle (1,0.9453);
\end{tikzpicture}
&
\begin{tikzpicture}[scale=2.8]
    \ImageTemplate{halftoning_bs_spp4_living-room-2_VDBSC0_DIN}
\end{tikzpicture}
&
\begin{tikzpicture}[scale=2.8]
    \ImageTemplate{halftoning_bs_spp4_living-room-2_VDBSC0_DHP8}
\end{tikzpicture}
\\[-0.5mm]
\rotatebox{90}{\hspace{5.6mm}$\ConfidenceBias=0.75$}
&
\begin{tikzpicture}[scale=2.8]
    \ImageTemplate{halftoning_bs_spp4_living-room-2_VDBSC1_REF}
\end{tikzpicture}
&
\begin{tikzpicture}[scale=2.8]
    \ImageTemplate{halftoning_bs_spp4_living-room-2_VDBSC1_DIN}
    \draw[red,thick] (0.004,0.205) rectangle (1,0.9453);
\end{tikzpicture}
&
\begin{tikzpicture}[scale=2.8]
    \ImageTemplate{halftoning_bs_spp4_living-room-2_VDBSC1_DHP8}
\end{tikzpicture}
\\[-0.5mm]
\rotatebox{90}{\hspace{6mm}$\ConfidenceBias=0.5$}
&
\begin{tikzpicture}[scale=2.8]
    \ImageTemplate{halftoning_bs_spp4_living-room-2_VDBSC2_REF}
\end{tikzpicture}
&
\begin{tikzpicture}[scale=2.8]
    \ImageTemplate{halftoning_bs_spp4_living-room-2_VDBSC2_DIN}
\end{tikzpicture}
&
\begin{tikzpicture}[scale=2.8]
    \ImageTemplate{halftoning_bs_spp4_living-room-2_VDBSC2_DHP8}
    \draw[red,thick] (0.004,0.205) rectangle (1,0.9453);
\end{tikzpicture}
\\[-0.5mm]
\rotatebox{90}{\hspace{7mm}$\ConfidenceBias=0$}
&
\begin{tikzpicture}[scale=2.8]
    \ImageTemplate{halftoning_bs_spp4_living-room-2_VDBSC4_REF}
\end{tikzpicture}
&
\begin{tikzpicture}[scale=2.8]
    \ImageTemplate{halftoning_bs_spp4_living-room-2_VDBSC4_DIN}
\end{tikzpicture}
&
\begin{tikzpicture}[scale=2.8]
    \ImageTemplate{halftoning_bs_spp4_living-room-2_VDBSC4_DHP8}
\end{tikzpicture}

\end{tabular}

%%%%%%%%%%%%%%%%%%%%%%%%%%%%%%%%%%%%

\let\BiasAnalysisVertTemplate\undefined
\let\ImageTemplate\undefined
    \vspace{-2.5mm}
    \caption{
        Balancing our iterative optimization between the surrogate (top row) and the initial estimate (bottom row) via the parameter $\ConfidenceBias$ from \cref{eq:energy-bias}. For high-quality surrogates (left and middle columns), the best result is achieved for values of $\ConfidenceBias$ close to 1. In contrast, strong structural artifacts (right column) call for lowering $\ConfidenceBias$ to avoid fitting too closely to the surrogate. The (subjectively) best image in each column is outlined in red.
    }
    \label{fig:bias-analysis}
\end{figure}
%%%%%%%%%%%%%%%%%%%%%%%%%%%%%%

%%%%%%%%%%%%%%%%%%%%%%%%%%%%%%%%%%%%%%%%%%%%%%%%%%%%%%%%%%%%
\section{Discussion}
\label{sec:Discussion}
%%%%%%%%%%%%%%%%%%%%%%%%%%%%%%%%%%%%%%%%%%%%%%%%%%%%%%%%%%%%

%%%%%%%%%%%%%%%%%%%%%%%%%%%%%%%%%%%%%%%%%%%%%%%%%%%%%%%%%%%%
\subsection{Bias towards surrogate}
%%%%%%%%%%%%%%%%%%%%%%%%%%%%%%%%%%%%%%%%%%%%%%%%%%%%%%%%%%%%

While ultimately we want to optimize \wrt the ground-truth image, in practice we have to rely on a surrogate. In our experiments we use reasonably high-quality surrogates, shown in \cref{fig:surrogates}, to best demonstrate the capabilities of our framework. But when using a surrogate of low quality, fitting too closely to it can produce an estimate with artifacts. In such cases less aggressive fitting may yield \emph{lower} perceptual error. To explore the trade-off, in \cref{sec:BiasConfidenceEnergy} we augment the perceptual error with a term that penalizes deviations from the initial estimate $\EstimatedImageInit$ (which case of \vertical optimization is obtained by averaging the input per-pixel estimates):
\begin{align}
    \label{eq:energy-bias}
    \Energy_\ConfidenceBias \,=\, (1 - \ConfidenceBias) \Norm{\Kernel}^2_1 \Norm{\EstimatedImage - \EstimatedImageInit}^2_2 \,+\, \ConfidenceBias\,\Energy.
\end{align}
The parameter $\ConfidenceBias \in [0,1]$ encodes our confidence in the surrogate quality. Setting $\ConfidenceBias=1$ reverts to the original formulation~\eqref{eq:energy_experiments}, while optimizing with $\ConfidenceBias=0$ yields the initial image estimate $\EstimatedImageInit$. Optimizing \wrt this energy can also be interpreted as projecting the surrogate onto the space of Monte Carlo estimates in $\SampleSetImageDomain$, with control over the fitting power of the projection via $\ConfidenceBias$.

In~\cref{fig:bias-analysis}, we plug the extended error formulation~\eqref{eq:energy-bias} into our \vertical iterative minimization. The results indicate that the visually best result is achieved for different values of $\ConfidenceBias$ depending on the surrogate quality. Specifically, when optimizing \wrt the ground truth, the fitting should be most aggressive: $\ConfidenceBias = 1$. Conversely, if the surrogate contains structural artifacts, the optimization should be made less biased to it, \eg by setting $\ConfidenceBias = 0.5$. Other ways to control this bias are using a more restricted search space (\eg non-power-set) and capping the number of minimization iterations of our methods. Note that the methods of \citet{Heitz2019} rely on implicit surrogates and energies and thus provide no control over this trade-off. We have found that their permutation method generally avoids tiling artifacts induced by their piecewise constant surrogate due to the retargeting step blurring the prediction image (shown in \cref{fig:horizontal-methods-second} zoom-ins); however, this blurring adds mispredictions which deteriorate the final image quality. Our methods provide better fits, target the error explicitly, and are much superior when the surrogate is good. With a bad surrogate, ours can be controlled to never do worse than theirs.

%%%%%%%%%%%%%%%%%%%%%%%%%%%%%%%%%%%%%%%%%%%%%%%%%%%%%%%%%%%%
\subsection{Denoising}
%%%%%%%%%%%%%%%%%%%%%%%%%%%%%%%%%%%%%%%%%%%%%%%%%%%%%%%%%%%%

Our images are optimized for eliminating error and preserving features when blurred with a given kernel. This blurring can be seen as a simple form of denoising, and it is reasonable to expect that the images are also better suited for general-purpose denoising than traditional white-noise renderings are~\citep{Heitz2019,Belcour:2021:Lessons}. However, we have found that obtaining such benefit is not straightforward.

\begin{figure}[t!]
    \centering
    %!TEX root = ../paper.tex

\newcommand{\LeftTemplate}[3]{
    \begin{scope}
        \clip (2.3,2.0) -- (3.2,2.0) -- (3.2,3.2) -- (2.3,3.2) -- cycle;
        \path[fill overzoom image=figures/denoising/#3] (0,0) rectangle (8.5,4.25);
    \end{scope}
}

\newcommand{\LeftSplitTemplate}[4]{
    \begin{scope}
        \clip (2.3,2.0) -- (3.2,2.0) -- (2.3,3.2) -- cycle;
        \path[fill overzoom image=figures/denoising/#3] (0,0) rectangle (8.5,4.25);
    \end{scope}
    \begin{scope}
        \clip (3.2,2.0) -- (3.2,3.2) -- (2.3,3.2) -- cycle;
        \path[fill overzoom image=figures/denoising/#4] (0,0) rectangle (8.5,4.25);
        \draw[draw=white,line width=0.3mm] (3.2,2.0) -- (2.3,3.2);
    \end{scope}
}

\newcommand{\RightTemplate}[3]{
    \begin{scope}
        \clip (4.5,1.5) -- (6.0,1.5) -- (6.0,3.5) -- (4.5,3.5) -- cycle;
        \path[fill overzoom image=figures/denoising/#3] (-0.1,0) rectangle ++(8.5,4.25);
    \end{scope}
}

\newcommand{\RightSplitTemplate}[4]{
    \begin{scope}
        \clip (4.5,1.5) -- (6.0,1.5) -- (4.5,3.5) -- cycle;
        \path[fill overzoom image=figures/denoising/#3] (-0.1,0) rectangle ++(8.5,4.25);
    \end{scope}
    \begin{scope}
        \clip (6.0,1.5) -- (6.0,3.5) -- (4.5,3.5) -- cycle;
        \path[fill overzoom image=figures/denoising/#4] (-0.1,0) rectangle ++(8.5,4.25);
        \draw[draw=white,line width=0.3mm] (6.0,1.5) -- (4.5,3.5);
    \end{scope}
}

\newcommand{\RightSplitTemplateTwo}[4]{
    \begin{scope}
        \clip (4.5,1.5) -- (6.0,1.5) -- (6.0,3.5) -- cycle;
        \path[fill overzoom image=figures/denoising/#3] (-0.1,0) rectangle ++(8.5,4.25);
    \end{scope}
    \begin{scope}
        \clip (4.5,1.5) -- (6.0,3.5) -- (4.5,3.5) -- cycle;
        \path[fill overzoom image=figures/denoising/#4] (-0.1,0) rectangle ++(8.5,4.25);
        \draw[draw=white,line width=0.3mm] (4.5,1.5) -- (6.0,3.5);
    \end{scope}
}

\tikzset{myarrow/.style={
    graph_outlinecolor, arrows={-Triangle[angle=75:4pt,graph_outlinecolor,fill=graph_outlinecolor]}, line width=0.45mm
}}
\tikzset{mycircle/.style={
    graph_outlinecolor,fill=graph_backcolor,line width=0.25mm,fill opacity=0.9
}}

\newcommand{\DrawGraph}[3]{
    \draw[myarrow]  (#1 - 2.42, #2 + 3.05) -- (#1 - 2.42, #2 + 2.76); % down
    \draw[myarrow]  (#1 - 1.86, #2 + 3.05) -- (#1 - 1.86, #2 + 2.76); % down
    \draw[myarrow]  (#1 - 2.28, #2 + 3.20) -- (#1 - 1.98, #2 + 3.20); % right
    \draw[myarrow]  (#1 - 2.32, #2 + 2.73) -- (#1 - 1.95, #2 + 3.11); % up-right
    \draw[mycircle] (#1 - 2.42, #2 + 3.20) circle (1.4mm) node {\textcolor{graph_textcolor}{\textsf{1}}};
    \draw[mycircle] (#1 - 2.42, #2 + 2.63) circle (1.4mm) node {\textcolor{graph_textcolor}{\textsf{2}}};
    \draw[mycircle] (#1 - 1.86, #2 + 3.20) circle (1.4mm) node {\textcolor{graph_textcolor}{\textsf{3}}};
    \draw[mycircle] (#1 - 1.86, #2 + 2.63) circle (1.4mm) node {\textcolor{graph_textcolor}{\textsf{4}}};
    \TikzLabelR[#3]{#1 - 2.58, #2 + 3.176}{4-spp input}
    \TikzLabelR[#3]{#1 - 2.58, #2 + 2.615}{Input denoised}
    \TikzLabelL[#3]{#1 - 1.73, #2 + 3.206}{\textbf{Ours}}
    \TikzLabelL[#3]{#1 - 1.73, #2 + 2.642}{\textbf{Ours denoised}}
    \TikzLabelR[#3]{#1 - 0.03, #2 + 5.5}{\fontsize{6.0}{7}\selectfont\textbf{Ours $\Kernel$-convolved}}
}

\definecolor{contour_color_left}{RGB}{40,40,40}
\definecolor{contour_color_right}{RGB}{67,42,38}
\definecolor{graph_outlinecolor}{RGB}{40,40,40}
\definecolor{graph_textcolor}{RGB}{0,0,0}
\definecolor{graph_backcolor}{RGB}{255,255,255}

% \footnotesize
\fontsize{6.5}{7}
\hspace*{-2.0mm}
\begin{tabular}{c@{\,}c@{\hspace{1mm}}c@{\,}c@{}}
\begin{tikzpicture}[scale=2.316]
   \LeftTemplate{4-spp average}{2.187}{halftoning_aa_spp4_living-room-2_average}
\end{tikzpicture}
&
\begin{tikzpicture}[scale=2.316]
    \LeftSplitTemplate{Ours}{2.187}
        {halftoning_aa_spp4_living-room-2_VDBS_DIN}
        {conv_aa_spp4_living-room-2_VDBS_DIN}
\end{tikzpicture}
&
\begin{tikzpicture}[scale=1.39]
    \RightTemplate{4-spp average}{3.5}{halftoning_aa_spp4_living-room-3_average}
\end{tikzpicture}
&
\begin{tikzpicture}[scale=1.39]
    \RightSplitTemplate{Ours}{3.5}
        {halftoning_aa_spp4_living-room-3_VDBS_DIN}
        {conv_aa_spp4_living-room-3_VDBS_DIN}
\end{tikzpicture}
\\ [-0.5mm]
\begin{tikzpicture}[scale=2.316]
   \LeftTemplate{Denoised}{3.205}{den_aa_spp4_living-room-2_average}
\end{tikzpicture}
&
\begin{tikzpicture}[scale=2.316]
   \LeftTemplate{Ours denoised}{3.205}{den_aa_spp4_living-room-2_VDBS_DIN}
\end{tikzpicture}
&
\begin{tikzpicture}[scale=1.39]
    \RightTemplate{4-spp avg.\ denoised}{1.81}{den_aa_spp4_living-room-3_average}
\end{tikzpicture}
&
\begin{tikzpicture}[scale=1.39]
    \RightTemplate{Ours denoised}{1.81}{den_aa_spp4_living-room-3_VDBS_DIN}
\end{tikzpicture}
\end{tabular}

\begin{tikzpicture}[overlay]
    \DrawGraph{0.003}{0.05}{contour_color_left}
    \DrawGraph{4.33}{0.05}{contour_color_right}
\end{tikzpicture}

\undefinecolor{contour_color_left}
\undefinecolor{contour_color_right}
\undefinecolor{graph_outlinecolor}
\undefinecolor{graph_textcolor}
\undefinecolor{graph_backcolor}
\let\LeftTemplate\undefined
\let\LeftSplitTemplate\undefined
\let\RightTemplate\undefined
\let\RightSplitTemplate\undefined
\let\RightSplitTemplateTwo\undefined
    \vspace{-3mm}
    \caption{
        By regularizing a noisy input, our optimization can help a denoiser avoid producing artifacts (left scene), even though it targets a different (perceptual) smoothing kernel $\Kernel$. However, it can also cause elimination of image features during denoising (right scene, the shadow).
    }
    \label{fig:artifacts-denoising-wide}
\end{figure}
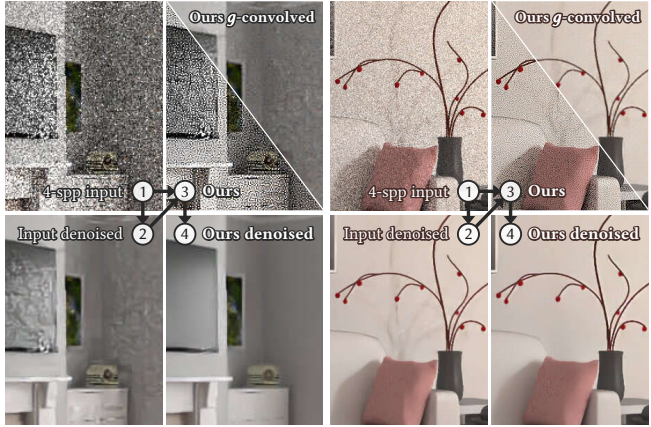

In \cref{fig:artifacts-denoising-wide} we run Intel Open Image Denoise on the results from our \vertical iterative minimization. On the left scene, the input samples~\CircleNumber{1} have white-noise image distribution with large magnitude; feeding their per-pixel averages to the denoiser, it cannot reliably separate the signal from the noise and produces conspicuous artifacts. Using this denoised image~\CircleNumber{2} as a surrogate for our optimization yields a ``regularized'' version~\CircleNumber{3} of the input that is easier for the denoiser to subsequently filter~\CircleNumber{4}. This process can be seen as projecting the initial denoised image back onto the space of exact per-pixel estimates (while minimizing the pMSE) whose subsequent denoising avoids artifacts. Note that obtaining this improved result requires no additional pixel sampling.

On the right scene in \cref{fig:artifacts-denoising-wide}, the moderate input-noise level is easy for the denoiser to clean while preserving the faint shadow on the wall. Our optimization subsequently produces an excellent result which yields a high-fidelity image when convolved with the optimization kernel $\Kernel$. Yet that same result is ruined by the denoiser which eradicates the shadow, even though subjectively its signal-to-noise ratio is higher than that of the input image. Overall, the denoiser blurs our result~\CircleNumber{3} aggressively on both scenes, eliminating not only the high-frequency noise but also lower-frequency signal not present in auxiliary input feature buffers (depth, normals, etc).

%%%%%%%%%%%%%%%%%%%%%%%%%%%%%%
%%%%%%%%%%%%%%%%%%%%%%%%%%%%%%

It should not be too surprising that an image optimized for one smoothing kernel does not always yield good results when filtered with other kernels. As an example, \cref{fig:Kernels} shows clearly that the optimal noise distribution varies significantly across different kernels. While our kernel $\Kernel$ has narrow support and fixed shape, denoising kernels vary wildly over the image and are inferred from the input in order to preserve features. Importantly, modern kernel-inference models (like in the used denoiser) are designed (or trained) to expect mutually uncorrelated pixel estimates~\citep{inteldenoiser}. This white-noise-error assumption can also yield wide smoothing kernels that are unnecessarily aggressive for blue-noise distributions; we suspect this is what hinders the denoiser from detecting features present in our optimized results whose pixels are highly correlated.

Our firm belief is that denoising could consistently benefit from error optimization, though that would require better coordination between the two. One avenue for future work would be to tailor the optimization to the kernels employed by a target denoiser. Conversely, denoising could be adapted to ingest correlated pixel estimates with high-frequency error distribution; this would enable the use of less aggressive smoothing kernels (see \cref{fig:blue-noise-steady-state}) and facilitate feature preservation. As a more immediate treatment, image features could be enhanced before or after our optimization to mitigate the risk of them being eliminated by denoising.

%%%%%%%%%%%%%%%%%%%%%%%%%%%%%%
\newcommand{\BestValue}[1]{\cellcolor{black!9}#1}
\begin{table*}
    \vspace{-2mm}
    \centering
    \small
    \caption{
        MSE and pMSE (\cref{sec:ExperimentSetup}) metrics for various methods (ours in bold) and scenes. For \horizontal methods we show the metrics for the 16\textsuperscript{th} frame of static-scene rendering. In each section we highlight the lowest error number per column. For the same number of samples per pixel (spp), our methods consistently outperform those of \citet{Heitz2019}---the current state of the art, except our dithering can do worse than their permutation method.
    }
    \label{tab:Errors}
    \vspace{-2.5mm}
    \setlength{\tabcolsep}{2.5pt}
    \begin{tabularx}{\textwidth}{X cc cc cc cc cc cc cc cc cc cc}
        \toprule
        \multirow{2}{*}{\textbf{Method}} & \multicolumn{2}{c}{\fontsize{7.8}{7}\selectfont\textbf{Bathroom}} & \multicolumn{2}{c}{\fontsize{7.8}{7}\selectfont\textbf{Classroom}} &
        \multicolumn{2}{c}{\fontsize{7.8}{7}\selectfont\!\textbf{Gray Room}\!} &
        \multicolumn{2}{c}{\fontsize{7.8}{7}\selectfont\!\textbf{Living Room}\!} &
        \multicolumn{2}{c}{\fontsize{7.8}{7}\selectfont\!\textbf{Modern Hall}\!} &
        \multicolumn{2}{c}{\fontsize{7.8}{7}\selectfont\!\textbf{San Miguel}\!} &
        \multicolumn{2}{c}{\fontsize{7.8}{7}\selectfont\textbf{Staircase}} &
        \multicolumn{2}{c}{\fontsize{7.8}{7}\selectfont\!\textbf{White Room}\!}
        \\
        \cmidrule(lr){2-3}\cmidrule(lr){4-5}\cmidrule(lr){6-7}\cmidrule(lr){8-9}\cmidrule(lr){10-11}\cmidrule(lr){12-13}\cmidrule(lr){14-15}\cmidrule(lr){16-17}
        & 
        \!MSE\! & \!pMSE\! & 
        \!MSE\! & \!pMSE\! & 
        \!MSE\! & \!pMSE\! & 
        \!MSE\! & \!pMSE\! & 
        \!MSE\! & \!pMSE\! & 
        \!MSE\! & \!pMSE\! & 
        \!MSE\! & \!pMSE\! & 
        \!MSE\! & \!pMSE\!
        \\
        &
        \scriptsize $\times 10^{-2}$ & \scriptsize $\times 10^{-3}$ & 
        \scriptsize $\times 10^{-2}$ & \scriptsize $\times 10^{-3}$ & 
        \scriptsize $\times 10^{-2}$ & \scriptsize $\times 10^{-2}$ &
        \scriptsize $\times 10^{-2}$ & \scriptsize $\times 10^{-3}$ &
        \scriptsize $\times 10^{-2}$ & \scriptsize $\times 10^{-2}$ &
        \scriptsize $\times 10^{-2}$ & \scriptsize $\times 10^{-3}$ &
        \scriptsize $\times 10^{-3}$ & \scriptsize $\times 10^{-3}$ &
        \scriptsize $\times 10^{-2}$ & \scriptsize $\times 10^{-3}$
        \\
        \midrule
        Random (4-spp average) &
         1.40 &  3.15 &
         3.13 & 7.91  &
         7.91 &  3.02 &
         3.37 &  5.61 &
         5.22 &  1.70 &
         3.58 &  8.92 &  
         8.88 & 5.60 &
         2.78 &  7.98
        \\%[0.92mm]
        \rowcolor{TableAltBackColor}
        \Vertical: Histogram \shortcite{Heitz2019} (\nicefrac{1}{4}\,spp) &
         3.58 &  6.29 &
         7.11 & 13.08 &
        11.49 &  6.67 &
         5.75 &  9.88 &
        11.43 &  3.60 &
        6.84  & 16.52 & 
        18.90 &  6.69 &
         5.75 & 14.09
        \\
        \textbf{\Vertical: Error diffusion} (\nicefrac{1}{4}\,spp) &
        \BestValue{1.22} & 2.27 &
        4.91 & 7.03  &
        8.76 & 2.82  &
        \BestValue{2.08} &  2.31 &
        4.86 & 1.33 &
        5.07 &  8.50 &  
        \BestValue{6.87} & 5.08 &
        \BestValue{2.19} &  5.16
        \\
        \rowcolor{TableAltBackColor}
        \textbf{\Vertical: Dithering} (\nicefrac{1}{4}\,spp) &
        1.31 & 3.31 &
        4.36 & 11.63 &
        8.46 &  5.07 &
        2.27 &  4.43 &
        5.25 &  1.80 &
        3.74 & 11.19 & 
        7.80 &  5.36 &
        2.51 &  7.95
        \\
        \textbf{\Vertical: Iterative} (\nicefrac{1}{4}\,spp) &
        2.32 &  2.02 &
        6.00 &  6.10 &
        9.07 &  2.97 &
        4.32 &  1.86 &
        7.15 &  1.29 &
        5.51 &  7.05 &  
        10.50 & 4.45 &
        3.98 &  5.00
        \\
        \rowcolor{TableAltBackColor}
        \textbf{\Vertical: Iterative} (power set, \nicefrac{1}{15}\,``spp'') &
        1.26 &  \BestValue{1.66} &
        \BestValue{3.12} &  \BestValue{4.91} &
        \BestValue{7.53} &  2.82 &
        2.46 & \BestValue{1.13} &
        \BestValue{4.55} &  \BestValue{1.18} &
        \BestValue{3.31} & \BestValue{5.85} & 
        7.08 & 4.31 &
        2.26 &	\BestValue{4.58}
        \\%[0.92mm]
        \Horizontal:\,Permut.\,\shortcite{Heitz2019}\,(frame\,16,\,4\,spp)\!\! &
        1.40 & 2.79 &
        3.15 & 7.25 &
        7.90 & 2.84 &
        3.38 & 3.14 &
        5.21 & 1.51 &
        3.59 & 8.51 &
        8.87 & 5.40 &
        2.72 & 6.73
        \\
        \rowcolor{TableAltBackColor}
        \textbf{\Horizontal: Iterative} (frame 16, 4\,spp) &
        1.52 & 2.06 &
        3.83 & 5.31 &
        8.34 & \BestValue{2.41} &
        3.59 & 1.59 &
        5.46 & 1.18 &
        3.94 & 7.31 &
        7.67 & \BestValue{4.30} &
        2.93 & 4.72
        \\
        \cmidrule(lr){1-17}
        Random (16-spp average) &
        0.49 & 1.47 &
        1.55 & 4.89 &
        \BestValue{3.77} & 1.04 &
        1.23 & 2.18 &
        2.14 & 0.80 &
        \BestValue{1.10} & 4.67 &
        3.39 & 3.78 &
        1.35  & 3.62
        \\%[0.92mm]
        \rowcolor{TableAltBackColor}
        \Vertical: Histogram \shortcite{Heitz2019} (\nicefrac{4}{16}\,spp) &
        1.40 & 2.37 &
        3.12 & 6.20 &
        7.88 & 2.72 &
        3.36 & 3.57 &
        5.23 & 1.48 &
        3.52 & 6.82 &
        7.13 & 4.09 &
        2.77 & 5.77
        \\
        \textbf{\Vertical: Error diffusion} (\nicefrac{4}{16}\,spp) &
        \BestValue{0.41} & 1.20 &
        \BestValue{0.94} & 3.85 &
        4.00          & 0.87 &
        \BestValue{0.86} & 1.07 &
        \BestValue{1.68} & 0.66 &
        1.33 & 4.70 &
        \BestValue{2.76} & 3.69 &
        \BestValue{0.73} & 2.13
        \\
        \rowcolor{TableAltBackColor}
        \textbf{\Vertical: Dithering} (\nicefrac{4}{16}\,spp) &
        0.50 & 1.52 &
        1.15 & 4.69 &
        4.12 & 1.36 &
        1.09 & 1.82 &
        1.93 & 0.83 &
        1.49 & 5.38 &
        3.09 & 3.73 &
        0.91  & 2.98
        \\
        \textbf{\Vertical: Iterative} (\nicefrac{4}{16}\,spp) &
        0.90 & \BestValue{1.10} &
        2.03 & \BestValue{3.35} &
        5.17 & \BestValue{0.84} &
        2.30 & \BestValue{0.84} &
        3.03 & \BestValue{0.64} &
        2.39 & \BestValue{4.02} &
        4.46 & \BestValue{3.14} &
        1.75 & \BestValue{1.99}
        \\
        \bottomrule
    \end{tabularx}
    \vspace{-1mm}
\end{table*}
%%%%%%%%%%%%%%%%%%%%%%%%%%%%%%

%%%%%%%%%%%%%%%%%%%%%%%%%%%%%%
\begin{table*}
    \centering
    \small
    \caption{
        Optimization run times (in seconds) for various methods (ours in bold) and scenes using 4 input samples per pixel (spp), excluding sampling and surrogate construction. For \horizontal methods we report the average time over 16 frames. Our error diffusion and dithering avoid sorting and are fastest; though dithering-based, \citeauthor{Heitz2019}'s approaches use sorting. Our iterative minimization methods are slowest (but can be sped up; see \cref{sec:Performance}).
    }
    \label{tab:Timings}
    \vspace{-2mm}
    \setlength{\tabcolsep}{2.5pt}
    \begin{tabularx}{\textwidth}{X c c c c c c c c}
        \toprule
        \multirow{1}{*}{\textbf{Method}} & 
        \multicolumn{1}{c}{\fontsize{7.8}{7}\selectfont\textbf{Bathroom}} &
        \multicolumn{1}{c}{\fontsize{7.8}{7}\selectfont\textbf{Classroom}} &
        \multicolumn{1}{c}{\fontsize{7.8}{7}\selectfont\textbf{Gray Room}} &
        \multicolumn{1}{c}{\fontsize{7.8}{7}\selectfont\textbf{Living Room}} &
        \multicolumn{1}{c}{\fontsize{7.8}{7}\selectfont\textbf{Modern Hall}} &
        \multicolumn{1}{c}{\fontsize{7.8}{7}\selectfont\textbf{San Miguel}} &
        \multicolumn{1}{c}{\fontsize{7.8}{7}\selectfont\textbf{Staircase}} &
        \multicolumn{1}{c}{\fontsize{7.8}{7}\selectfont\textbf{White Room}}
        \\
        \midrule
        \Vertical: Histogram \shortcite{Heitz2019} (\nicefrac{1}{4}\,spp) &
        0.06 & 0.07 & 0.11 & 0.06 & 0.02 & 0.09 & 0.08 & 0.06 \\
        \rowcolor{TableAltBackColor}
        \textbf{\Vertical: Error diffusion} (\nicefrac{1}{4}\,spp) &
        \BestValue{0.04} &  \BestValue{0.03} & \BestValue{0.04} & \BestValue{0.04} & \BestValue{0.01} & 0.06 & \BestValue{0.04} & \BestValue{0.04} \\
        \textbf{\Vertical: Dithering} (\nicefrac{1}{4}\,spp) &
        \BestValue{0.04} & \BestValue{0.03} & \BestValue{0.04} & \BestValue{0.04} & \BestValue{0.01} & \BestValue{0.05} & \BestValue{0.04} & \BestValue{0.04} \\
        \rowcolor{TableAltBackColor}
        \textbf{\Vertical: Iterative} (\nicefrac{1}{4}\,spp) &
        18.44 & 111.41 & 12.82 & 15.26 & 5.43 & 29.09 & 15.21 & 19.45 \\
        \textbf{\Vertical: Iterative} (power set, \nicefrac{1}{15}\,``spp'') &
        95.09 & 404.12 & 59.69 & 83.41 & 23.93 & 137.89 & 35.39 & 102.05
        \\%[0.92mm]
        \rowcolor{TableAltBackColor}
        \Horizontal: Permutation\,\shortcite{Heitz2019} (frame 16) &
        0.10 & 0.10 & 0.10 & 0.11 & 0.03 & 0.21 & 0.10 & 0.14 \\
        \textbf{\Horizontal: Iterative} (frame 16) &
        23.04 & 21.57 & 22.00 & 30.08 & 8.48 & 36.36 & 23.78 & 22.76 \\
        \bottomrule
    \end{tabularx}
    \vspace{-1mm}
\end{table*}
\let\BestValue\undefined
%%%%%%%%%%%%%%%%%%%%%%%%%%%%%%

%%%%%%%%%%%%%%%%%%%%%%%%%%%%%%%%%%%%%%%%%%%%%%%%%%%%%%%%%%%%
\subsection{Performance and utility}
\label{sec:Performance}
%%%%%%%%%%%%%%%%%%%%%%%%%%%%%%%%%%%%%%%%%%%%%%%%%%%%%%%%%%%%

Throughout our experiments, we have found that the tested algorithms rank in the following order in terms of increasing ability to minimize perceptual error on static scenes at equal sampling cost: histogram sampling, our dithering, permutation, our error diffusion, our \horizontal iterative, our \vertical iterative. The three lowest-ranked methods employ some form of dithering which by design assumes (a) constant image signal and (b) equi-spaced quantization levels shared by all pixels. The latter assumption is severely broken in the rendering setting where the ``quantization levels'' arise from (random) pixel estimation. Our \vertical methods (dithering, error diffusion, iterative) are more practical than the histogram sampling of \citet{Heitz2019} as they can achieve high fidelity with a much lower input-sample count. \Horizontal algorithms are harder to control due to their mispredictions which are further exacerbated when reusing estimates across frames in dynamic scenes.

Our iterative minimizations can best adapt to the input and also directly benefit from the extensions in \cref{sec:Extensions} (unlike all others). However, they are also the slowest, as evident in \cref{tab:Timings}. Fortunately, they can be sped up by several orders of magnitude through additional optimizations from halftoning literature~\citep{Allebach1992,DBS_GPU}; we discuss these optimizations in the context of our rendering setting in supplemental 
\ifdefined\arXiv
Section 3.
\else
\cref{sec:supp_iterative_minimization_optimization}.
\fi

Error diffusion is often on par with \vertical iterative minimization in quality and with dithering-based methods in run time. In a single-threaded implementation it can outperform all others; parallel error-diffusion variants exist too~\citep{parallel_error_diffusion}.

%%%%%%%%%%%%%%%%%%%%%%%%%%%%%%
\definecolor{SurrogateOutlineColor}{gray}{0.2}
\begin{figure}[t!]
    \centering
    \footnotesize
    \begin{overpic}[abs,unit=0.25mm,scale=0.125]{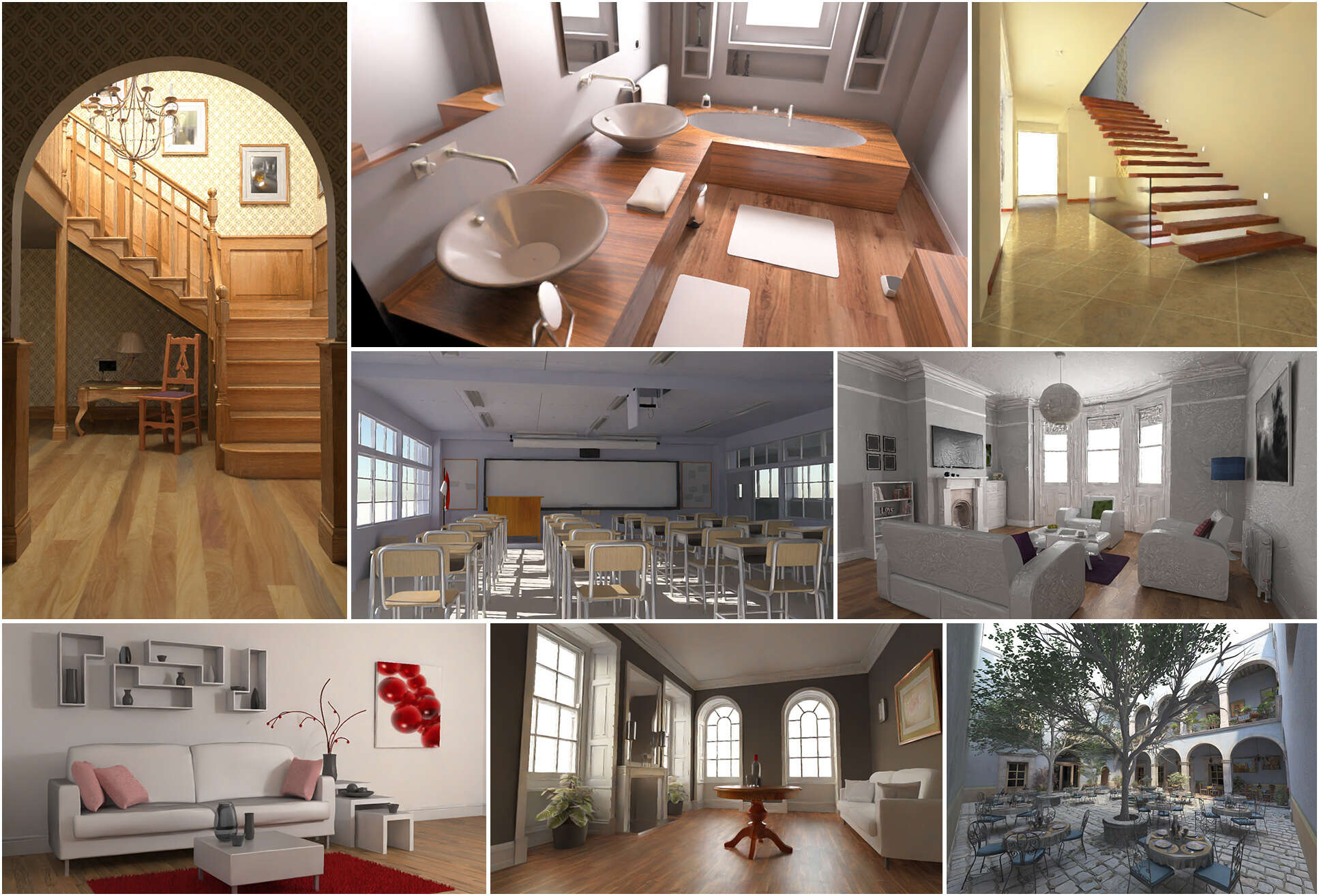}
        \put(4,4){\Outlined[white][SurrogateOutlineColor]{Modern living room}}
        \put(130,4){\Outlined[white][SurrogateOutlineColor]{Grey \& white room}}
        \put(249,4){\Outlined[white][SurrogateOutlineColor]{San Miguel}}
        \put(4,75){\Outlined[white][SurrogateOutlineColor]{Wooden staircase}}
        \put(94,75){\Outlined[white][SurrogateOutlineColor]{Japanese classroom}}
        \put(220,75){\Outlined[white][SurrogateOutlineColor]{White room}}
        \put(94,145){\Outlined[white][SurrogateOutlineColor]{Bathroom}}
        \put(255,145){\Outlined[white][SurrogateOutlineColor]{Modern hall}}
    \end{overpic}
    \vspace{-5mm}
    \caption{
        Collage of the surrogates used in our experiments, obtained by denoising the input estimates using Intel Open Image Denoise~\citep{inteldenoiser}. 
    }
    \label{fig:surrogates}
\end{figure}
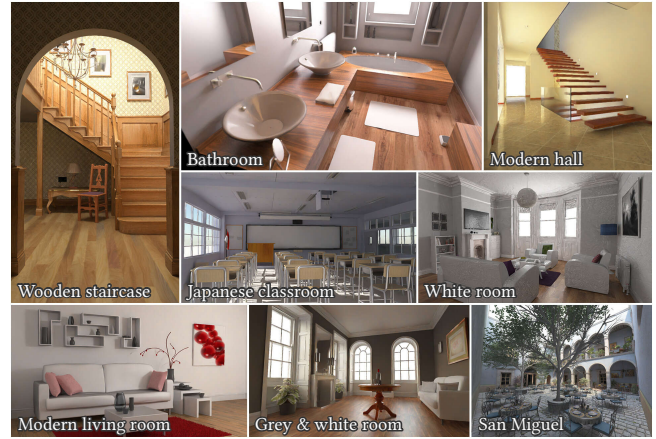
\undefinecolor{SurrogateOutlineColor}
%%%%%%%%%%%%%%%%%%%%%%%%%%%%%%

%%%%%%%%%%%%%%%%%%%%%%%%%%%%%%
\Paragraph{Practical utility}
%%%%%%%%%%%%%%%%%%%%%%%%%%%%%%

Our methods can enhance the perceptual fidelity of static and dynamic renderings as demonstrated by our experiments. For best results and maximum flexibility, we suggest using our \vertical iterative optimization, optionally with the efficiency improvements mentioned above. \Cref{fig:bias-analysis} illustrates that in practical scenarios (middle and right columns) this method can improve upon both the surrogate (top row) and the input-estimate average (bottom row) for a suitable value of the confidence parameter $\ConfidenceBias$. For maximum efficiency we recommend using our \vertical error diffusion. To obtain a surrogate, we recommend regularizing the input estimates via fast denoising or more basic bilateral or non-local-means filtering. Our optimization can then be interpreted as reducing bias or artifacts in such denoised images~(see \cref{fig:bias-analysis}). Simple denoising of the result may yield better quality than traditional aggressive denoising of the input samples.

%%%%%%%%%%%%%%%%%%%%%%%%%%%%%%
\Paragraph{Progressive rendering}
%%%%%%%%%%%%%%%%%%%%%%%%%%%%%%

Our optimization methods produce biased pixel estimates through manipulating the input samples; this is true even for a-priori methods where the sampling is completely deterministic. Nevertheless, consistency can be achieved through a simple progressive-rendering scheme: For each pixel, newly generated samples are cumulatively averaged into a fixed set of per-pixel estimates that are periodically passed to the optimization to obtain an updated image. Each individual estimate will converge to the true pixel value, thus the optimized image will also approach the ground truth---with bounded memory footprint. Interestingly, convergence is guaranteed regardless of the optimization method and surrogate used, though better methods and surrogates will yield better starting points. Lastly, adaptive sampling is naturally supported by \vertical methods as they are agnostic of differences in sample counts between pixels.

%%%%%%%%%%%%%%%%%%%%%%%%%%%%%%%%%%%%%%%%%%%%%%%%%%%%%%%%%%%%
\section{Conclusion}
\label{sec:conclusion}
%%%%%%%%%%%%%%%%%%%%%%%%%%%%%%%%%%%%%%%%%%%%%%%%%%%%%%%%%%%%

We devise a formal treatment of image-space error distribution in Monte Carlo rendering from both quantitative and perceptual aspects. Our formulation bridges the gap between halftoning and rendering by interpreting the error distribution problem as an extension of non-uniform multi-tone energy minimization halftoning. To guide the distribution of rendering error, we employ a perceptual kernel-based model whose practical optimization can deliver improvements not achievable by prior methods given the same sampling data. Our model provides valuable insights as well as a framework to further study the problem and its solutions.

A promising avenue for future research is to adapt even stronger perceptual error models. Prior work has already demonstrated a strong potential in reducing Monte Carlo noise visibility error using visual masking~\citep{Bolin98, Ramasubramanian99}. Robust metrics, other than squared $\Lnorm{2}$ norm, can also be considered with possible nonlinear relationships.

Our framework could conceivably be extended beyond the human visual system, \ie for optimizing the inputs to other types of image processing such as denoising. For such tasks, one could consider lifting the assumption of a fixed kernel to obtain an even more general problem where the kernel and sample distribution are optimized simultaneously (or alternatingly).

%%%%%%%%%%%%%%%%%%%%%%%%%%%%%%%%%%%%%%%%%%%%%%%%%%%%%%%%%%%
% Acknowledgments
%%%%%%%%%%%%%%%%%%%%%%%%%%%%%%%%%%%%%%%%%%%%%%%%%%%%%%%%%%%

\begin{acks}
    Our results show scenes (summarized in \cref{fig:surrogates}) coming from third parties. We acknowledge the \href{https://www.pbrt.org/scenes-v3}{PBRT scene repository} for \textit{San Miguel} and \textit{Bathroom}. \textit{Wooden staircase}, \textit{Modern hall}, \textit{Modern living room}, \textit{Japanese classroom}, \textit{White room}, \textit{Grey \& white room}, and \textit{Utah teapot} have been provided by \href{https://benedikt-bitterli.me/resources/}{Benedikt Bitterli}. The first author is funded from the European Research Council (ERC) under the European Union's Horizon 2020 research and innovation program (grant agreement \textnumero741215, ERC Advanced Grant INCOVID).
\end{acks}

%%%%%%%%%%%%%%%%%%%%%%%%%%%%%%%%%%%%%%%%%%%%%%%%%%%%%%%%%%%%

\bibliographystyle{ACM-Reference-Format}
\bibliography{paper}

%%%%%%%%%%%%%%%%%%%%%%%%%%%%%%%%%%%%%%%%%%%%%%%%%%%%%%%%%%%%

\appendix

%%%%%%%%%%%%%%%%%%%%%%%%%%%%%%%%%%%%%%%%%%%%%%%%%%%%%%%%%%%%
\section{Error decomposition for \horizontal optimization}
\label{app:error_decomposition}
%%%%%%%%%%%%%%%%%%%%%%%%%%%%%%%%%%%%%%%%%%%%%%%%%%%%%%%%%%%%

Substituting $\EstimatedImage(\Permutation(\SampleSetImage)) = \Permutation(\EstimatedImage(\SampleSetImage)) + \ErrorDifference$ into \cref{eq:HorizontalOptimization}, we bound the perceptual error using Jensen's inequality and the discrete Young convolution inequality~\citep{YoungTheoremBook}:
\begin{subequations}
\begin{align}
    \Energy(\Permutation(\SampleSetImage))
    &= 
    \Norm{\Kernel * (\Permutation(\EstimatedImage(\SampleSetImage)) - \ReferenceImage \, + \, \ErrorDifference)}^2_2 \\
    &=
    4\Norm{0.5\,\Kernel * (\Permutation(\EstimatedImage(\SampleSetImage)) - \ReferenceImage) \, + \, 0.5\,\Kernel *\ErrorDifference)}^2_2 \\
    &\leq
    \label{eq:HorizontalOptimizationEnergybApp}
    2\Norm{\,\Kernel * (\Permutation(\EstimatedImage(\SampleSetImage)) - \ReferenceImage)}^2_2 + 2\|\Kernel\|^2_1\Norm{ \ErrorDifference}^2_2.
\end{align}
\end{subequations}
The first term in \cref{eq:HorizontalOptimizationEnergybApp} involves pixel permutations in the readily available estimated image $\EstimatedImage(\SampleSetImage)$. In the second term we make an approximation that avoids rendering invocations: $\|\ErrorDifference\|^2_2 \approx \sum_{\PixelIndex}\DissimilarityMetric(\PixelIndex,\Permutation(\PixelIndex)\!)$, where $\DissimilarityMetric(\PixelIndex,\PixelIndexTwo)$ measures the dissimilarity between the light-transport integrals of pixels $\PixelIndex$ and $\PixelIndexTwo$. Dropping the constant $2$, we take the resulting bound as the optimization energy $\EnergyApprox$ in \cref{eq:HorizontalOptimizationEnergyApprox}.

%%%%%%%%%%%%%%%%%%%%%%%%%%%%%%%%%%%%%%%%%%%%%%%%%%%%%%%%%%%%
\section{Surrogate confidence control}
\label{sec:BiasConfidenceEnergy}
%%%%%%%%%%%%%%%%%%%%%%%%%%%%%%%%%%%%%%%%%%%%%%%%%%%%%%%%%%%%

Here we extend our perceptual error formulation to account for deviations of the surrogate image $\SurrogateImage$ from the ground truth $\ReferenceImage$. We introduce a parameter $\ConfidenceBias \in [0,1]$ that encodes our confidence in the quality of the surrogate and instructs the optimization how closely to fit to it. Given an initial image estimate $\EstimatedImageInit$ (the per-pixel estimate average for \vertical optimization), we look to optimize for $\EstimatedImage$. We begin with an artificial expansion:
\begin{subequations}
\begin{align}
    \sqrt{\Energy}
        &= \Norm{\Kernel * \EstimatedImage - \KernelRef * \ReferenceImage}_2 \\
        \nonumber
        &= \|(1-\ConfidenceBias)(\Kernel*\EstimatedImage-\Kernel*\EstimatedImageInit) + \ConfidenceBias(\Kernel * \EstimatedImage-\KernelRef*\SurrogateImage) \, +\\
        &\quad\;\; (1-\ConfidenceBias)(\Kernel*\EstimatedImageInit-\KernelRef*\ReferenceImage) + \ConfidenceBias(\KernelRef*\SurrogateImage-\KernelRef*\ReferenceImage)\|_2.
\end{align}
\end{subequations}
Using the triangle inequality we then obtain the following bound:
\begin{align}
    \nonumber
    \sqrt{\Energy} \leq
    \;& \|(1-\ConfidenceBias)(\Kernel*\EstimatedImage-\Kernel*\EstimatedImageInit) + \ConfidenceBias(\Kernel * \EstimatedImage-\KernelRef*\SurrogateImage)\|_2 \, + \\
    & \|(1-\ConfidenceBias)(\Kernel*\EstimatedImageInit-\KernelRef*\ReferenceImage) + \ConfidenceBias(\KernelRef*\SurrogateImage-\KernelRef*\ReferenceImage)\|_2.
\end{align}
The second term on the right-hand side can be dropped as it is independent of the optimization variable $\EstimatedImage$. We bound the square of the first term using Jensen's and Young's convolution inequalities:
\begin{subequations}
\begin{align}
    \|(1-\ConfidenceBias)(\Kernel*\EstimatedImage-\Kernel*\EstimatedImageInit) &+ \ConfidenceBias(\Kernel * \EstimatedImage-\KernelRef*\SurrogateImage)\|^2_2\leq \\
    (1-\ConfidenceBias)\|\Kernel\|^2_1\|\EstimatedImage-\EstimatedImageInit\|^2_2 &+ \ConfidenceBias\|\Kernel * \EstimatedImage-\KernelRef*\SurrogateImage\|^2_2.
\end{align}
\end{subequations}
We take this bound to be our optimization energy in \cref{eq:energy-bias}, noting that the squared norm in the second term is the original energy with the surrogate $\SurrogateImage$ substituted for the ground truth $\ReferenceImage$.

If a \emph{confidence map}\, $\pmb{\ConfidenceBias}$ is available (\eg as a byproduct of denoising), the minimization can be done with per-pixel control:
\begin{equation}
    E_{\pmb{\ConfidenceBias}} = \|\Kernel\|^2_1\|\sqrt{\pmb{1}-\pmb{\ConfidenceBias}}\odot(\EstimatedImage-\EstimatedImageInit)\|^2_2 + \|\sqrt{\pmb{\ConfidenceBias}}\odot(\Kernel * \EstimatedImage-\KernelRef*\SurrogateImage)\|^2_2.
\end{equation}

%%%%%%%%%%%%%%%%%%%%%%%%%%%%%%%%%%%%%%%%%%%%%%%%%%%%%%%%%%%

\end{document}

\hypersetup{colorlinks,
    linkcolor=ACMDarkBlue,
    citecolor=ACMDarkBlue,
    urlcolor=ACMDarkBlue,
    filecolor=ACMDarkBlue}

%\settopmatter{printacmref=false}
%\ifdefined\arXiv
\settopmatter{printacmref=false} % Removes citation information below abstract
%\renewcommand\footnotetextcopyrightpermission[1]{} % removes footnote with conference information in first column
%\pagestyle{plain} % removes running headers
%\fi

%%%%%%%%%%%%%%%%%%%%%%%%%%%%%%%%%%%%%%%%%%%%%%%%%%%%%%%%%%%%
% Title & authors
%%%%%%%%%%%%%%%%%%%%%%%%%%%%%%%%%%%%%%%%%%%%%%%%%%%%%%%%%%%%

\title[Perceptual error optimization for Monte Carlo rendering]{Supplementary material for:\\ Perceptual error optimization for Monte Carlo rendering}

\author{Vassillen Chizhov}
\orcid{0000-0001-7082-6367}
\affiliation{%
  \institution{MIA Group, Saarland University,}
  \institution{Max-Planck-Institut f{\"u}r Informatik}
  \city{Saarbr{\"u}cken}
  \country{Germany}
}

\author{Iliyan Georgiev}
\orcid{0000-0002-9655-2138}
\affiliation{%
  \institution{Autodesk}
  \country{United Kingdom}
}

\author{Karol Myszkowski}
\orcid{0000-0002-8505-4141}
\affiliation{%
  \institution{Max-Planck-Institut f{\"u}r Informatik}
  \city{Saarbr{\"u}cken}
  \country{Germany}
}

\author{Gurprit Singh}
\orcid{0000-0003-0970-5835}
\affiliation{%
  \institution{Max-Planck-Institut f{\"u}r Informatik}
  \city{Saarbr{\"u}cken}
  \country{Germany}
}

\authorsaddresses{}

%%%%%%%%%%%%%%%%%%%%%%%%%%%%%%%%%%%%%%%%%%%%%%%%%%%%%%%%%%%%
% Dummy teaser for some vertical spacing
%%%%%%%%%%%%%%%%%%%%%%%%%%%%%%%%%%%%%%%%%%%%%%%%%%%%%%%%%%%%

\begin{teaserfigure}
    \vspace{0mm}
\end{teaserfigure}

%%%%%%%%%%%%%%%%%%%%%%%%%%%%%%%%%%%%%%%%%%%%%%%%%%%%%%%%%%%%
% Abstract
%%%%%%%%%%%%%%%%%%%%%%%%%%%%%%%%%%%%%%%%%%%%%%%%%%%%%%%%%%%%

\begin{abstract}

In this supplemental document we discuss various details related to our general formulation from the main paper. We start with a description of the extension of our framework to the \emph{a-priori} setting (\cref{sec:supp_our_apriori_approach}). In \cref{sec:supp_our_demodulation} we describe a way in which textures can be accounted for in our \emph{\horizontal} approach, so that mispredictions due to multiplicative (and additive) factors are eliminated. In \cref{sec:supp_iterative_minimization_optimization} we describe ways in which the runtime of iterative energy minimization methods can be improved considerably. Notably, an expression is derived allowing the energy difference due to trial swaps to be evaluated in constant time (no scaling with image size or kernel size). In the remaining sections we analyze how current \emph{a-posteriori} \cite{Heitz2019} (\cref{sec:supp_aposteriori_approaches}) and \emph{a-priori} \cite{georgiev16bluenoise, Heitz2019b} (\cref{sec:supp_apriori_approaches}) state of the art approaches can be related to our framework. Interpretations are discussed, major sources of error are identified, and the assumptions of the algorithms are made explicit. 

\end{abstract}

%%%%%%%%%%%%%%%%%%%%%%%%%%%%%%%%%%%%%%%%%%%%%%%%%%%%%%%%%%%%

\maketitle

\thispagestyle{empty}

%%%%%%%%%%%%%%%%%%%%%%%%%%%%%%%%%%%%%%%%%%%%%%%%%%%%%%%%%%%%
\section{A-priori optimization}
%%%%%%%%%%%%%%%%%%%%%%%%%%%%%%%%%%%%%%%%%%%%%%%%%%%%%%%%%%%%
\label{sec:supp_our_apriori_approach}

We extend our theory to the \emph{a-priori} setting and discuss the main factors affecting the quality. The quality of \emph{a-priori} approaches is determined mainly by three factors: the energy, the search space, and the optimization strategy. We discuss each of those briefly in the following paragraphs.

%%%%%%%%%%%%%%%%%%%%%%%%%%%%%%
\Paragraph{Our energy}
%%%%%%%%%%%%%%%%%%%%%%%%%%%%%%

We extend the \emph{a-posteriori} energy from the main paper in order to handle multiple estimators involving different integrands: $\EstimatedImage_1,\ldots,\EstimatedImage_T$, with associated weights $w_1,\ldots,w_T$:

\begin{equation}
    \label{eq:supp_our_apriori_energy}
    E(\SampleSetImage) = \sum_{t=1}^{T}w_t\|\Kernel * \EstimatedImage_t(\SampleSetImage) - \ReferenceImage_t\|^2.
\end{equation}

In the above $\Kernel$ would typically be a low-pass kernel (\eg Gaussian), and $\ReferenceImage_t$ is the integral of the function used in the estimator $\EstimatedImage_t$. Through this energy a whole set of functions can be optimized for, in order for the sequence to be more robust to different scenes and estimators, that do not fit any of the considered integrands exactly. We note that the derived optimization in \cref{sec:supp_iterative_minimization_optimization} below is also applicable to the minimization of the proposed energy.

%%%%%%%%%%%%%%%%%%%%%%%%%%%%%%
\Paragraph{Search space}
%%%%%%%%%%%%%%%%%%%%%%%%%%%%%%

The search space plays an important role for the qualities which the optimized sequences exhibit. A more restricted search space provides more robustness and may help avoid over-fitting to the considered set of integrands. 

For instance, sample sets may be generated randomly within each pixel. Then, their assignment to pixels may be optimized over the space of all possible permutations. This is the setting of \emph{\horizontal} methods. If additionally this assignment is done within each dimension separately it allows for an even better fit to the integrands in the energy (but may degrade the general integration properties of the sequence). The scrambling keys' search space in \cite{Heitz2019b} is a special case of the latter applied for the Sobol sequence.

Constraining the search space to points generated from low-discrepancy sequences provides further robustness and guarantees desirable integration properties of the considered sequences. Similarly to \cite{Heitz2019b}, we can consider a search space of Sobol scrambling keys in order for the optimized sequence to have a low discrepancy. 

Ideally, such integration properties should arise directly from the energy. However, in practice the scene integrand cannot be expected to exactly match the set of considered integrands, thus extra robustness is gained through the restriction. Additionally, optimizing for many dimensions at the same time is costly as noted in \cite{Heitz2019b}, thus imposing low-discrepancy properties also helps in that regard.

Finally, by imposing strict search space constraints a severe restriction on the distribution of the error is imposed. This can be alleviated by imposing the restrictions through soft penalty terms in the energy. This can allow for a trade-off between blue noise distribution and integration quality for example.

%%%%%%%%%%%%%%%%%%%%%%%%%%%%%%
\Paragraph{Progressive rendering}
%%%%%%%%%%%%%%%%%%%%%%%%%%%%%%

In order to make the sequence applicable to progressive rendering, subsets of samples should be considered in the optimization. Given a sample set $\SampleSetPixel_{\PixelIndex}$ for pixel $\PixelIndex$ we can decompose it in sample sets of $1,\ldots,N$ samples: $\SampleSetPixel_{\PixelIndex,1} \subset \ldots \subset \SampleSetPixel_{\PixelIndex,N} \equiv \SampleSetPixel_{\PixelIndex}$. We denote the respective images of sample sets $\SampleSetImage_1,\ldots,\SampleSetImage_N$.

Then an energy that also optimizes for the distribution of the error at each sample count is:

\begin{equation}
    \label{eq:our_apriori_progressive}
    E(\SampleSetImage) = \sum_{t=1}^{T}\sum_{\PixelIndexThree=1}^{N}w_{t,\PixelIndexThree}\|\Kernel * \EstimatedImage_t(\SampleSetImage_{\PixelIndexThree}) - \ReferenceImage_t\|^2.
\end{equation}

If $w_{\PixelIndex,\PixelIndexThree}$ are set to zero for $\PixelIndexThree<N$ then the original formulation is recovered. The more general formulation imposes additional constraints on the samples, thus the quality at the full sample count may be compromised if we also require a good quality at lower sample counts.

Choosing samples from $\SampleSetPixel_{\PixelIndex}$ for $\SampleSetPixel_{\PixelIndex,1},\ldots,\SampleSetPixel_{\PixelIndex,N-1}$ (in each dimension) constitutes a \emph{\vertical} search space analogous to the one discussed in the main paper for \emph{a-posteriori} methods. The ranking keys' optimization in \cite{Heitz2019b} is a special case of this search space for the Sobol sequence.

Adaptive sampling can be handled by allowing a varying number of samples per pixel, and a corresponding energy derived from the one above. Note that this poses further restrictions on the achievable distribution.

%%%%%%%%%%%%%%%%%%%%%%%%%%%%%%
\Paragraph{Optimization strategies}
%%%%%%%%%%%%%%%%%%%%%%%%%%%%%%

Typically the energies for \emph{a-priori} methods have been optimized through simulated annealing \cite{georgiev16bluenoise, Heitz2019b}. Metaheuristics can lead to very good minima especially if the runtime is not of great concern, which is the case since the sequences are precomputed. Nevertheless, the computation still needs to be tractable. The energies in previous works are generally not cheap to evaluate. On the other hand, our energies, especially if the optimizations in \cref{sec:supp_iterative_minimization_optimization} are considered, can be evaluated very efficiently. This is beneficial for keeping the runtime of metaheuristics manageable, allowing for more complex search spaces to be considered.

%%%%%%%%%%%%%%%%%%%%%%%%%%%%%%
\Paragraph{Implementation details}
%%%%%%%%%%%%%%%%%%%%%%%%%%%%%%

Implementation decisions for a renderer, such as how samples are consumed, or how those are mapped to the hemisphere and light sources, affect the estimator $\EstimatedImage$. This is important, especially when choosing $\EstimatedImage$ for the described energies to optimize a sequence. It is possible that very small implementation changes make a previously ideal sequence useless for a specific renderer. It is important to keep this in mind when optimizing sequences by using the proposed energies and when those are used in a renderer.

%%%%%%%%%%%%%%%%%%%%%%%%%%%%%%%%%%%%%%%%%%%%%%%%%%%%%%%%%%%%
\section{Texture demodulation for \horizontal optimization}
\label{sec:supp_our_demodulation}
%%%%%%%%%%%%%%%%%%%%%%%%%%%%%%%%%%%%%%%%%%%%%%%%%%%%%%%%%%%%

Our iterative energy minimization algorithms \ifdefined\arXiv
%(\textcolor{red}{MISSING})
(Alg. 1, Alg. 2, main paper)
\else 
(\cref{alg:VerticalOptimization}, \cref{alg:HorizontalOptimization})
\fi directly work with the original energy formulation, unlike error diffusion and dither matrix halftoning which only approximately minimize the energy. This allows textures to be handled more robustly compared to the permutation approach of \citeauthor{Heitz2019}. 

%%%%%%%%%%%%%%%%%%%%%%%%%%%%%%
\Paragraph{Reducing misprediction errors}
%%%%%%%%%%%%%%%%%%%%%%%%%%%%%%

Our \horizontal approach relies on a dissimilarity metric $d(\cdot,\cdot)$ which approximates terms involving the difference $\ErrorDifference$ due to swapping sample sets instead of pixels. This difference can be decreased, so that $d$ is a better approximation, if additional information is factored out in the energy: screen-space varying multiplicative and additive terms. 
Specifically, if we have a spatially varying multiplicative image $\pmb{\alpha}$, and a spatially varying additive image $\pmb{\beta}$:
\begin{gather}
    \EstimatedImage = \pmb{\alpha}\EstimatedImage' + \pmb{\beta} \\
    \ErrorDifference'(\Permutation) = \pmb{\alpha}\odot\EstimatedImage'(\Permutation(\SampleSetImage))-\pmb{\alpha}\odot\Permutation(\EstimatedImage'(\SampleSetImage)) \\
    \begin{gathered}
    \ErrorDifference(\Permutation) = \EstimatedImage(\Permutation(\SampleSetImage))-\Permutation(\EstimatedImage(\SampleSetImage)) = \\ \pmb{\alpha}\odot\EstimatedImage'(\Permutation(\SampleSetImage)) + \pmb{\beta} - \Permutation(\pmb{\alpha}\odot\EstimatedImage'(\SampleSetImage)+\pmb{\beta}),
    \end{gathered}
\end{gather}
we can make use of this in our formulation:
\begin{gather}
     E(\Permutation) = \|\Kernel * \EstimatedImage(\Permutation(\SampleSetImage)) - \KernelReference*\ReferenceImage\|^2_2 \\
     \sqrt{E(\Permutation)} \leq \Norm{\Kernel * (\pmb{\alpha} \odot \Permutation(\EstimatedImage'(\SampleSetImage)) + \pmb{\beta}) - \KernelReference*\ReferenceImage}_2  + \Norm{\, \Kernel \,}_1\|\ErrorDifference'\|_2.
\end{gather}
Contrast this to the original formulation where $\pmb{\alpha}$ and $\pmb{\beta}$ are not factored out:
\begin{equation}
    \sqrt{E(\Permutation)} \leq \Norm{\Kernel * \Permutation\left(\pmb{\alpha} \odot\EstimatedImage'(\SampleSetImage)+\pmb{\beta}\right) - \KernelReference*\ReferenceImage}_2  + \Norm{\, \Kernel \,}_1\|\ErrorDifference\|_2.
\end{equation}
With the new formulation it is sufficient that $\EstimatedImage'(\Permutation(\SampleSetImage))=\Permutation(\EstimatedImage'(\SampleSetImage))$ for $\ErrorDifference'$ to be zero, while originally both $\pmb{\alpha}$ and $\pmb{\beta}$ play a role in $\ErrorDifference$ becoming zero. Intuitively this means that screen space integrand differences due to additive and multiplicative factors do not result in mispredictions with the new formulation, if the integrand is assumed to be the same (locally) in screen space.

\Paragraph{Comparison to demodulation} In the method of \citeauthor{Heitz2019} the permutation is applied on the albedo demodulated image. This preserves the property that the global minimum of the implicit energy can be found through sorting. Translated to our framework this can be formulated as ($\BlueNoiseImage$ is a blue noise mask optimized for a kernel $\Kernel$):
\begin{equation}
    E_{HBP}(\Permutation) = \|\Permutation(\EstimatedImage'(\SampleSetImage))-\SurrogateImage - \BlueNoiseImage\|^2_2 \approx \|\Kernel * \Permutation(\EstimatedImage'(\SampleSetImage))-\Kernel*\SurrogateImage\|^2_2.
\end{equation}
We have assumed that $\pmb{\beta}$ is zero, but we can also extend the method to handle an additive term $\pmb{\beta}$ as in our case. The more important distinction is that while the albedo demodulated image $\EstimatedImage'$ is used in the permutation, it is never re-modulated ($\pmb{\alpha}\odot\cdot$ is missing). Thus, this does not allow for proper handling of textures, even if it allows for modest improvements in practice. An example of a fail case consists of an image $\pmb{\alpha}$ that is close to white noise. Then the error distribution will also be close to white noise due to the missing $\pmb{\alpha}\odot\cdot$ factor. More precisely, even if $\Permutation(\EstimatedImage'(\SampleSetImage))-\SurrogateImage$ is distributed as $\BlueNoiseImage$, this does not imply that $\pmb{\alpha}\odot\Permutation(\EstimatedImage'(\SampleSetImage))-\SurrogateImage$ will be distributed similarly. Dropping $\pmb{\alpha}\odot\cdot$ is, however, a reasonable option if one is restricted to sorting as an optimization strategy.

We propose a modification of the original approach (and energy) such that not only the demodulated estimator values are used, but the blue noise mask $\BlueNoiseImage$ is also  \ifdefined\arXiv
demodulated.
\else
demodulated (\cref{fig:supp_cornellbox-textured}).
\fi
To better understand how it is derived (and how $\pmb{\beta}$ may be integrated) we study a bound based on the assumption that $\alpha_{\PixelIndex}\in[0,1]$, and $\ErrorDifference' = 0$
\begin{gather}
\label{eq:supp_energy_factor_linear}
     E(\Permutation) = \Norm{\Kernel * (\pmb{\alpha} \odot \Permutation(\EstimatedImage'(\SampleSetImage)) + \pmb{\beta}) - \Kernel*\SurrogateImage}^2_2 \approx \\
     \| \pmb{\alpha} \odot \Permutation(\EstimatedImage'(\SampleSetImage)) + \pmb{\beta} - \SurrogateImage - \BlueNoiseImage\|^2_2 =  \\
     \sum_{\PixelIndex}\alpha^2_{\PixelIndex}\left((\Permutation(\EstimatedImage'(\SampleSetImage)))_{\PixelIndex} + \frac{\beta_{\PixelIndex} - \SurrogatePixel_{\PixelIndex} - \BlueNoisePixel_{\PixelIndex}}{\alpha_{\PixelIndex}}\right)^2 \leq \\
     \left\| \Permutation(\EstimatedImage'(\SampleSetImage)) + \frac{\pmb{\beta} - \SurrogateImage - \BlueNoiseImage}{\pmb{\alpha}}\right\|^2_2.
\end{gather}
The global minimum of the last energy (w.r.t. $\Permutation$) can be found through sorting also, since there is no spatially varying multiplicative factor $\pmb{\alpha}$ in front of the permutation.

\todo{Compare texture demodulation to our texture demodulation -> remodulation}

%%%%%%%%%%%%%%%%%%%%%%%%%%%%%%
\Paragraph{Sinusoidal textures}
%%%%%%%%%%%%%%%%%%%%%%%%%%%%%%

%%%%%%%%%%%%%%%%%%%%%%%%%%%%%%
\begin{figure}[t!]
    \centering
    %!TEX root = ../supplemental.tex

\newcommand{\ImageWithSpectrum}[2]{%
    \begin{tikzpicture}
      \node[anchor=south west,inner sep=0] (image) at (0,0)
      {
        \pdfliteral{ 1 w}\includegraphics[width=0.65in,page=1]{figures/supp_sine-texture/#1}
      };
      \begin{scope}[x={(image.south east)},y={(image.north west)}]
      \node[anchor=south west,inner sep=0] (lines) at (0.0,0.0)
      {
        \pdfliteral{ 1 w}\includegraphics[width=0.375in,page=1]{figures/supp_sine-texture/#2}
      };
      \end{scope}
    \end{tikzpicture}%
}

\footnotesize
\hspace*{-2.2em}
\begin{tabular}{c@{\;}c@{\;}c@{\;}c@{\;}c@{\;}c@{}}
& Input & Ours & \multicolumn{3}{c}{\citet{Heitz2019}} 
\\[0.2mm]
\rotatebox{90}{\; multiplicative}
&
\ImageWithSpectrum{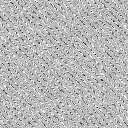}{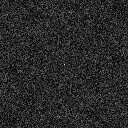}
&
\ImageWithSpectrum{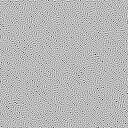}{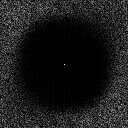}
&
\ImageWithSpectrum{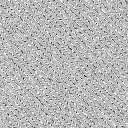}{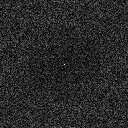}
&
\ImageWithSpectrum{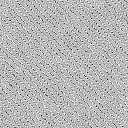}{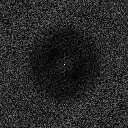}
&
\ImageWithSpectrum{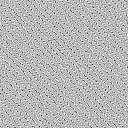}{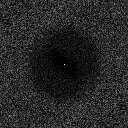}
\\
\rotatebox{90}{\quad\; additive}
&
\ImageWithSpectrum{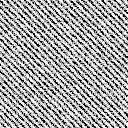}{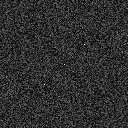}
&
\ImageWithSpectrum{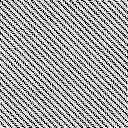}{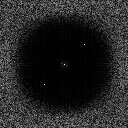}
&
\ImageWithSpectrum{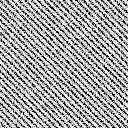}{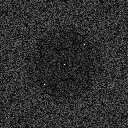}
&
\ImageWithSpectrum{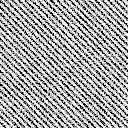}{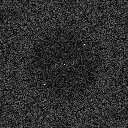}
&
\ImageWithSpectrum{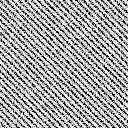}{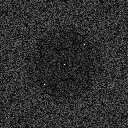}
\\
& & & No & Demodulation & Demodulation
\\
& & &  demodulation & w/ tilesize 8 & w/o tiling 
\end{tabular}%

\let\ImageWithSpectrum\SomeUndefinedMacroBrotha

    \vspace{-1em}
    \caption{
        We demonstrate the importance of the extension presented in \cref{sec:supp_our_demodulation}. A high-frequency sinusoidal texture is corrupted by white noise (leftmost column) multiplicatively (\textbf{top row}) and additively (\textbf{bottom row}). Contrary to~\citeauthor{Heitz2019}'s method, our optimization distributes error as a high-quality blue-noise distribution (see the power-spectrum insets). The reference images for the top/bottom image are respectively a flat grey and a sinusoidal image.
    }
    \label{fig:supp_SineTexture}
\end{figure}
%%%%%%%%%%%%%%%%%%%%%%%%%%%%%%

To demonstrate texture handling (multiplicative term $\pmb{\alpha}$), in the top row of~\cref{fig:supp_SineTexture}, a white-noise texture $W$ is multiplied with a sine-wave input signal: $f(x,y) = 0.5*{(1.0+\sin(x+y))}*W(x,y)$. The reference is a constant image at $0.5$.
 \citeauthor{Heitz2019} proposed to handle such textures by applying their method on the albedo-demodulated image. While this strategy may lead to a modest improvement, it ignores the fact that the image is produced by re-modulating the albedo, which can negate that improvement. Instead, our \horizontal iterative minimization algorithm can incorporate the albedo explicitly using the discussed energy. 

The bottom row demonstrates the effect of a non-flat signal on the error distribution (additive term $\pmb{\beta}$). Here $W$ is added to a sine-wave input signal: $f(x,y)= 0.5*{(1.0+\sin(x+y))}+W(x,y)$. The reference image is $0.5*{(1+\sin(x+y))}$. Our optimization is closer to the reference suggesting that our method can greatly outperform the current state of the art by properly accounting for auxiliary information, especially in regions with high-frequency textures.

%%%%%%%%%%%%%%%%%%%%%%%%%%%%%%
\Paragraph{Dimensional decomposition}
%%%%%%%%%%%%%%%%%%%%%%%%%%%%%%

The additive factor $\pmb{\beta}$ can be used to motivate splitting the optimization over several dimensions, since the Liouville–Neumann expansion of the rendering equation is additive \cite{Kajiya86}. If some dimensions are smooth (\eg lower dimensions), then a screen space local integrand similarity assumption can be encoded in $d(\cdot,\cdot)$ and it will approximate $\ErrorDifference$ better for smoother dimensions. If the optimization is applied over all dimensions at the same time, this may result in many mispredictions due to the assumption being violated for dimensions in which the integrand is less smooth in screen space (\eg higher dimensions). We propose splitting the optimization problem starting from lower dimensions and sequentially optimizing higher dimensions while encoding a local smoothness (in screen space) assumption on the integrand in $d(\cdot,\cdot)$ (\eg swaps limited to a small neighborhood around the pixel). This requires solving several optimization problems, but potentially reduces the amount of mispredictions. Note that it does not require more rendering operations than usual.

%%%%%%%%%%%%%%%%%%%%%%%%%%%%%%%%%%%%%%%%%%%%%%%%%%%%%%%%%%%%
\section{Improving iterative-optimization performance}
\label{sec:supp_iterative_minimization_optimization}
%%%%%%%%%%%%%%%%%%%%%%%%%%%%%%%%%%%%%%%%%%%%%%%%%%%%%%%%%%%%

The main cost of iterative minimization methods is computing the energy for each trial swap, more specifically the required convolution and the subsequent norm computation. In the work of \citeauthor{Allebach1992} an optimization has been proposed to efficiently evaluate such trial swaps, without recomputing a convolution or norm at each step, yielding a speed up of more than 10 times. The optimization relies on the assumption that the kernel $\Kernel$ is the same in screen space (the above optimization is not applicable for spatially varying kernels).
We extend the described optimization to a more general case, also including spatially varying kernels. We also note some details not mentioned in the original paper.

%%%%%%%%%%%%%%%%%%%%%%%%%%%%%%%%%%%%%%%%%%%%%%%%%%%%%%%%%%%%
\subsection{\Horizontal swaps}
%%%%%%%%%%%%%%%%%%%%%%%%%%%%%%%%%%%%%%%%%%%%%%%%%%%%%%%%%%%%

We will assume the most general case: instead of just swapping pixels, we consider swapping sample sets from which values are generated through $\EstimatedImage$. It subsumes both swapping pixel values and swapping pixel values in the presence of a multiplicative factor $\pmb{\alpha}$. 

\Paragraph{Single swap} The main goal is to evaluate the change of the energy $\EnergyChange$ due to a swap between the sample sets of some pixels $a,b$. More precisely, if the original sample set image is $\SampleSetImage$ then the new sample set image is $\SampleSetImage'$ such that $\SampleSetPixel'_a = \SampleSetPixel_b, \SampleSetPixel'_b = \SampleSetPixel_a$, and $\SampleSetPixel'_\PixelIndex = \SampleSetPixel_\PixelIndex$ everywhere else. This corresponds to images: $\EstimatedImage = \EstimatedImage(\SampleSetImage)$ and $\EstimatedImage' = \EstimatedImage(\SampleSetImage')$. The two images differ only in the pixels with indices $a$ and $b$. Let:
\begin{gather}
    \delta_a = \EstimatedPixel'_a-\EstimatedPixel_a = \EstimatedPixel_a(\SampleSetPixel_b)-\EstimatedPixel_a(\SampleSetPixel_a) \\
    \delta_b = \EstimatedPixel'_b-\EstimatedPixel_b = \EstimatedPixel_b(\SampleSetPixel_a)-\EstimatedPixel_b(\SampleSetPixel_b).
\end{gather}
We will also denote the convolved images $\tilde{\EstimatedImage} = \Kernel * \EstimatedImage$ and $\tilde{\EstimatedImage}' = \Kernel * \EstimatedImage'$, and also $\ErrorImage = \tilde{\EstimatedImage} - \ReferenceImage$. Specifically:
\begin{gather}
\tilde{\EstimatedPixel}_{\PixelIndex} = \sum_{\PixelIndexTwo \in \mathbb{Z}^2}\EstimatedPixel_{\PixelIndexTwo}\KernelPixel_{\PixelIndex-\PixelIndexTwo}, \,\, \tilde{\EstimatedPixel}'_{\PixelIndex} = \tilde{\EstimatedPixel}_{\PixelIndex} + \delta_a\KernelPixel_{\PixelIndex-a} + \delta_b\KernelPixel_{\PixelIndex-b}.
\end{gather}
We want to be able to efficiently evaluate $\delta = \|\tilde{\EstimatedImage}'-\ReferenceImage\|^2 - \|\tilde{\EstimatedImage}-\ReferenceImage\|^2$, since in the iterative minimization algorithms the candidate with the minimum $\delta$ is kept. Using the above expressions for $\tilde{\EstimatedPixel}'_{\PixelIndex}$ we rewrite $\delta$ as:
\begin{gather}
    \label{eq:supp_dbs_opt_kernel_error_cross_correlation}
    \delta = \|\tilde{\EstimatedImage}'-\ReferenceImage\|^2 - \|\tilde{\EstimatedImage}-\ReferenceImage\|^2 = \\
    \sum_{\PixelIndex\in \mathbb{Z}^2}\|\tilde{\EstimatedPixel}_{\PixelIndex}-\ReferencePixel_{\PixelIndex} + \delta_a\KernelPixel_{\PixelIndex-a} + \delta_b\KernelPixel_{\PixelIndex-b}\|^2 - \|\tilde{\EstimatedImage}-\ReferenceImage\|^2 = \\
    2\sum_{\PixelIndex \in \mathbb{Z}^2}\InnerProd{\tilde{\EstimatedPixel}_{\PixelIndex}-\ReferencePixel_{\PixelIndex}, \delta_a\KernelPixel_{\PixelIndex-a} + \delta_b\KernelPixel_{\PixelIndex-b}} + \sum_{\PixelIndex \in \mathbb{Z}^2}\|\delta_a\KernelPixel_{\PixelIndex-a} + \delta_b\KernelPixel_{\PixelIndex-b}\|^2 = \\
    \begin{gathered}
    2\InnerProd{\delta_a,\sum_{\PixelIndex \in \mathbb{Z}^2}\ErrorPixel_{\PixelIndex}\KernelPixel_{\PixelIndex-a}} + 
    2\InnerProd{\delta_b,\sum_{\PixelIndex \in \mathbb{Z}^2}\ErrorPixel_{\PixelIndex}\KernelPixel_{\PixelIndex-b}} + \\
    \InnerProd{\delta^2_a,\sum_{\PixelIndex \in \mathbb{Z}^2}\KernelPixel_{\PixelIndex-a}\KernelPixel_{\PixelIndex-a}}  +
    \InnerProd{\delta^2_b,\sum_{\PixelIndex \in \mathbb{Z}^2}\KernelPixel_{\PixelIndex-b}\KernelPixel_{\PixelIndex-b}} + \\
    2\InnerProd{\delta_a\delta_b,\sum_{\PixelIndex \in \mathbb{Z}^2}\KernelPixel_{\PixelIndex-a}\KernelPixel_{\PixelIndex-b}} =
    \end{gathered} \\
    \begin{gathered}
    2\InnerProd{\delta_a,C_{\Kernel,\ErrorImage}(a)} + 2\InnerProd{\delta_b,C_{\Kernel,\ErrorImage}(b)} + \\
    \InnerProd{(\delta^2_a+\delta^2_b),C_{\Kernel,\Kernel}(0)} +
    2\InnerProd{\delta_a\delta_b,C_{\Kernel,\Kernel}(b-a)},
    \end{gathered}
\end{gather}
where $C_{f,h}(x) = \sum_{\PixelIndex \in \mathbb{Z}^2}f(\PixelIndex-x)h(\PixelIndex)$ is the cross-correlation of $f$ and $h$. We have reduced the computation of $\delta$ to the sum of only $4$ terms. Assuming that $C_{\Kernel,\Kernel}$ is known (it can be precomputed once for a known kernel) and that $C_{\Kernel,\ErrorImage}$ is known (it can be recomputed after a sufficient amount of swaps have been accepted), then evaluating a trial swap takes constant time (it does not scale in the size of the image or the size of the kernel). 

\Paragraph{Multiple accepted swaps} It may be desirable to avoid recomputing $C_{\Kernel,\ErrorImage}$ even upon accepting a trial swap. For that purpose we extend the strategy from \cite{Allebach1992} for computing $C_{\Kernel,
\ErrorImage^n}$, where $\ErrorImage^n$ is the error image after $n$ swaps have been accepted:
\begin{equation}
    \{(\delta_{a^1},\delta_{b^1}),\ldots,(\delta_{a^{n}},\delta_{b^{n}})\}.
\end{equation}
This implies: $\tilde{\EstimatedPixel}^n_{\PixelIndex} = \tilde{\EstimatedPixel} + \sum_{\PixelIndexThree=1}^{n}(\delta_{a^{\PixelIndexThree}}\KernelPixel_{i-a^{\PixelIndexThree}}+\delta_{b^{\PixelIndexThree}}\KernelPixel_{i-b^{\PixelIndexThree}})$, and consequently: 
\begin{gather}
    C_{\Kernel,\ErrorImage^n}(x) = \\ \sum_{\PixelIndex\in\mathbb{Z}^2}\left(\tilde{\EstimatedPixel}_{\PixelIndex}-\ReferencePixel_{\PixelIndex} + \sum_{\PixelIndexThree=1}^{n}(\delta_{a^{\PixelIndexThree}}\KernelPixel_{i-a^{\PixelIndexThree}}+\delta_{b^{\PixelIndexThree}}\KernelPixel_{i-b^{\PixelIndexThree}})\right)\KernelPixel_{\PixelIndex-x} = \\
    C_{\Kernel,\ErrorImage}(x) + \sum_{\PixelIndexThree=1}^{n}(\delta_{a^{\PixelIndexThree}}C_{\Kernel,\Kernel}(x-a^{\PixelIndexThree}) + \delta_{b^{\PixelIndexThree}}C_{\Kernel,\Kernel}(x-b^{\PixelIndexThree})).
\end{gather}
This allows avoiding the recomputation of $C_{\Kernel,\ErrorImage}$ after every accepted swap, and instead, the delta on the $n+1$-st swap with trial differences $\delta_a,\delta_b$ is:
\begin{gather}
    \delta^{n+1} = \|\EstimatedImage^{n+1}-\ReferenceImage\|^2 - \|\EstimatedImage^{n}-\ReferenceImage\|^2 = \\
    \begin{gathered}
    2\InnerProd{\delta_{a},C_{\Kernel,\ErrorImage^n}(a)} + 2\InnerProd{\delta_{b},C_{\Kernel,\ErrorImage^n}(b)} + \\
    \InnerProd{(\delta^2_{a}+\delta^2_{b}),C_{\Kernel,\Kernel}(0)} +
    2\InnerProd{\delta_{a}\delta_{b},C_{\Kernel,\Kernel}(b-a)},
    \end{gathered}
\end{gather}
where $C_{\Kernel,\ErrorImage^n}$ is computed from $C_{\Kernel,\ErrorImage}$ and $C_{\Kernel,\Kernel}$ as derived in \cref{eq:supp_dbs_opt_kernel_error_cross_correlation}. This computation scales only in the number of accepted swaps since the last recomputation of $C_{\Kernel,\ErrorImage}$. We also note that $C_{\Kernel,\Kernel}(x-y)$ evaluates to zero if $x-y$ is outside of the support of $C_{\Kernel,\Kernel}$. Additional optimizations have been devised due to this fact \cite{Allebach1992}.

%%%%%%%%%%%%%%%%%%%%%%%%%%%%%%%%%%%%%%%%%%%%%%%%%%%%%%%%%%%%
\subsection{\Vertical swaps}
%%%%%%%%%%%%%%%%%%%%%%%%%%%%%%%%%%%%%%%%%%%%%%%%%%%%%%%%%%%%

In the \emph{\vertical} setting swaps happen only within the pixel itself, that is: $\delta_a = \EstimatedPixel_a(\SampleSetPixel'_a)-\EstimatedPixel_a(\SampleSetPixel_a)$. Consequently, $\tilde{\EstimatedPixel}'_{\PixelIndex} = \tilde{\EstimatedPixel}_{\PixelIndex} + \delta_a\KernelPixel_{\PixelIndex-a}$. Computing the difference in the energies for the $n+1$-st swap:
\begin{gather}
    \delta^{n+1} = \|\tilde{\EstimatedImage}^{n+1}-\ReferenceImage\|^2 - \|\tilde{\EstimatedImage}^n-\ReferenceImage\|^2 = \\
    \sum_{\PixelIndex\in \mathbb{Z}^2}\|\tilde{\EstimatedPixel}^{n}_{\PixelIndex}-\ReferencePixel_{\PixelIndex} + \delta_a\KernelPixel_{\PixelIndex-a}\|^2 - \|\tilde{\EstimatedImage}^n-\ReferenceImage\|^2 = \\
    2\sum_{\PixelIndex \in \mathbb{Z}^2}\InnerProd{\tilde{\EstimatedPixel}^n_{\PixelIndex}-\ReferencePixel_{\PixelIndex}, \delta_a\KernelPixel_{\PixelIndex-a}} + \sum_{\PixelIndex \in \mathbb{Z}^2}\|\delta_a\KernelPixel_{\PixelIndex-a}\|^2 = \\
    \begin{gathered}
    2\InnerProd{\delta_a,\sum_{\PixelIndex \in \mathbb{Z}^2}\ErrorPixel^n_{\PixelIndex}\KernelPixel_{\PixelIndex-a}} + 
    \InnerProd{\delta^2_a,\sum_{\PixelIndex \in \mathbb{Z}^2}\KernelPixel_{\PixelIndex-a}\KernelPixel_{\PixelIndex-a}} =
    \end{gathered} \\
    \begin{gathered}
    2\InnerProd{\delta_a,C_{\Kernel,\ErrorImage^n}(a)} + 
    \InnerProd{\delta^2_a,C_{\Kernel,\Kernel}(0)}.
    \end{gathered}
\end{gather}
The corresponding expression for $C_{\Kernel,\ErrorImage^n}$ is:
\begin{equation}
    C_{\Kernel,\ErrorImage^n}(x) = C_{\Kernel,\ErrorImage}(x) + \sum_{\PixelIndexThree=1}^{n}\delta_{a^{\PixelIndexThree}}C_{\Kernel,\Kernel}(x-a^{\PixelIndexThree}).
\end{equation}

%%%%%%%%%%%%%%%%%%%%%%%%%%%%%%%%%%%%%%%%%%%%%%%%%%%%%%%%%%%%
\subsection{Multiple simultaneous updates}
%%%%%%%%%%%%%%%%%%%%%%%%%%%%%%%%%%%%%%%%%%%%%%%%%%%%%%%%%%%%

If the search space is ignored and the formulation is analyzed in an abstract setting it becomes obvious that the \emph{\vertical} approach corresponds to an update of a single pixel, while the \emph{\horizontal} approach corresponds to an update of two pixels at the same time. This can be generalized further. Let $N$ different pixels be updated per trial, and let there be $n$ trials that have been accepted since $C_{\Kernel,\ErrorImage}$ has been updated. Let the pixels to be updated in the current trial be: $a^{n+1}_1,\ldots,a^{n+1}_N$, and the accepted update at step $\PixelIndexThree$ be at pixels: $a^{\PixelIndexThree}_1,\ldots,a^{\PixelIndexThree}_N$. Let $\EstimatedImage^0 = \EstimatedImage$ be the original image. We define the sequence of images: $\EstimatedImage^{\PixelIndexThree}:\EstimatedPixel^{\PixelIndexThree}_{\PixelIndex} = \EstimatedPixel^{\PixelIndexThree-1}_{\PixelIndex}, \PixelIndex \not\in \{a^{\PixelIndexThree}_1,\ldots,a^{\PixelIndexThree}_N\}$ and otherwise let $\EstimatedPixel^{\PixelIndexThree}_{a^{\PixelIndexThree}_{\PixelIndex}}$ be given. Let $\delta^{\PixelIndexThree}_{\PixelIndex} = \EstimatedPixel^{\PixelIndexThree}_{a^{\PixelIndexThree}_\PixelIndex} - \EstimatedPixel^{\PixelIndexThree-1}_{a^{\PixelIndexThree}_\PixelIndex}$. Using the above notation we arrive at an expression for $C_{\Kernel,\ErrorImage^n}$:

\begin{equation}
    C_{\Kernel,\ErrorImage^n}(x) = C_{\Kernel,\ErrorImage}(x) + \sum_{\PixelIndexThree=1}^n\sum_{\PixelIndex=1}^N\delta^{\PixelIndexThree}_{\PixelIndex}C_{\Kernel,\Kernel}(x-a^{\PixelIndexThree}_\PixelIndex).
\end{equation}

The change in the energy due to the $n+1$-st update is:

\begin{gather}
    \delta^{n+1} =  \|\tilde{\EstimatedImage}^{n+1}-\ReferenceImage\|^2 - \|\tilde{\EstimatedImage}^n-\ReferenceImage\|^2 = \\
    \sum_{\PixelIndex\in \mathbb{Z}^2}\|\tilde{\EstimatedPixel}^{n}_{\PixelIndex}-\ReferencePixel_{\PixelIndex} + \sum_{\PixelIndexTwo=1}^{N}\delta^{n+1}_{\PixelIndexTwo}\KernelPixel_{\PixelIndex-a^{n+1}_{\PixelIndexTwo}}\|^2 - \|\tilde{\EstimatedImage}^n-\ReferenceImage\|^2 = \\
    2\sum_{\PixelIndex \in \mathbb{Z}^2}\InnerProd{\tilde{\EstimatedPixel}^n_{\PixelIndex}-\ReferencePixel_{\PixelIndex}, \sum_{\PixelIndexTwo=1}^{N}\delta^{n+1}_{\PixelIndexTwo}\KernelPixel_{\PixelIndex-a^{n+1}_{\PixelIndexTwo}}} + \sum_{\PixelIndex \in \mathbb{Z}^2}\Norm{\sum_{\PixelIndexTwo=1}^{N}\delta^{n+1}_{\PixelIndexTwo}\KernelPixel_{\PixelIndex-a^{n+1}_{\PixelIndexTwo}}}^2 = \\
    \begin{gathered}
    2\sum_{\PixelIndexTwo=1}^{N}\InnerProd{\delta^{n+1}_{\PixelIndexTwo},\sum_{\PixelIndex \in \mathbb{Z}^2}\ErrorPixel^n_{\PixelIndex}\KernelPixel_{\PixelIndex-a^{n+1}_{\PixelIndexTwo}}} + \\
    \sum_{\PixelIndexTwo=1}^{N}\sum_{\PixelIndexThree=1}^{N}\InnerProd{\delta^{n+1}_{\PixelIndexTwo}\delta^{n+1}_{\PixelIndexThree},\sum_{\PixelIndex \in \mathbb{Z}^2}\KernelPixel_{\PixelIndex-a^{n+1}_{\PixelIndexTwo}}\KernelPixel_{\PixelIndex-a^{n+1}_{\PixelIndexThree}}} =
    \end{gathered} \\
    \begin{gathered}
    2\sum_{\PixelIndexTwo=1}^{N}\InnerProd{\delta^{n+1}_{\PixelIndexTwo},C_{\Kernel,\ErrorImage^n}(a^{n+1}_{\PixelIndexTwo})} + \\
    \sum_{\PixelIndexTwo=1}^{N}\sum_{\PixelIndexThree=1}^{N}\InnerProd{\delta^{n+1}_{\PixelIndexTwo}\delta^{n+1}_{\PixelIndexThree},C_{\Kernel,\Kernel}(a^{n+1}_{\PixelIndexTwo}-a^{n+1}_{\PixelIndexThree})}.
    \end{gathered} 
\end{gather}

%%%%%%%%%%%%%%%%%%%%%%%%%%%%%%%%%%%%%%%%%%%%%%%%%%%%%%%%%%%%
\subsection{Implementation details}
%%%%%%%%%%%%%%%%%%%%%%%%%%%%%%%%%%%%%%%%%%%%%%%%%%%%%%%%%%%%

%%%%%%%%%%%%%%%%%%%%%%%%%%%%%%
\Paragraph{Leaky energy}
%%%%%%%%%%%%%%%%%%%%%%%%%%%%%%

Similar to the original paper \cite{Allebach1992}, in our extension $\delta$ was computed for a "leaky energy" which extended the support of the image by convolution. That is reflected in the fact that the sums are over $\mathbb{Z}^2$. To rectify this, the sum needs to be limited to the support of $\ReferenceImage$. This would require clamped sums of the cross-correlation to be evaluated, which can also be precomputed but requires extra memory. The same holds for the cross-correlation with $\ErrorImage$, where clamped terms are required near the image boundary.

%%%%%%%%%%%%%%%%%%%%%%%%%%%%%%
\Paragraph{Reflecting boundary conditions}
%%%%%%%%%%%%%%%%%%%%%%%%%%%%%%

Another desirable property may be a convolution such that it acts on the image extended to be reflected at the boundaries - this avoids artifacts near the borders. This can be achieved by including the relevant terms including pixels for which the kernel is partially outside of the support of $\ReferenceImage$. Care must be taken when expressing $\tilde{\EstimatedPixel}_{\PixelIndex}$, however, since it may include the same updated pixel numerous times (especially if it is near the border). The same ideas apply for a toroidally extended convolution.

%%%%%%%%%%%%%%%%%%%%%%%%%%%%%%
\Paragraph{Further optimizations}
%%%%%%%%%%%%%%%%%%%%%%%%%%%%%%

Various other strategies have been proposed in the literature for improving the runtime of iterative error minimization approaches for halftoning.

In our algorithms we usually use a randomized initial state, however, it is possibly to initialize the algorithms with the result of a dither matrix halftoning algorithm or error diffusion algorithm which would result in faster convergence \cite{Allebach1992}. 

Another strategy involves partitioning the image in blocks. Instead of updating the pixels in raster or serpentine order, the blocks are updated simultaneously by keeping only the best update per block in each iteration. This has been reported to run 10+ times faster \cite{Allebach1997}. In the same paper \cite{Allebach1997}, approximating the kernel with box functions has been proposed, yielding a speed up of 6 times. Similarly, if the kernel is separable or can be approximated by a separable kernel, the convolution can also be made considerably faster. A speed-up of an additional 30 times has been reported in \cite{DBS_GPU} through the usage of a GPU.

Finally, several heuristics related to the order in which pixels are iterated over have been proposed in \cite{DBS_heuristics}.

%%%%%%%%%%%%%%%%%%%%%%%%%%%%%%%%%%%%%%%%%%%%%%%%%%%%%%%%%%%%
\subsection{Spatially varying kernels}
%%%%%%%%%%%%%%%%%%%%%%%%%%%%%%%%%%%%%%%%%%%%%%%%%%%%%%%%%%%%

We propose an optimization for spatially varying kernels also. Let kernel $\Kernel_{\PixelIndex}$ be associated with pixel $\PixelIndex$. Let pixel $a$ be updated to a new value $\EstimatedPixel'_{a}$, while everywhere else the images match: $\EstimatedPixel'_{\PixelIndex} = \EstimatedPixel_{\PixelIndex}$, and $\delta_a = \EstimatedPixel'_{a} - \EstimatedPixel_{a}$. We denote $\tilde{\EstimatedPixel}_{\PixelIndex} = \InnerProd{\Kernel_{\PixelIndex},\EstimatedImage}$, $\tilde{\EstimatedPixel}'_{\PixelIndex} = \InnerProd{\Kernel_{\PixelIndex},\EstimatedImage'} = \tilde{\EstimatedPixel}_{\PixelIndex} + \KernelPixel_{\PixelIndex,a}\delta_a$. Our goal is to evaluate the change in the energy due to the update:
\begin{gather}
    \delta = \|\tilde{\EstimatedImage}'-\ReferenceImage\|^2 - \|\tilde{\EstimatedImage}-\ReferenceImage\|^2 = \\
    \sum_{\PixelIndex \in \mathbb{Z}^2}\|\tilde{\EstimatedPixel}_{\PixelIndex}-\ReferencePixel_{\PixelIndex}+\KernelPixel_{\PixelIndex,a}\delta_a\|^2 - \|\tilde{\EstimatedImage}-\ReferenceImage\|^2 = \\
    2\sum_{\PixelIndex \in \mathbb{Z}^2}\InnerProd{\ErrorPixel_{\PixelIndex},\KernelPixel_{\PixelIndex,a}\delta_a} + \sum_{\PixelIndex \in \mathbb{Z}^2}\|\KernelPixel_{\PixelIndex,a}\delta_a\|^2 = \\
    2\InnerProd{\delta_a,\sum_{\PixelIndex \in \mathbb{Z}^2}\ErrorPixel_{\PixelIndex}\KernelPixel_{\PixelIndex,a}} + \InnerProd{\delta^2_a,\sum_{\PixelIndex \in \mathbb{Z}^2}\KernelPixel_{\PixelIndex,a}\KernelPixel_{\PixelIndex,a}}.
\end{gather}
In the above $C_{\Kernel,\Kernel}(a)=\sum_{\PixelIndex \in \mathbb{Z}^2}\KernelPixel_{\PixelIndex,a}\KernelPixel_{\PixelIndex,a}$ may be precomputed for every $a$, which yields a function with support $\operatorname{supp}(C_{\Kernel,\Kernel})=\bigcup_{\PixelIndex}\operatorname{supp}(\Kernel_{\PixelIndex})$, and $C_{\Kernel,\ErrorImage}(a)=\sum_{\PixelIndex \in \mathbb{Z}^2}\ErrorPixel_{\PixelIndex}\KernelPixel_{\PixelIndex,a}$ can also be recomputed after enough updates have been accepted.

%%%%%%%%%%%%%%%%%%%%%%%%%%%%%%
\Paragraph{Multiple accepted updates}
%%%%%%%%%%%%%%%%%%%%%%%%%%%%%%

Let a set of accepted updates results in the differences: $\{\delta_{a^1},\ldots,\delta_{a^n}\}$. And let $\ErrorImage^n$ be the error image after the updates. We derive an expression for the efficient evaluation of $C_{\Kernel,\ErrorImage^n}$:

\begin{gather}
    C_{\Kernel,\ErrorImage^n}(x) = \sum_{\PixelIndex \in \mathbb{Z}^2}\ErrorPixel^n_{\PixelIndex}\KernelPixel_{\PixelIndex,x} = 
    C_{\Kernel,\ErrorImage}(x) + \sum_{\PixelIndexThree=1}^{n}\delta_{a^{\PixelIndexThree}}\sum_{\PixelIndex \in \mathbb{Z}^2}\KernelPixel_{\PixelIndex,a^{\PixelIndexThree}}\KernelPixel_{\PixelIndex,x}.
\end{gather}

An efficient computation of $C_{\Kernel,\ErrorImage^n}$ can then be achieved if the function $C_{\Kernel,\Kernel}(x,y) = \sum_{\PixelIndex \in \mathbb{Z}^2}\KernelPixel_{\PixelIndex,x}\KernelPixel_{\PixelIndex,y}$ is precomputed. Then, at step $n+1$ the change in energy is:

\begin{gather}
    \delta^{n+1} = \|\tilde{\EstimatedImage}^{n+1}-\ReferenceImage\|^2 - \|\tilde{\EstimatedImage}^n-\ReferenceImage\|^2 = \\
    2\InnerProd{\delta_{a^{n+1}},C_{\Kernel,\ErrorImage^n}(a^{n+1})} + \InnerProd{\delta^2_{a^{n+1}},C_{\Kernel,\Kernel}(a^{n+1})}.
\end{gather}

%%%%%%%%%%%%%%%%%%%%%%%%%%%%%%
\Paragraph{Multiple simultaneous updates}
%%%%%%%%%%%%%%%%%%%%%%%%%%%%%%

We derive an expression where an update consists of changing $N$ pixels simultaneously, and we assume that $n$ such updates have been accepted previously. We denote the differences of the pixels in update $\PixelIndexThree$:  $\{\delta^{\PixelIndexThree}_1,\ldots,\delta^{\PixelIndexThree}_N\}$. The expression for the change in the energy is given as:

\begin{gather}
    \delta^{n+1} = \|\tilde{\EstimatedImage}^{n+1}-\ReferenceImage\|^2 - \|\tilde{\EstimatedImage}^n-\ReferenceImage\|^2 = \\
    \sum_{\PixelIndex \in \mathbb{Z}^2}\|\tilde{\EstimatedPixel}^{n}_{\PixelIndex} - \ReferencePixel_{\PixelIndex} + \sum_{\PixelIndexTwo=1}^{N}\delta^{n+1}_{\PixelIndexTwo}\KernelPixel_{\PixelIndex,a^{n+1}_{\PixelIndexTwo}}\|^2 - \|\tilde{\EstimatedImage}^n-\ReferenceImage\|^2 = \\
    2\sum_{\PixelIndexTwo=1}^{N}\InnerProd{\delta^{n+1}_{\PixelIndexTwo},C_{\Kernel,\ErrorImage^n}(a^{n+1}_{\PixelIndexTwo})} + \sum_{\PixelIndex=1}^{N}\sum_{\PixelIndexTwo=1}^{N}\InnerProd{\delta^{n+1}_{\PixelIndex}\delta^{n+1}_{\PixelIndexTwo},C_{\Kernel,\Kernel}(a^{n+1}_{\PixelIndex},a^{n+1}_{\PixelIndexTwo}}.
\end{gather}

Where $C_{\Kernel,\Kernel}(x,y) = \sum_{\PixelIndex \in \mathbb{Z}^2}\KernelPixel_{\PixelIndex,x}\KernelPixel_{\PixelIndex,y}$ is assumed to be precomputed, and $C_{\Kernel,\ErrorImage^n}$ can be computed as:

\begin{equation}
    C_{\Kernel,\ErrorImage^n}(x) = 
    C_{\Kernel,\ErrorImage}(x) + \sum_{\PixelIndexThree=1}^{n}\sum_{\PixelIndexTwo=1}^{N}\delta_{a^{\PixelIndexThree}_{\PixelIndexTwo}}C_{\Kernel,\Kernel}(a^{\PixelIndexThree}_{\PixelIndexTwo},x).
\end{equation}

%%%%%%%%%%%%%%%%%%%%%%%%%%%%%%%%%%%%%%%%%%%%%%%%%%%%%%%%%%%%
\section{Relationship to previous work}
\label{sec:supp_relationship_to_previous_work}
%%%%%%%%%%%%%%%%%%%%%%%%%%%%%%%%%%%%%%%%%%%%%%%%%%%%%%%%%%%%

We show that the recent publications \cite{georgiev16bluenoise, Heitz2019b, Heitz2019} on blue noise error distribution for path tracing, can be seen as special cases in our framework. This allows for a novel analysis and interpretation of the results in the aforementioned works. We also state the necessary assumptions and approximations necessary to get from our general formulation to the algorithms presented in the papers.

%%%%%%%%%%%%%%%%%%%%%%%%%%%%%%
\Paragraph{Classification}
%%%%%%%%%%%%%%%%%%%%%%%%%%%%%%

The proposed techniques can be divided into \emph{a-priori} \cite{georgiev16bluenoise, Heitz2019b} and \emph{a-posteriori} \cite{Heitz2019}. The main difference is that for \emph{a-priori} techniques broad assumptions are made on the integrand without relying on information from renderings of the current scene. The cited \emph{a-priori} approaches describe ways for constructing offline optimized point sets/sequences.
We denote the method in \cite{georgiev16bluenoise} as BNDS (blue-noise dithered sampling), the method in \cite{Heitz2019b} as HBS (Heitz-Belcour Sobol), and the histogram and permutation method in \cite{Heitz2019} as HBH and HBP respectively (Heitz-Belcour histogram/permutation). 

%%%%%%%%%%%%%%%%%%%%%%%%%%%%%%
\Paragraph{Energy}
%%%%%%%%%%%%%%%%%%%%%%%%%%%%%%

HBH/HBP both rely on a blue noise dither matrix optimized while using a Gaussian kernel (through void-and-cluster \cite{Ulichney1993Void}). This kernel corresponds to the kernel in our framework $\Kernel$. The optimization of this dither matrix happens offline unlike in our iterative energy minimization algorithms. This imposes multiple restrictions while allowing for a lower runtime. On the other hand, the dither matrices in HBS and BNDS are optimized with respect to empirically motivated energies that cannot be related directly to what is used as energy in HBH and HBP. In the case of BNDS the energy does not even introduce an implicit integrand, and instead it is devised to represent a whole class of integrands. We propose to substitute those empirically motivated energies with a modified version of our energy. This allows an intuitive interpretation and relating \emph{a-posteriori} approaches to \emph{a-priori} approaches.

%%%%%%%%%%%%%%%%%%%%%%%%%%%%%%
\Paragraph{Search space}
%%%%%%%%%%%%%%%%%%%%%%%%%%%%%%

Another notable difference constitute the search spaces on which the different approaches operate. HBH selects a subset from a set of precomputed samples in each pixel, HBP permutes the assignment of sample sets to pixels, BNDS directly modifies the set of samples in each pixel, and HBS considers a search space made up of scrambling and ranking keys for a Sobol sequence. Working on the space of scrambling and ranking keys guarantees the preservation of the desirable integration qualities of the Sobol sequence used, and it should be clear that other methods can also be restricted to such a space. Clearly, a search space restriction diminishes the achievable blue noise quality. On the other hand, it makes sequences more robust to integrands for which those were not optimized.

\section{A-posteriori approaches}
\label{sec:supp_aposteriori_approaches}
%%%%%%%%%%%%%%%%%%%%%%%%%%%%%%%%%%%%%%%%%%%%%%%%%%%%%%%%%%%%

In this section we analyze the permutation based approach (HBP) and the histogram sampling approach (HBH) proposed in \cite{Heitz2019}. The two methods can be classified as dither matrix halftoning methods in our framework, that operate on a \emph{\horizontal} and \emph{\vertical} search space respectively. We make the approximations and assumptions necessary to get from our general formulation to HBP/HBH explicit.

We also note that \emph{a-posteriori} methods lead to solutions that adapt to the current render by exploiting known information (e.g. previously rendered data, auxiliary buffers). They can generally produce better results than \emph{a-priori} methods.

Both HBP and HBH rely on a blue noise dither matrix $\BlueNoiseImage$. Let $\BlueNoiseImage$ be the optimized blue noise dither matrix resulting from the minimization of $E(\BlueNoiseImage) = \Norm{\Kernel * \BlueNoiseImage}_2^2$ over a suitable search space. The kernel $\Kernel$ is the one used to generate the blue noise images for HBP/HBH. That is, the Gaussian kernel in the void-and-cluster method~\cite{Ulichney1993Void}. Our analysis does not rely on the kernel being a Gaussian, or on the void-and-cluster optimization, this is simply the setting of the HBP/HBH method. In the more general setting any kernel is admissible.

%%%%%%%%%%%%%%%%%%%%%%%%%%%%%%%%%%%%%%%%%%%%%%%%%%%%%%%%%%%%
\subsection{Sorting step for the permutation approach}
%%%%%%%%%%%%%%%%%%%%%%%%%%%%%%%%%%%%%%%%%%%%%%%%%%%%%%%%%%%%

The permutation approach \cite{Heitz2019} consists of two main parts: sorting (optimization), and retargeting (correcting for mispredictions). The sorting step in HBP can be interpreted as minimizing the energy:
\begin{equation}
    \label{eq:supp_HBP_energy}
    E_{HBP}(\Permutation) = \Norm{\Permutation(\EstimatedImage) - f_2(\BlueNoiseImage)}_2^2, \forall f_2 : a < b \implies f_2(a) < f_2(b).
\end{equation}
A global minimum of the above energy is achieved for a permutation $\Permutation$ that matches the order statistics of $\EstimatedImage$ and $\BlueNoiseImage$. Thus our goal would be to get from the minimization of:
\begin{equation}
\label{eq:supp_original_energy}
    E(\Permutation) = \Norm{\Kernel * (\EstimatedImage(\Permutation(\SampleSetImage)) - \ReferenceImage)}_2^2 = \Norm{\Kernel * \ErrorImage(\Permutation(\SampleSetImage))}_2^2,
\end{equation}
to the minimization of~\cref{eq:supp_HBP_energy} over a suitable search space (in practice it is limited to permutations within tiles).

We successively bound the error, while introducing the assumptions implicit to the HBP method. The bounds are not tight, however, the different error terms that we consider illustrate the major sources of error due to the approximation of the more general energy (\cref{eq:supp_original_energy}) with a simpler one (\cref{eq:supp_HBP_energy}). The substitution of the kernel convolution $\Kernel*\cdot$ by a difference with a blue noise mask $\BlueNoiseImage$ restricts the many possible blue noise error distributions towards which $\ErrorImage(\Permutation(\SampleSetImage))$ can go with a single one: $\BlueNoiseImage$. A global minimizer of the new simplified energy can thus be found by just sorting. 

The closer the distributions of $\ErrorImage(\Permutation(\SampleSetImage))$ and $\alpha\BlueNoiseImage, \alpha > 0$ are locally, the lower this restriction error can be made. Notably, for a close to linear relationship between the samples and the integrand, and sufficiently many pixels, $\ErrorImage(\Permutation(\SampleSetImage))$ and $\alpha\BlueNoiseImage$ can be matched closely in practice. A different way to reduce the approximation error is to introduce a sufficient amount of different blue noise images and pick the one that minimizes the error. \todo{Study the error behaviour due to the substitution of $\Kernel * \ErrorImage$ with $\ErrorImage - f_2(\pmb{B}^*)$}We start with the original energy (\cref{eq:supp_original_energy}) and bound it through terms that capture the main assumptions on which the model relies:
\begin{equation}
\begin{gathered}
    \label{eq:supp_HBP_blue_noise_image_bound}
    \Norm{\Kernel * \ErrorImage(\Permutation(\SampleSetImage))}_2 = \\
    \min_{f_2} \Norm{\Kernel * (\ErrorImage(\Permutation(\SampleSetImage)) - f_2(\BlueNoiseImage) + f_2(\BlueNoiseImage))}_2 \leq \\
    \min_{\alpha>0, f_2} \Norm{\Kernel}_1\Norm{ \ErrorImage(\Permutation(\SampleSetImage)) - f_2(\BlueNoiseImage)}_2  \\
    + \Norm{\Kernel * (f_2(\BlueNoiseImage) - \alpha\BlueNoiseImage + \alpha\BlueNoiseImage)}_2 \leq \\
    \min_{\alpha>0, f_2} \Norm{\Kernel}_1\Norm{ \ErrorImage(\Permutation(\SampleSetImage)) - f_2(\BlueNoiseImage)}_2 \\
    + \Norm{\Kernel}_1\Norm{(f_2(\BlueNoiseImage) - \alpha\BlueNoiseImage)}_2 + \alpha\Norm{\Kernel * \BlueNoiseImage}_2.
\end{gathered}
\end{equation}
In the above, $f_2$ is taken over the space of all strictly monotonically increasing functions, and $\alpha>0$ is a real value used to provide an amplitude matching between $\ErrorImage(\Permutation(\SampleSetImage))$ and $\BlueNoiseImage$ (this allows for the second term to go to zero as the pointwise error goes to zero).

%%%%%%%%%%%%%%%%%%%%%%%%%%%%%%
\subsubsection{Third error term}
%%%%%%%%%%%%%%%%%%%%%%%%%%%%%%

We note that $\BlueNoiseImage$ is precomputed offline in order to approximately minimize $E(\BlueNoiseImage) = \Norm{\Kernel * \BlueNoiseImage}_2$. Thus, the third term reflects the quality of the blue noise achieved with respect to $\Kernel$ in the offline minimization. This error can be made small without a performance penalty since the optimization is performed offline. We factor out a multiplicative scaling factor $\alpha>0$ in the blue noise quality term, to allow for the second term to go to zero. With this change, we can consider $\BlueNoiseImage$ to be normalized in the range $[-1,1]$ and we can encode the scaling in $\alpha$.

%%%%%%%%%%%%%%%%%%%%%%%%%%%%%%
\subsubsection{Second error term}
\label{ss:supp_HBP_2nd_error_term}
%%%%%%%%%%%%%%%%%%%%%%%%%%%%%%

The second term reflects the error introduced by substituting a large search space (many local minima) with a small search space. It introduces the first implicit assumption of HBP by relating the first and third error terms (by using $f_2$ and $\alpha$ respectively) through the second error term. The assumption is that there exists a permutation for which $\ErrorImage(\Permutation(\SampleSetImage))$ can be made close to $\alpha\BlueNoiseImage$, which would make the second term small. This holds in practice if the pixel-wise error is zero on average (unbiased estimator within each pixel), and we have a sufficiently large resolution/tiles: which results in a higher probability that pixels from $\ErrorImage(\Permutation(\SampleSetImage))$ can match $\BlueNoiseImage$ well. Then the term $\Norm{\Kernel}_1\Norm{f_2(\BlueNoiseImage) - \alpha\BlueNoiseImage}_2$ can be made small. We note that this is a generalization of the third optimality condition in \cite{Heitz2019} (\emph{correlation-preserving integrand}) since an integrand linear in the samples can also better match $\BlueNoiseImage$ provided enough pixels. For a linear integrand the optimal $f_2$ is also a linear function (ideal correlation between samples and integrand). The main difference between a linear integrand and a nonlinear/discontinuous one, is the amount of sample sets/pixels necessary to match $f_2(\BlueNoiseImage)$ well, given an initial white noise samples' distribution. So in practice there are 4 factors directly affecting the magnitude of the second term: the number of considered blue noise images, the size of the tiles, the correlation between samples and integrand (accounted for by $f_2$), the bias/consistency of the estimators. \todo{Study behaviour with respect to number of blue noise images used, resolution, linearity - partially done through the comparison of the effect of tiling on the problem.} 

We note that the number of considered pixels depends on the tile size in HBP, and the practical significance of this has been demonstrated through a canonical experiment in the main paper.

%%%%%%%%%%%%%%%%%%%%%%%%%%%%%%
\subsubsection{First error term}
%%%%%%%%%%%%%%%%%%%%%%%%%%%%%%

Before we proceed we need to further bound the first error term by substituting $\EstimatedImage(\Permutation(\pmb{S}))$ by $\Permutation(\EstimatedImage(\pmb{S}))$. As discussed in the main paper, this is achieved by introducing a difference term $\ErrorDifference(\Permutation) = \EstimatedImage(\Permutation(\pmb{S}))-\Permutation(\EstimatedImage(\pmb{S}))$, and then $\sqrt{E_{HBP}}$ is recovered. The error there can be made arbitrarily small through $f_2$ (it is accounted for in the second term). Thus we only need to study the remaining error due to $\ErrorDifference$. In the case of HBP, $\ErrorDifference$ is approximated by non-overlapping characteristic functions in each tile ($d(x,y)=\infty$, for $x,y$ in different tiles). This means that the approximation error is zero within each tile if the integrands are the same within the tile and permutations act only within the tile, since $\ErrorDifference(\Permutation)=\pmb{0}$. On the other hand, if this assumption is violated, mispredictions occur, usually resulting in white noise.

%%%%%%%%%%%%%%%%%%%%%%%%%%%%%%
\subsubsection{$\ErrorDifference$ term}
%%%%%%%%%%%%%%%%%%%%%%%%%%%%%%

HBP partitions screen space into a several tiles $\mathcal{R}_1$, $\ldots$, $\mathcal{R}_K$, and permutations are only over the pixel values in a tile. Having the partition induced by the tiling we can bound the first term:
\begin{equation}
    \Norm{\ErrorImage(\Permutation(\SampleSetImage)) - f_2(\BlueNoiseImage)}_2 \leq \sum_{k=1}^{K}\Norm{\ErrorImage_k(\Permutation_k(\SampleSetImage_k)) - f_2(\BlueNoiseImage)}_2.
\end{equation}
Since additionally the permutations are optimized for the pixel values instead of the sample sets (which saves re-rendering operations), then there is an assumption that within each tile $\mathcal{R}_{\PixelIndexThree}$ the following holds (we denote $\pmb{A}_{\PixelIndexThree} = \pmb{A}\restrict{\mathcal{R}_{\PixelIndexThree}}):$
\begin{equation}
    \EstimatedImage_{\PixelIndexThree}(\Permutation_{\PixelIndexThree}(\SampleSetImage_{\PixelIndexThree})) = \Permutation_{\PixelIndexThree}(\EstimatedImage_{\PixelIndexThree}(\SampleSetImage_{\PixelIndexThree})).
\end{equation}
Consequently it follows that $\ReferencePixel_{\PixelIndex} = \ReferencePixel_{\PixelIndexTwo}, \forall \PixelIndex,\PixelIndexTwo \in \mathcal{R}_{\PixelIndexThree}$.

This assumption can be identified with the 4-th optimality condition proposed in \cite{Heitz2019}: \emph{screen-space coherence}. As discussed, the search space restriction to the tiles corresponds to an approximation of the $\ErrorDifference$ term in our framework by characteristic functions: $d_k(x,y) = \infty, x \in \mathcal{R}_k, y\not\in \mathcal{R}_k$ and $d_k(x,y) = 0, x,y \in \mathcal{R}_k$. To account for the actual error when the assumption is violated we introduce an additional error term per tile: $\ErrorDifference_k=\EstimatedImage_k(\Permutation_k(\SampleSetImage_k)) - \Permutation_k(\EstimatedImage_k(\SampleSetImage_k))$, then we have the bound:
\begin{equation}
\begin{gathered}
    \label{eq:supp_HBR_final_energy}
    \Norm{\ErrorImage_k(\Permutation_k(\SampleSetImage_k)) - f_2(\BlueNoiseImage_k)}_2 = \\
    \Norm{\Permutation_k(\EstimatedImage_k(\SampleSetImage_k)) - \ReferenceImage_k - f_2(\BlueNoiseImage_k) + \ErrorDifference_k}_2 \leq \\ \Norm{\Permutation_k(\EstimatedImage_k(\SampleSetImage_k)) - \ReferenceImage_k - f_2(\BlueNoiseImage_k)}_2 + \Norm{\ErrorDifference_k}_2.
\end{gathered}
\end{equation}
This means that even if all of the previous error terms are made small, including $\Norm{\Permutation_k(\ErrorImage_k(\SampleSetImage_k)) - f_2(\BlueNoiseImage_k)}_2$, the error may still be large due to $\Norm{\ErrorDifference}_2$. We refer to a large error due to the delta term as \emph{misprediction} - that is, a mismatch between the predicted error distribution from the minimization of $\Norm{\Permutation_k(\ErrorImage_k(\SampleSetImage_k)) - f_2(\BlueNoiseImage_k)}_2$ and the actual error distribution resulting from the above permutation applied to $\ErrorImage_k(\Permutation_k(\SampleSetImage_k))$. The best way to identify mispredictions is to compare the predicted image $\Permutation_k(\EstimatedImage_k(\SampleSetImage_k))$ and the image rendered with the same permutation for the sample sets $\EstimatedImage_k(\Permutation_k(\SampleSetImage_k))$. A misprediction occurring means that the assumption made to approximate $\ErrorDifference$ was incorrect ( $\ErrorDifference_k \ne \pmb{0}$ for some tile $\mathcal{R}_{\PixelIndexThree}$), equivalently the optimality condition of \emph{screen-space coherence} is not satisfied. \todo{Showcase predictions and mispredictions at regions in screen space where the assumption holds and where it does not.}

%%%%%%%%%%%%%%%%%%%%%%%%%%%%%%
\Paragraph{Avoiding mispredictions}
%%%%%%%%%%%%%%%%%%%%%%%%%%%%%%

In practice mispredictions often occur for larger tile sizes, since it is hard to guarantee that the integrand remains similar over each tile. On the other hand, larger tiles allow for a better blue noise as long as $\ErrorDifference_k=0$ in each tile, thus larger tiles are desirable. The method fails even more often near edges, since even for small tile sizes it allows swapping pixels over an edge. A straightforward improvement involves partitioning the domain by respecting edges. More involved methods may take into account normals, depth, textures, etc. 

\subsubsection{$E_{HBP}$ error term} The final step involves the minimization of the energy in \cref{eq:supp_HBR_final_energy}. Since different tiles do not affect each other the minimization can be performed per tile (we adopt the assumption from HBP $\ErrorDifference_k = \pmb{0}$):
\begin{equation}
\begin{gathered}
    \Permutation^*_k \in \arg\min_{\Permutation_k}\Norm{\Permutation_k(\EstimatedImage_k(\SampleSetImage_k)) - \ReferenceImage_k - f_2(\BlueNoiseImage_k)}_2 = \\
    \arg\min_{\Permutation_k}\Norm{\Permutation_k(\EstimatedImage_k(\SampleSetImage_k)) - f_2(\BlueNoiseImage_k)}^2_2.
\end{gathered}    
\end{equation}
We have dropped the term $\ReferenceImage_k$ since it does not affect the set of minimizers ($\ReferenceImage_k$ is assumed constant in each tile). As discussed in \cref{eq:supp_HBP_energy}, a global minimum is given by matching the order statistics of $\EstimatedImage_k$ to the order statistics of $f_2(\BlueNoiseImage))$ (we note that the order statistics of $\BlueNoiseImage_k$ do not change from the application of $f_2$ since it is a strictly increasing function). This is equivalent to performing the sorting pass described in \cite{Heitz2019}. A minor optimization would be to pre-sort $\BlueNoiseImage$ and instead store the sorted indices.

%%%%%%%%%%%%%%%%%%%%%%%%%%%%%%
\begin{figure}[t!]
    \centering
    %!TEX root = ../supplemental.tex

\footnotesize
\hspace*{-4.5mm}
\begin{tabular}{c@{\;}c@{\;}c@{\;}c@{\;}c@{\;}c@{\;}c@{}}
    & & Power & & Power & Heitz and & Power
    \\
    & Ours ($R=1$) & spectrum & Ours ($R=2$) & spectrum & \!Belcour~\shortcite{Heitz2019}\! & spectrum
    \\
    \rotatebox{90}{\quad tile size 2}
    &
    \includegraphics[width=0.535in,page=1]{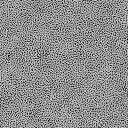}
    &
    \includegraphics[width=0.535in,page=1]{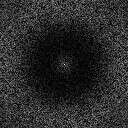}
    &
    \includegraphics[width=0.535in,page=1]{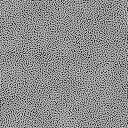}
    &
    \includegraphics[width=0.535in,page=1]{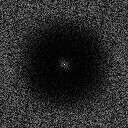}
    &
    \includegraphics[width=0.535in,page=1]{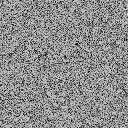}
    &
    \includegraphics[width=0.535in,page=1]{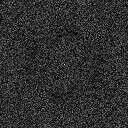}
    \\
    \rotatebox{90}{\quad tile size 4}
    &
    \includegraphics[width=0.535in,page=1]{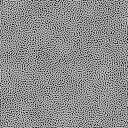}
    &
    \includegraphics[width=0.535in,page=1]{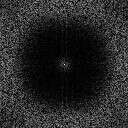}
    &
    \includegraphics[width=0.535in,page=1]{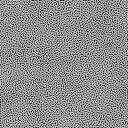}
    &
    \includegraphics[width=0.535in,page=1]{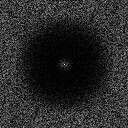}
    &
    \includegraphics[width=0.535in,page=1]{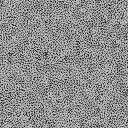}
    &
    \includegraphics[width=0.535in,page=1]{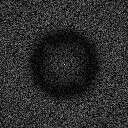}
    \\
    \rotatebox{90}{\quad tile size 8}
    &
    \includegraphics[width=0.535in,page=1]{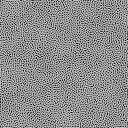}
    &
    \includegraphics[width=0.535in,page=1]{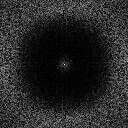}
    &
    \includegraphics[width=0.535in,page=1]{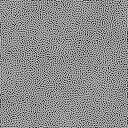}
    &
    \includegraphics[width=0.535in,page=1]{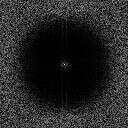}
    &
    \includegraphics[width=0.535in,page=1]{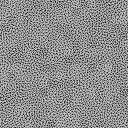}
    &
    \includegraphics[width=0.535in,page=1]{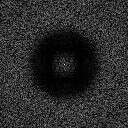}
\end{tabular}%

    \vspace{-2mm}
    \caption{
        Here we showcase the effect of tile size on the quality of blue noise. We also demonstrate the effect of a larger search neighborhood $R$ in our  optimization \ifdefined\arXiv
        %\textcolor{red}{MISSING}.
        Alg. 2 (main paper).
        \else
        \cref{alg:HorizontalOptimization}.
        \fi
        For our case, we consider disk neighborhoods so that they are contained within \citeauthor{Heitz2019}'s tiles in terms of size, but they can also overlap due to our formulation. From left-to-right, the input white noise texture is optimized using our relocation algorithm. The last two columns are from~\citeauthor{Heitz2019}'s~\shortcite{Heitz2019} method. The corresponding power spectra of these optimized images ($128 \times 128$) are also shown.
    }
    \label{fig:supp_tiling-effect}
\end{figure}
%%%%%%%%%%%%%%%%%%%%%%%%%%%%%%

%%%%%%%%%%%%%%%%%%%%%%%%%%%%%%
\Paragraph{Tiling effect}
%%%%%%%%%%%%%%%%%%%%%%%%%%%%%%

%\iliyan{This paragraph seems totally out of place. Is this specific aspect really that interesting, given all the other comparisons we have? We're frying a bigger fish now in this revision. The paper is dense enough already that it seems best to move this to supplemental too. We can summarize the main message of this experiment in the experiments section and refer to supplemental.}\vassillen{It is important for understanding why Heitz gets mostly white noisy results. It can be moved to the supplemental however, since I do the Heitz methods' analysis there.}
In \cref{fig:supp_tiling-effect} we compare the effect of the tile size. In our approach, the ``tiles'' can be  defined per pixel, can have arbitrary shapes, and are overlapping, the last being crucial for achieving a good blue noise distribution. We consider white-noise with mean 0.5 (which is an ideal scenario for~\citeauthor{Heitz2019}'s method) and compare various tile sizes.
For a fair comparison, our tile radius $r$ corresponds similar tile-size in the permutation~\shortcite{Heitz2019} approach.
%in~\citeauthor{Heitz2019}'s method, where for a fair comparison we set the $r$ parameter of our dissimilarity metric (\cref{sec:relocation_based}) to yield a disk with similar size to the tiles used in the permutation approach. 
%Notice that our smallest radius tile of 1 produces a better pMSE than~\citeauthor{Heitz2019}'s method at a tile of 32x32, and in general the pMSE for our method is about 6$\times$ lower at comparable tile sizes. 
The power-spectrum profiles confirm the better performance of our method. Retargeting~\shortcite{Heitz2019} cannot improve the quality of the permutation approach either, since no misprediction can occur ($\ErrorDifference = 0$). The adverse effect of tiling is exacerbated in practice since, for images which are not smooth enough in screen space, tiles of smaller sizes need to be considered.

%%%%%%%%%%%%%%%%%%%%%%%%%%%%%%
\Paragraph{Custom surrogate}
%%%%%%%%%%%%%%%%%%%%%%%%%%%%%%

The $\ReferenceImage_k$ term does not need to be assumed constant in fact. If it is assumed constant, that is equivalent to picking a tile-constant surrogate, however, a custom surrogate may be provided instead. Then one would simply minimize the energy:

\begin{equation}
\Norm{\Permutation_k(\EstimatedImage_k(\SampleSetImage_k)) - (\BlueNoiseImage_k+\ReferenceImage_k)}^2_2.
\end{equation}

The energy has a different minimizer than the original HBP energy, but the global minimum can be found efficiently through sorting once again.

%%%%%%%%%%%%%%%%%%%%%%%%%%%%%%%%%%%%%%%%%%%%%%%%%%%%%%%%%%%%
\subsection{HBP retargeting}
%%%%%%%%%%%%%%%%%%%%%%%%%%%%%%%%%%%%%%%%%%%%%%%%%%%%%%%%%%%%

The retargeting pass in HBP achieves two things. It introduces new possible target solutions through new blue noise images, and it corrects for mispredictions. The first is not so much a result of the retargeting, as it is of varying the blue noise image every frame. Ideally several blue noise images would be considered in a single frame, and the best image would be chosen per tile (in that case one must make sure that there are no discontinuities between the blue noise images' tiles) in order to minimize the second term in \cref{eq:supp_HBP_blue_noise_image_bound}. Instead, in HBP this is amortized over several frames. 

The more important effect of retargeting is correcting for mispredictions, by transferring the recomputed correspondence between sample set and pixel value (achieved through rerendering) to the next frame. This allows reducing the error due to the approximation of $\ErrorDifference$ (when the piecewise-tile constancy assumption on the integrand is violated). Note however, that this is inappropriate if there is a large temporal discontinuity between the two frames. 

%%%%%%%%%%%%%%%%%%%%%%%%%%%%%%
\Paragraph{Implementation details}
%%%%%%%%%%%%%%%%%%%%%%%%%%%%%%

Retargeting requires a permutation that transforms the blue noise image in the current frame into the blue noise image of the next frame \cite{Heitz2019}. This permutation is applied on the optimized seeds to transfer the learned correspondence between sample sets and pixel values to the next frame. Implicitly, this transforming permutation also relies on a screen space integrand similarity assumption, since there is no guarantee that the corresponding values from the swap will match, possibly incurring a misprediction once again (it can be modeled by an additional $\ErrorDifference$ term). In HBP \cite{Heitz2019} the maximum radius of travel of each pixel in the permutation is set to 6 pixels. This has a direct effect on the approximation of $\pmb{\Delta}$, as the travel distance of a pixel is allowed to extend beyond the original tile bounds. In the worst case scenario a pixel may allowed to travel a distance of $\sqrt{t_x^2 + t_y^2} + 6$ pixels, where $t_x, t_y$ are the dimensions of the tiles. 
An additional error is introduced since the retargeting pass does not produce the exact blue noise image used in the next frame, but some image that is close to it \cite{Heitz2019}. This seems to be done purely from memory considerations since it allows one blue noise image to be reused by translating it toroidally each frame to produce the blue noise image for the next frame.

%%%%%%%%%%%%%%%%%%%%%%%%%%%%%%
\Paragraph{Relationship to our \horizontal approach}
%%%%%%%%%%%%%%%%%%%%%%%%%%%%%%

Our \horizontal approach does not require a retargeting pass. It can directly continue with the optimized sample sets and pixel values from last frame. There is also no additional travel distance for a matching permutation as in retargeting, which further minimizes the probability of misprediction. Thus, it inherently and automatically produces all of the advantages of retargeting while retaining none of its disadvantages.

\subsection{Histogram sampling approach}
\label{sec:supp_histogram_sampling}
%%%%%%%%%%%%%%%%%%%%%%%%%%%%%%%%%%%%%%%%%%%%%%%%%%%%%%%%%%%%

The histogram sampling approach from \citeauthor{Heitz2019} can be interpreted as both a dithering and a sampling method. We study the dithering aspect to better understand the quality of blue noise achievable by the method. 

%%%%%%%%%%%%%%%%%%%%%%%%%%%%%%
\Paragraph{Algorithm analysis}
%%%%%%%%%%%%%%%%%%%%%%%%%%%%%%

The sampling of an estimate in each pixel by using the corresponding mask value to the pixel can be interpreted as performing a mapping of the mask's range and then quantizing to the closest estimate. In HBH each estimate is equally likely to be sampled (if a random mask is used), which implies a transformation that maps equal parts of the range to each estimate. Let $\EstimatedPixel_{\PixelIndexThree,1}$, $\ldots$, $\EstimatedPixel_{\PixelIndexThree,N}$ be the greyscale estimates in pixel $\PixelIndexThree$ sorted in ascending order. Let the range of the blue noise mask be in [0,1]. Then the range is split into $N$ equal subintervals: $[0,\frac{1}{N}),\ldots,[\frac{N-1}{N},1]$ which respectively map to $[\EstimatedPixel_1, \frac{\EstimatedPixel_1+\EstimatedPixel_2}{2}),\ldots,[\frac{\EstimatedPixel_{i-1}+\EstimatedPixel_i}{2},\frac{\EstimatedPixel_{i}+\EstimatedPixel_{i+1}}{2}),\ldots,[\frac{\EstimatedPixel_{N-1}+\EstimatedPixel_N}{2}, \EstimatedPixel_N]$. If the quantization rounds to the closest estimate, then the above mapping guarantees the desired behavior. We note that since the estimates in each pixel can have different values, the mapping for each pixel may be different. We will denote the above mapping through $\pmb{f}$. Then the mapping plus quantization problem in a pixel $\PixelIndexThree$ may be formulated as:

\begin{equation}
    \min_{\PixelIndex \in \{1,...,N\}}|\EstimatedPixel_{\PixelIndexThree,\PixelIndex} - f_{\PixelIndexThree}(\BlueNoisePixel_{\PixelIndexThree})|.
\end{equation}

Note that the minimization in each pixel is independent, and it aims to minimize the distance between the estimates and the remapped value from the blue noise mask. If the set of estimates are assumed to be the same across pixels, and are also assumed to be spaced regularly, then $f$ is only a linear remapping, which effectively transfers the spectral properties of $\BlueNoiseImage$ onto the optimized image. Notably, the former is the \emph{screen-space coherence} assumption from HBP, while the latter is the \emph{correlation-preserving integrand} assumption. Thus we have seen that for optimal results the HBH method relies on exactly the same assumption as the HBP method (while our \emph{\vertical} iterative minimization approach lifts both assumptions).

%%%%%%%%%%%%%%%%%%%%%%%%%%%%%%
\Paragraph{Disadvantages}
%%%%%%%%%%%%%%%%%%%%%%%%%%%%%%

One of the key points is that the error distribution and not the signal itself ought to ideally be shaped as $\BlueNoiseImage$. This is actually the case even in the above energy. From the way $\pmb{f}$ was chosen it follows that the surrogate is equivalent to $\pmb{f}(0.5)$ which can be identified as the image made of the median of the sorted estimates within each pixel. This is the case since if the target surrogate of $\BlueNoiseImage$ (during the offline optimization) was assumed to be $0.5$, then after the mapping it is $\pmb{f}(0.5)$. Generally, this is a very bad surrogate in the context of rendering, and it generally increases the error compared to the averaged image, making the method impractical. 

Another notable disadvantage is that all estimates are considered with an equal weight. This means that outliers are as likely to be picked as estimates closer to the surrogate. This results in fireflies appearing even when those were not present in the averaged imaged. Compared to classical halftoning, where only the \emph{closest} lower and upper quantization levels are considered, HBH does not minimize the magnitude of the error to the surrogate.

Finally, the two assumptions of: \emph{screen-space coherence} and \emph{correlation}-\emph{preserving integrand}, generally do not hold in practice. Estimates cannot be assumed to match between pixels (especially if samples are taken at random), and they cannot be assumed to be uniformly distributed, which implies that $\pmb{f}$ is not linear. This greatly impacts the quality of the result, especially if it is compared to \emph{adaptive} approaches such as our \emph{\vertical} error diffusion approach and our iterative minimization techniques (see the experiments in the main paper).

%%%%%%%%%%%%%%%%%%%%%%%%%%%%%%
\Paragraph{Generalization}
%%%%%%%%%%%%%%%%%%%%%%%%%%%%%%

The method can be generalized to take a custom surrogate instead of the one constructed by the median of the estimates within each pixel. This is achieved by splitting the per pixel set of estimates into two parts: (greyscale) estimates greater than the value of the (greyscale) surrogate in the current pixel, and estimates lower than it. Then the mapping $f_k$ for the current pixel $k$ maps values in $[0,0.5)$ to the lower set, and values in $[0.5,1]$ to the higher set, such that $f_k(0.5) = \ReferencePixel_k$. The original method is recovered if the surrogate is chosen to be the implicit one for the original histogram sampling method and if the appropriate corresponding mapping $\pmb{f}$ is kept.

The approach can be extended further by setting different probabilities for the different estimates. The original histogram sampling method correspond to setting the same probability for sampling every estimate, equivalently: equal sized sub-intervals from $[0,1]$ map to each estimate. Classical dither matrix halftoning can be interpreted as setting an equal probability for the closest to the surrogate upper and lower estimates, while every other estimate gets a zero probability. Equivalently: equal sub-intervals from $[0,1]$ map to the two aforementioned estimates while no part of the interval maps to the remaining estimates. Generally a custom probability can be assigned to each estimate: $p_1,...,p_N$, by having the intervals $[0,p_1),...,[\sum_{k=1}^{N-1}p_k,1]$ map to $Q_1,...,Q_N$ (after quantization). We note that an unbiased image can be recovered only if there is a map to every estimate.

%The analysis for the permutation approach (HBP) is also valid %for the histogram approach (HBH) with some small changes. The %same bound \cref{eq:HBP_blue_noise_image_bound} can be used to %illustrate the errors introduced by HBH. The main difference is %that: a) $\ErrorDifference$ is always zero, b) there are no %(explicit) tiles where the integrand is assumed to be the same. %We will see that even if no tiles are assumed the method still %relies on a local integrand similarity assumption. In %comparison, our subset selection method does not require such an %assumption.
%
%In HBH each pixel is rendered using several different sample %sets, resulting in several different estimates: %$\EstimatedPixel_i(\SampleSetPixel_{i,1}), \ldots, %\EstimatedPixel_i(\SampleSetPixel_{i,N})$. We will also assume %that the above indexing has the estimates sorted from darkest to %brightest (HBH works on the luminance/brightness). A precomputed %blue noise image $\pmb{B}^*$ is used to choose one of the %estimates by remapping the range of $\pmb{B}^*$ to $[0,N+1)$ and %then quantizing the values to get the index of the estimate in %each pixel. Similar to the permutation approach, this matches %the order statistics of $\pmb{B}^*_i$ and the vector

\todo{Compare with subset selection}

%%%%%%%%%%%%%%%%%%%%%%%%%%%%%%%%%%%%%%%%%%%%%%%%%%%%%%%%%%%%
\section{A-priori approaches}
\label{sec:supp_apriori_approaches}
%%%%%%%%%%%%%%%%%%%%%%%%%%%%%%%%%%%%%%%%%%%%%%%%%%%%%%%%%%%%

We discuss current state of the art \emph{a-priori} approaches \cite{georgiev16bluenoise, Heitz2019b} and their relation to our framework, as well as insights regarding those.

%%%%%%%%%%%%%%%%%%%%%%%%%%%%%%%%%%%%%%%%%%%%%%%%%%%%%%%%%%%%
\subsection{HBS}
%%%%%%%%%%%%%%%%%%%%%%%%%%%%%%%%%%%%%%%%%%%%%%%%%%%%%%%%%%%%

In \citeauthor{Heitz2019b}'s work, a scrambling energy and a ranking energy have been proposed (note that those energies are maximized and not minimized):

\begin{gather}
    E_s = \sum_{\PixelIndex,\PixelIndexTwo} \exp\left(-\frac{\|\PixelIndex-\PixelIndexTwo\|^2_2}{2\sigma^2}\right)\|E_{\PixelIndex}-E_{\PixelIndexTwo}\|^2_2 \\
    E_r = \sum_{\PixelIndex,\PixelIndexTwo} \exp\left(-\frac{\|\PixelIndex-\PixelIndexTwo\|^2_2}{2\sigma^2}\right)(\|E_{i}^{1}-E_{j}^{1}\|^2_2+\|E_{i}^{2}-E_{j}^{2}\|^2_2) \\
    E_{\PixelIndex} = (e_{1,\PixelIndex},\ldots,e_{T,\PixelIndex}) \\
    e_{t,\PixelIndex}(\SampleSetPixel_{\PixelIndex}) = \frac{1}{|\SampleSetPixel_{\PixelIndex}|}\sum_{\PixelIndexThree=1}^{|\SampleSetPixel_{\PixelIndex}|}f_t(\SampleSetPoint_{\PixelIndex,\PixelIndexThree}) - \int_{[0,1]^D}f_t(x)\,dx \\
    \SampleSetPixel_{\PixelIndex} = \{\SampleSetPoint_{\PixelIndex,1},\ldots,\SampleSetPoint_{\PixelIndex,M_i}\}.
\end{gather}

The upper indices in $E^1_{\PixelIndex},E^2_{\PixelIndex}$ indicate that the two energies are evaluated with different subsets of the sample set $\SampleSetPixel_{\PixelIndex}$ in the pixel $\PixelIndex$. The $f_t$ are taken from an arbitrary set of functions (in the original paper those are random Heaviside functions). The described form of the energies has been partially motivated by the energy in \cite{georgiev16bluenoise}. This does not allow for a straightforward interpretation or a direct relation to the (implicit) energy used for \emph{a-posteriori} approaches in \cite{Heitz2019}. 

%%%%%%%%%%%%%%%%%%%%%%%%%%%%%%
\Paragraph{Scrambling energy}
%%%%%%%%%%%%%%%%%%%%%%%%%%%%%%

We modify $E_s$ in order to relate it to the energy in our framework and to provide a meaningful interpretation:

\begin{gather}
    E_s' = \sum_{t=1}^{T}w_t\|\Kernel * \EstimatedImage_t(\SampleSetImage) - \ReferenceImage_t\|^2_2, \\
    \EstimatedPixel_{t,\PixelIndex}(\SampleSetImage) = \frac{1}{|\SampleSetPixel_{\PixelIndex}|}\sum_{\PixelIndexThree=1}^{|\SampleSetPixel_{\PixelIndex}|}f_t(\SampleSetPoint_{\PixelIndex,\PixelIndexThree}), \,\,
    \ReferencePixel_{t,\PixelIndex} = \int_{[0,1]^D}f_t(x)\,dx.
\end{gather}

We have relaxed the Gaussian kernel to an arbitrary kernel $\Kernel$ and absorbed it into the norm. More importantly we have removed the heuristic dependence of error terms on their neighbors, and instead the coupling happens through the kernel itself. Finally, we have introduced weights $w_1,\ldots,w_T$ that allow assigning different importance to different integrands. Thus, this is a weighted average of our original energy applied to several different integrands, matching our \emph{a-priori} approach (\cref{eq:supp_our_apriori_energy}). Through this formulation a direct relationship to the \emph{a-posteriori} methods can be established, and it can be motivated in the context of both the human visual system and denoising. Particularly, the scrambling energy $E_s'$ is over the space of scrambling keys, which allow permuting the assignment of sample sets. This is in fact the \emph{\horizontal} setting from our formulation in the main paper. The space can be extended further if the scrambling keys in each dimension are different (as in HBS). The same can be done in \emph{a-posteriori} methods, if the optimization is performed in each dimension as discussed in \cref{sec:supp_our_demodulation}.

%%%%%%%%%%%%%%%%%%%%%%%%%%%%%%
\Paragraph{Ranking energy}
%%%%%%%%%%%%%%%%%%%%%%%%%%%%%%

The ranking keys in HBS describe the order in which samples are consumed. This is useful for constructing progressive \emph{a-priori} methods. Notably, the order in which samples will be introduced can be optimized. Having a sequence of sample sets in each pixel: $\SampleSetPixel_{\PixelIndex,1} \subset \ldots \subset \SampleSetPixel_{\PixelIndex,M} \equiv \SampleSetPixel_{\PixelIndex}$ and respectively the images formed by those: $\SampleSetImage_1,\ldots,\SampleSetImage_M$, the progressive energy may be constructed as:

\begin{gather}
    E_r' = \sum_{\PixelIndexThree=1}^{M}w_{\PixelIndexThree}\|\Kernel * \EstimatedImage(\SampleSetImage_{\PixelIndexThree}) - \ReferenceImage\|^2_2.
\end{gather}

The quality at a specific sample count corresponding to $\SampleSetImage_{\PixelIndexThree}$ is controlled through the weight $w_{\PixelIndexThree}$. The original energy maximizing the quality of the full set is retrieved for $(w_1,\ldots,w_{M-1},w_M) = (0,\ldots,0,1)$. Since the sample sets $\SampleSetPixel_{\PixelIndex}, \ldots, \SampleSetPixel_{\PixelIndex,M}$ are optimized by choosing samples from $\SampleSetPixel_{\PixelIndex}$ this can be seen as a \emph{\vertical} method. Finally, the ranking keys can also be defined per dimension, which can be related to \emph{a-posteriori} methods through the suggested dimensional decomposition in \cref{sec:supp_our_demodulation}. \todo{Discuss progressive and adaptive in the main paper.}
\vassillen{Possibly repetition with the section our a-priori approaches, but it is probably fine to repeat it in other words, especially in the context of Heitz's method.}

%%%%%%%%%%%%%%%%%%%%%%%%%%%%%%
\begin{figure*}[t!]
    \centering
    %!TEX root = ../supplemental.tex

\newcommand{\AprioriMethodTemplate}[2]{
    \begin{scope}
        \clip (0,0) -- (0.5,0) -- (0.5,1) -- (0,1) -- cycle;
        \path[fill overzoom image=figures/supp_apriori-teapot/#1] (0,0) rectangle (1,1);
    \end{scope}
    \begin{scope}
        \clip (0.5,0) -- (1,0) -- (1,1) -- (0.5,1) -- cycle;
        \path[fill overzoom image=figures/supp_apriori-teapot/#2] (0,0) rectangle (1,1);
    \end{scope}
}

\small
\hspace*{-2.5mm}
\begin{tabular}{c@{\;}c@{\;}c@{\;}c@{\;}c@{}}
    \begin{tikzpicture}[scale=3.54]
        \AprioriMethodTemplate{halftoning_ar_spp4_simplified_teapot_diffuse_Random}{ps_ar_spp4_simplified_teapot_diffuse_Random}
    \end{tikzpicture}
    &
    \begin{tikzpicture}[scale=3.54]
        \AprioriMethodTemplate{halftoning_ar_spp4_simplified_teapot_diffuse_BNDS_RSeq}{ps_ar_spp4_simplified_teapot_diffuse_BNDS_RSeq}
    \end{tikzpicture}
    &
    \begin{tikzpicture}[scale=3.54]
        \AprioriMethodTemplate{halftoning_ar_spp4_simplified_teapot_diffuse_HeitzSobolBlue}{ps_ar_spp4_simplified_teapot_diffuse_HeitzSobolBlue}
    \end{tikzpicture}
    &
    \begin{tikzpicture}[scale=3.54]
        \AprioriMethodTemplate{halftoning_ar_spp4_simplified_teapot_diffuse_BN_RSeq}{ps_ar_spp4_simplified_teapot_diffuse_BN_RSeq}
    \end{tikzpicture}
    &
    \begin{tikzpicture}[scale=3.54]
        \AprioriMethodTemplate{halftoning_ar_spp4_simplified_teapot_diffuse_Sobol}{ps_ar_spp4_simplified_teapot_diffuse_Sobol}
    \end{tikzpicture}
    \\[-0.3mm]
    Random & \citet{georgiev16bluenoise} & \citet{Heitz2019b} & Ours & Sobol
    \\
    MSE: $0.118636$ & 
    $0.0921076$ & 
    $0.0787028$ &
    $0.117336$  &
    $0.178861$ 
    \\
    pMSE: $0.0170958$ &
    $0.011277$ &
    $0.00869183$ &
    $0.0119757$ &
    $0.0126065$
\end{tabular}

\let\AprioriMethodTemplate\undefined
    \vspace{-2mm}
    \caption{
        A comparison illustrating that even a sampling sequence formed by a stack of blue noise images (Ours) yields a good distribution (note the tiled error spectra). The integration error is higher however, degrading the quality. This is the case because the assumed integrand is far from linear in each dimension (see \emph{Extension} in \cref{ss:supp_BDNS_algorithm}). The images use 4 samples per pixel, and the degradation of the spectral properties with the number of samples is clear for \cite{georgiev16bluenoise} and even \cite{Heitz2019b}, while it is not so much the case for Ours. This demonstrates that different methods offer a different trade-off between integration error and distribution for arbitrary integrands. Constraining the search space to using toroidal shifts or scrambling and ranking keys restricts the achievable blue noise distribution.
    }
    \label{fig:supp_apriori-teapot}
\end{figure*}
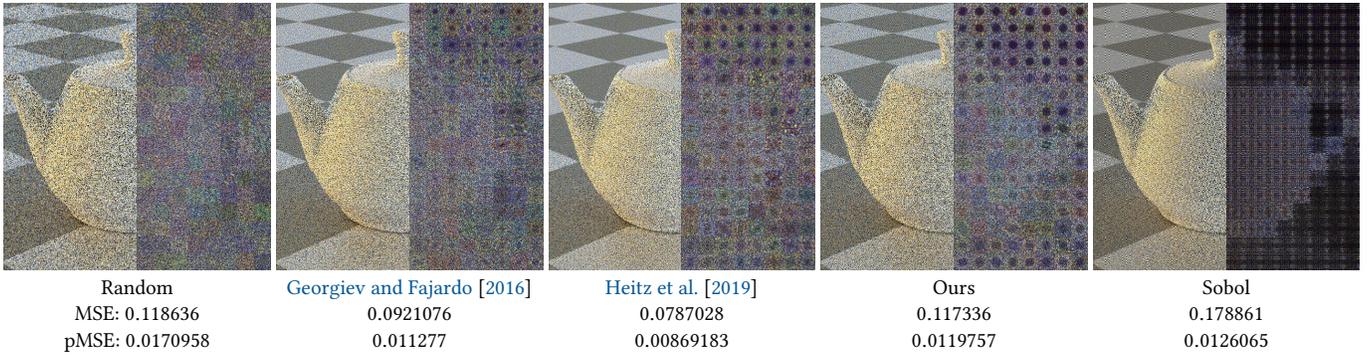
%%%%%%%%%%%%%%%%%%%%%%%%%%%%%%

%\Paragraph{idk}
%A theoretical analysis of the technique in \cite{Heitz2019b} %is not straightforward due to two main reasons: a) The %method is tied to a Sobol sampler where the optimization %space is made up of the scrambling and ranking keys, b) The %optimization is done with respect to $2^{16}$ randomized %step functions.
%
%Nevertheless, we have empirically studied the behavior of %the sampler in order to verify whether it fulfills the %"optimality" criteria outlined for BNDS. We have confirmed %that single index samples from an (optimized) dimension %indeed form blue noise images. Additionally, sums over %several samples also form images with improving blue noise %characteristics, which explains the scaling behavior. The %above results suggests that the sampler is close to optimal %for the linear integrand discussed in the BNDS section, and %that it may be reasonable to perform the optimization only %with respect to a linear integrand instead of $2^{16}$ step %functions.
%
%It is unclear whether the optimization over a random family %of step functions yields advantages in the average case in %practice, compared to optimizing wrt a linear integrand. In %our framework, choosing a different family of functions can %be seen as simply making a different assumption on the %integrand.

%%%%%%%%%%%%%%%%%%%%%%%%%%%%%%%%%%%%%%%%%%%%%%%%%%%%%%%%%%%%
\subsection{Blue-noise dithered sampling energy}
%%%%%%%%%%%%%%%%%%%%%%%%%%%%%%%%%%%%%%%%%%%%%%%%%%%%%%%%%%%%

In \citeauthor{georgiev16bluenoise}'s work, in order to get an optimized (multi-channel) blue noise mask, the following energy has been proposed:

\begin{equation}
    \label{eq:supp_BNDS_energy}
    E(p_1,\ldots,p_N) = \sum_{\PixelIndex\ne\PixelIndexTwo} \exp\left(-\frac{\|\PixelIndex-\PixelIndexTwo\|^2}{\sigma^2}\right)\exp\left(-\frac{\|p_{\PixelIndex}-p_{\PixelIndexTwo}\|^{d/2}}{\sigma^2_s}\right),
\end{equation}

which bears some similarity to the weights of a bilateral filter. In the above $\PixelIndex,\PixelIndexTwo$ are pixel coordinates, and $p_{\PixelIndex},p_{\PixelIndexTwo}$ are $d$-dimensional vectors associated with $\PixelIndex,\PixelIndexTwo$. Let the image formed by those vectors be $\SampleSetImage$. The energy aims to make samples $p_{\PixelIndex},p_{\PixelIndexTwo}$ distant ($\|p_{\PixelIndex}-p_{\PixelIndexTwo}\|$ must be large) if they are associated with pixels which are close ($\|\PixelIndex-\PixelIndexTwo\|$ is small).

%%%%%%%%%%%%%%%%%%%%%%%%%%%%%%
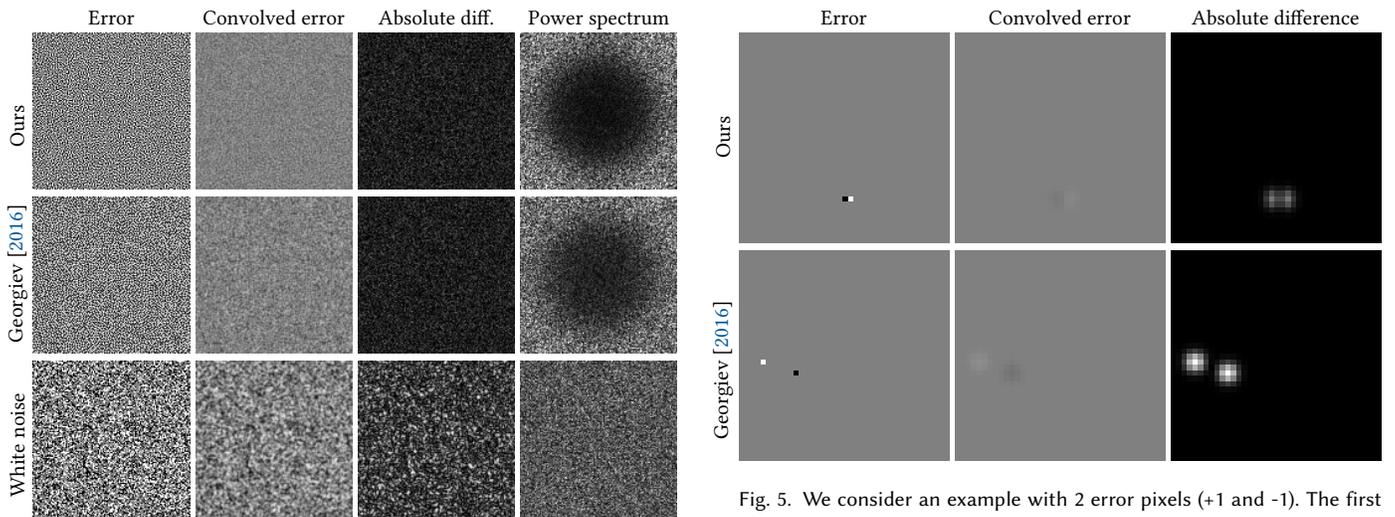
\begin{figure}[t!]
    \centering
    %!TEX root = ../supplemental.tex

\newcommand{\VanTemplate}[1]{
    \begin{scope}
        \clip (0,0) -- (5,0) -- (5,5) -- (0,5) -- cycle;
        \path[fill overzoom image=figures/supp_unit-test-metrics/#1.png] (0,0) rectangle (5cm,5cm);
    \end{scope}
}

\small
\hspace*{-5.5mm}
\begin{tabular}{c@{\;}c@{\;}c@{\;}c@{\;}c@{}}
    & Error & Convolved error & Absolute diff.\ & Power spectrum
    \\
    \rotatebox{90}{\qquad Ours}
    &
    \begin{tikzpicture}[scale=0.416]
        \VanTemplate{error_Ours}
    \end{tikzpicture}
    &
    \begin{tikzpicture}[scale=0.416]
        \VanTemplate{conv_error_Ours}
    \end{tikzpicture}
    &
    \begin{tikzpicture}[scale=0.416]
        \VanTemplate{abs_conv_signal_Ours}
    \end{tikzpicture}
    &
    \begin{tikzpicture}[scale=0.416]
        \VanTemplate{ps_error_Ours}
    \end{tikzpicture}
    \\
    \rotatebox{90}{\; Georgiev~\citeyearpar{georgiev16bluenoise}}
    &
    \begin{tikzpicture}[scale=0.416]
        \VanTemplate{error_Georgiev}
    \end{tikzpicture}
    &
    \begin{tikzpicture}[scale=0.416]
        \VanTemplate{conv_error_Georgiev}
    \end{tikzpicture}
    &
    \begin{tikzpicture}[scale=0.416]
        \VanTemplate{abs_conv_signal_Ours}
    \end{tikzpicture}
    &
    \begin{tikzpicture}[scale=0.416]
        \VanTemplate{ps_error_Georgiev}
    \end{tikzpicture}
    \\
    \rotatebox{90}{\quad White noise}
    &
    \begin{tikzpicture}[scale=0.416]
        \VanTemplate{error_white}
    \end{tikzpicture}
    &
    \begin{tikzpicture}[scale=0.416]
        \VanTemplate{conv_error_white}
    \end{tikzpicture}
    &
    \begin{tikzpicture}[scale=0.416]
        \VanTemplate{abs_conv_signal_white}
    \end{tikzpicture}
    &
    \begin{tikzpicture}[scale=0.416]
        \VanTemplate{ps_error_white}
    \end{tikzpicture}
\end{tabular}

\let\VanTemplate\undefined
    \vspace{-2mm}
    \caption{
        We show an example demonstrating how our energy (top row) forms clusters where required so that the convolved error (second column) produces the best cancellation effect.  
        The first column shows error images. Ours would converge to a grey (reference) image faster compared to the one using the energy in~\cref{eq:supp_BNDS_energy}. The convolved images in the second column show the same behavior. The third column shows the absolute difference between the convolved error and the reference grey image (darker is better). The fourth column shows the error power spectra, with ours showing much better blue-noise characteristics than others.}
    \label{fig:supp_unit_test}
\end{figure}
%%%%%%%%%%%%%%%%%%%%%%%%%%%%%%

%%%%%%%%%%%%%%%%%%%%%%%%%%%%%%
\Paragraph{Relation to our framework}
%%%%%%%%%%%%%%%%%%%%%%%%%%%%%%

Even though the energy is heuristically motivated, we can very roughly relate it to our framework. The above energy implicitly assumes classes of integrands $\EstimatedImage_1$, ..., $\EstimatedImage_T$, such that close samples $p_{\PixelIndex}$, $p_{\PixelIndexTwo}$ are mapped to close values $\EstimatedPixel_{\PixelIndex,t}(p_{\PixelIndex})$, $\EstimatedPixel_{\PixelIndexTwo,t}(p_{\PixelIndexTwo})$, and distant samples are mapped to distant values. Notably, the form of the energy does not change over screen-space, so the same can be implied about the integrands. One such class is the class of bi-Lipschitz functions. The bound can be used to relate a modified version of the original energy, to an energy of the form:
\begin{equation}
    E_{\EstimatedImage_t} = \sum_{\PixelIndex\ne\PixelIndexTwo} \exp\left(-\frac{\|\PixelIndex-\PixelIndexTwo\|^2}{\sigma^2}\right)\exp\left(-\frac{C\|\EstimatedPixel_{\PixelIndex,t}(p_{\PixelIndex})-\EstimatedPixel_{\PixelIndexTwo,t}(p_{\PixelIndexTwo})\|^{d/2}}{\sigma^2_s}\right).
\end{equation}
Thus, the original energy can indeed be interpreted as reasonable for a whole class of sufficiently smooth integrands, instead of an energy that works very well with one specific integrand.

A similar thing can be achieved in our framework, if the weighted energy is considered:
\begin{equation}
     E'(\SampleSetImage) = \sum_{t=1}^{T}w_{t}\|\Kernel * \EstimatedImage_{t}(\SampleSetImage)-\ReferenceImage_{t}\|^2.
\end{equation}
The kernel $\Kernel$ can be a Gaussian with standard deviation $\sigma$, as in the original energy, or it can be relaxed to an arbitrary desired kernel. $\EstimatedImage_1,\ldots,\EstimatedImage_T$ are representative integrands that satisfy the discussed smoothness requirements, and $w_t$ are associated weights assigning different importance to the integrands. Finally, the reference images are given by the integrals $\ReferenceImage_t = \int_{[0,1]^d}\EstimatedImage_t(x)\,dx$.

It should be clear that this is a weighted average constructed from the standard energy in our framework applied to a set of integrands. There are a number of benefits of such an explicit formulation. Most importantly, it allows for \emph{a-priori} methods to be studied in the same framework as \emph{a-posteriori} approaches. Additionally, explicit control is provided over the set of integrands and the kernel in a manner that allows for a straightforward interpretation.

%\vassillen{Reptition with our a-priori. Worth keeping?}\gurprit{It is okay to be redundant here.}

%%%%%%%%%%%%%%%%%%%%%%%%%%%%%%
\Paragraph{Perceptual quality trade-off}
%%%%%%%%%%%%%%%%%%%%%%%%%%%%%%

While the energy of \citeauthor{georgiev16bluenoise} is able to account for many different integrands, this is achieved at the cost of the perceptual quality of the produced patterns. We illustrate this in~\cref{fig:supp_unit_test} by considering a constraint where 25\% of all pixels have an error of +1 and 25\% of all pixels have an 
%error of -1~\Cref{fig:georgiev_perceptual}.
error of -1.

%%%%%%%%%%%%%%%%%%%%%%%%%%%%%%
\begin{figure}[t!]
    \centering
    %!TEX root = ../supplemental.tex

\newcommand{\TwoPixelTemplate}[1]{
    \begin{scope}
        \clip (0,2) -- (3,2) -- (3,5) -- (0,5) -- cycle;
        \path[fill overzoom image=figures/supp_unit-test-two-points/#1.png] (0,0) rectangle (5cm,5cm);
    \end{scope}
}

\small
\hspace*{-6mm}
\begin{tabular}{c@{\;}c@{\;}c@{\;}c@{}}
    & Error & Convolved error & Absolute difference %\absConvDiffImage
    \\
    \rotatebox{90}{\qquad \qquad Ours}
    &
    \begin{tikzpicture}[scale=0.93]
        \TwoPixelTemplate{error_Ours}
    \end{tikzpicture}
    &
    \begin{tikzpicture}[scale=0.93]
        \TwoPixelTemplate{conv_error_Ours}
    \end{tikzpicture}
    &
    \begin{tikzpicture}[scale=0.93]
        \TwoPixelTemplate{abs_conv_signal_Ours}
    \end{tikzpicture}
    \\
    \rotatebox{90}{\quad Georgiev~\citeyearpar{georgiev16bluenoise}}
    &
    \begin{tikzpicture}[scale=0.93]
        \TwoPixelTemplate{error_Georgiev}
    \end{tikzpicture}
    &
    \begin{tikzpicture}[scale=0.93]
        \TwoPixelTemplate{conv_error_Georgiev}
    \end{tikzpicture}
    &
    \begin{tikzpicture}[scale=0.93]
        \TwoPixelTemplate{abs_conv_signal_Georgiev}
    \end{tikzpicture}
\end{tabular}
    \vspace{-2mm}
    \caption{
        We consider an example with 2 error pixels (+1 and -1). The first column shows the error images, the second column shows this error convolved with a gaussian kernel, and the third column shows the difference between the convolved error and the \emph{reference} (constant) greyscale image. In the top row, our energy clusters these pixels such that they can cancel out each other's contribution under convolution. \citeauthor{georgiev16bluenoise}'s energy in the bottom row pushes these pixels farther away. The corresponding absolute difference (convolved error $-$ constant grey image) images in the third column demonstrate that our energy makes the error converge faster to the constant greyscale image (darker is better). 
        %not because it is perceptually optimal but to be able to potentially handle other integrands except for a linear one.
    }
    \label{fig:supp_unit_test_two_pixels}
\end{figure}
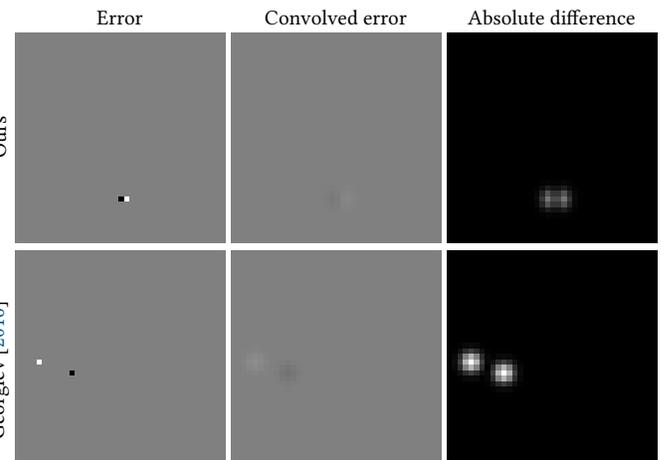
%%%%%%%%%%%%%%%%%%%%%%%%%%%%%%

% \begin{figure}
% \resizebox{\columnwidth}{!}{%
% \begin{tabular}{cc}
% \includegraphics[]{figures/supplemental_georgiev_energy/1_sqrt2_geo.png} & \includegraphics[]{figures/supplemental_georgiev_energy/1_sqrt2_sig.png}
% %\\ \citeauthor{georgiev16bluenoise}'s energy & our  energy
% \\ \includegraphics[]{figures/supplemental_georgiev_energy/1_sqrt2_geo_conv.png} & \includegraphics[]{figures/supplemental_georgiev_energy/1_sqrt2_sig_conv.png} 
% %\\ above image convolved & above image convolved
% \end{tabular}
% }
% \caption{An example illustrating that our energy (right) forms clusters where required so that the convolution (bottom) produces the best cancellation effect. This does not occur for \citeauthor{georgiev16bluenoise}'s energy (left).}
% \label{fig:georgiev_perceptual}
% \end{figure}

For the experiment an initial white noise image is permuted using a brute force optimization with our energy from the main paper and the energy of \citeauthor{georgiev16bluenoise}. One can see that the pattern resulting from our energy always decays faster under convolution. This can be explained by the fact that the bilateral filter-like energy forces nearby pixels to be as different as possible. This doesn't necessarily lead to the best results under convolution illustrated 
%by~\Cref{fig:georgiev_perceptual_1p}, 
by~\Cref{fig:supp_unit_test_two_pixels},
but it is necessary in the setting of \emph{a-priori} methods since not much information is assumed regarding the integrand.

\begin{figure}[t!]
    \centering
    %!TEX root = ../supplemental.tex

\newcommand{\SineTemplate}[1]{
    \begin{scope}
        \clip (0,0) -- (5,0) -- (5,5) -- (0,5) -- cycle;
        \path[fill overzoom image=figures/supp_unit-test-sine/#1.png] (0,0) rectangle (5cm,5cm);
    \end{scope}
}

\small
\hspace*{-2.2mm}
\begin{tabular}{c@{\;}c@{\;}c@{\;}c@{}}
Initial & Georgiev~\citeyearpar{georgiev16bluenoise} & Ours & Reference
\\
\begin{tikzpicture}[scale=0.417]
    \SineTemplate{signal_white}
\end{tikzpicture}
&
\begin{tikzpicture}[scale=0.417]
    \SineTemplate{signal_Georgiev}
\end{tikzpicture}
&
\begin{tikzpicture}[scale=0.417]
    \SineTemplate{signal_Ours}
\end{tikzpicture}
&
\begin{tikzpicture}[scale=0.417]
    \SineTemplate{original_image}
\end{tikzpicture}
\end{tabular}
    \vspace{-2mm}
    \caption{
        A more realistic test with kernel $\Kernel$ using $\sigma=1/\sqrt{2}$. The signal is a sine function that increases in frequency along the vertical axis. Our method handles tone mapping and preserves well both the lower and higher frequencies present in the signal.
    }
    \label{fig:supp_unit_test_sine}
\end{figure}
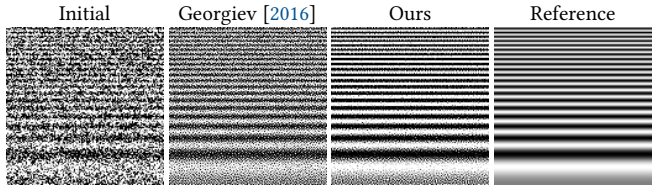

We consider a more realistic example in~\cref{fig:supp_unit_test_sine} where the underlying signal is a sine function with
vertically increasing frequency. We first degrade the signal with uniform white noise. To optimize
the error distribution, we use our Kronecker kernel energy extension (eq. 11 from the main paper
where $h = \delta$) that is given by:
\begin{align}
    \Energy(\EstimatedImage)
        \,=\, \Norm{\Kernel * \ToneMappingOp(\EstimatedImage) - \ToneMappingOp(\ReferenceImage)}_2^2,
\end{align}
where $\ToneMappingOp$ simply clamps values to $[0,1]$. The result with our energy function matches better the original
signal. This is perfectly in line with all of our results on realistic scenes presented in the main paper and the supplemental HTML.

%%%%%%%%%%%%%%%%%%%%%%%%%%%%%%
%%%%%%%%%%%%%%%%%%%%%%%%%%%%%%

% \begin{figure}
% \resizebox{\columnwidth}{!}{%
% \begin{tabular}{cc}
% \includegraphics[scale=14]{figures/supplemental_georgiev_energy/1p_18_geo.png} & \includegraphics[scale=14]{figures/supplemental_georgiev_energy/1p_18_sig.png}
% %\\ \citeauthor{georgiev16bluenoise}'s energy & our  energy
% \\ \includegraphics[scale=14]{figures/supplemental_georgiev_energy/1p_18_geo_conv.png} & \includegraphics[scale=14]{figures/supplemental_georgiev_energy/1p_18_sig_conv.png} 
% %\\ above image convolved & above image convolved
% \end{tabular}
% }
% \caption{An example with 2 error pixels (+1 and -1). Our energy clusters them such that they cancel out as much as possible each other's contribution under convolution. \citeauthor{georgiev16bluenoise}'s energy (left) pushes them farther away not because it is perceptually optimal but to be able to potentially handle other integrands except for a linear one.}
% \label{fig:georgiev_perceptual    _1p}
% \end{figure}

%%%%%%%%%%%%%%%%%%%%%%%%%%%%%%%%%%%%%%%%%%%%%%%%%%%%%%%%%%%%
\subsection{Blue-noise dithered sampling algorithm}
\label{ss:supp_BDNS_algorithm}
%%%%%%%%%%%%%%%%%%%%%%%%%%%%%%%%%%%%%%%%%%%%%%%%%%%%%%%%%%%%

The second contribution of \citeauthor{georgiev16bluenoise}'s work is a sampler which relies on an image optimized with \cref{eq:supp_BNDS_energy} and uses it to achieve a blue noise distribution of the rendering error. We summarize the algorithm and discuss some details related to it. 

%%%%%%%%%%%%%%%%%%%%%%%%%%%%%%
\Paragraph{Algorithm}
%%%%%%%%%%%%%%%%%%%%%%%%%%%%%%

Let $\BlueNoiseImage$ be an image (with $d$-channels) optimized by minimizing \cref{eq:supp_BNDS_energy} over a suitable search space. Let $\mathcal{P}=\{p_1,\ldots,p_N\}$ be a sequence of $d$-dimensional points. Within each pixel $\PixelIndex$ the sample set $\SampleSetPixel_{\PixelIndex}$ is constructed, such that
\begin{equation}
    p_{\PixelIndexTwo} \in \mathcal{P} \implies p_{\PixelIndex,\PixelIndexTwo} \in \SampleSetPixel_{\PixelIndex} : p_{\PixelIndex,\PixelIndexTwo} = (p_{\PixelIndexTwo}+\BlueNoisePixel_{\PixelIndex}) \, \operatorname{mod} \, 1.
\end{equation}
The sequence $\mathcal{P}$ can be constructed by using various samplers (\eg random, low-discrepancy, blue-noise, etc.). The construction of the new points for pixel $\PixelIndex$ can be interpreted either as toroidally shifting the sequence $\mathcal{P}$ by $\BlueNoisePixel_{\PixelIndex}$ or equivalently as toroidally shifting the sequence $\{\BlueNoisePixel_{\PixelIndex},\ldots,\BlueNoisePixel_{\PixelIndex}\}$ by $\mathcal{P}$.

The sequences constructed within each pixel are used to estimate the integral in the usual manner. Since a finite number of dimensions $d$ are optimized the suggestion is to distribute the constructed sequences over smoother dimensions, while other dimensions may use a standard sampler.

%%%%%%%%%%%%%%%%%%%%%%%%%%%%%%
\Paragraph{Effect of the toroidal shift}
%%%%%%%%%%%%%%%%%%%%%%%%%%%%%%

Let us consider a linear one-dimensional integrand $f(q) = \alpha q+\beta$ that does not vary in screen space, and a sequence $\mathcal{P}$ with a single point $p$. Furthermore, if we assume $p=0$, then the error is given by:
\begin{equation}
    \EstimatedImage(\BlueNoiseImage)-\ReferenceImage = \alpha\BlueNoiseImage+\pmb{\beta}-\ReferenceImage.
\end{equation}
Since $\EstimatedImage$ does not vary in screen space, then $\ReferenceImage$ also does not. Then the power spectrum of the error (excluding the DC) matches the power spectrum of $\pmb{B}$ up to the multiplicative factor $\alpha^2$. Then, under the assumption that the integrand is linear, does not vary in screen space, and there is no toroidal shift, the power spectral properties of $\BlueNoiseImage$ are transferred ideally to the error.

On the other hand, if $p$ is chosen to be non-zero, then the spectral characteristics of the image $((\BlueNoiseImage+p) \, \operatorname{mod} \, 1)$ will be transferred instead. We have empirically verified that even with a very good quality blue noise image $\BlueNoiseImage$ the toroidal shift degrades its quality due to the introduced discontinuities. Thus, even in the ideal case of a constant in screen space linear $1$-D integrand, toroidal shifts degrade the quality.

\Paragraph{Effect of using multiple samples}
Let us consider the same integrand $f(q) = \alpha q+\beta$, which we have identified as being ideal for transferring the spectral characteristics of $\BlueNoiseImage$ to the error. And let us assume that we are given several samples: $\mathcal{P}=\{p_1,...,p_N\}$, and we have constructed the sample set image $\SampleSetImage$ through toroidal shifts with $\BlueNoiseImage$. Then the error is:
\begin{equation}
    \EstimatedPixel_{\PixelIndex}(\SampleSetPixel_{\PixelIndex}) - \ReferencePixel_{\PixelIndex} = \frac{\alpha}{N}\sum_{\PixelIndexThree=1}^{N}p_{\PixelIndexThree,\PixelIndex} + \beta - \ReferencePixel_{\PixelIndex}.
\end{equation}
The power spectrum of the error thus matches the power spectrum of the image $A_{\PixelIndex} = \sum_{\PixelIndexThree=1}^{N}p_{\PixelIndexThree,\PixelIndex}$ (excluding the DC) up to a multiplicative factor. For a random point sequence $\mathcal{P}$ the more points are considered, the closer to white noise $\pmb{A}$ becomes. This is further exacerbated by the discussed discontinuities introduced by the toroidal shifts.

%%%%%%%%%%%%%%%%%%%%%%%%%%%%%%
\Paragraph{Extension}
%%%%%%%%%%%%%%%%%%%%%%%%%%%%%%

We have argued that both toroidal shifts and increasing the number of samples has a negative effect on transferring the spectral properties of $\BlueNoiseImage$ even in an ideal scenario. Naturally the question arises whether this can be improved. Our proposal is the direct optimization of point sets without the application of a toroidal shift.

For the discussed example this entails constructing a sequence of $N$ images $\BlueNoiseImage_1,\ldots,\BlueNoiseImage_N$ such that $\pmb{A}_{\PixelIndexThree} = \sum_{\PixelIndexTwo=1}^{\PixelIndexThree}\BlueNoiseImage_{\PixelIndexTwo}$ is a blue noise image. Then the error has the (blue noise) spectral characteristics of $\pmb{A}_{\PixelIndexThree}$ at each sample count (\cref{fig:supp_apriori-teapot}):

\begin{equation}
    \EstimatedPixel_{\PixelIndex}(\BlueNoisePixel_{1,\PixelIndex},\ldots,\BlueNoisePixel_{\PixelIndexThree,\PixelIndex})-\ReferencePixel_{\PixelIndex} = \frac{\alpha}{\PixelIndexThree}\sum_{\PixelIndexTwo=1}^{\PixelIndexThree}\BlueNoisePixel_{\PixelIndexTwo,\PixelIndex} + \beta - \ReferencePixel_{\PixelIndex}.
\end{equation}

\ifdefined\arXiv
\else
%%%%%%%%%%%%%%%%%%%%%%%%%%%%%%%%%%%%%%%%%%%%%%%%%%%%%%%%%%%%
\section{Additional results}
%%%%%%%%%%%%%%%%%%%%%%%%%%%%%%%%%%%%%%%%%%%%%%%%%%%%%%%%%%%%

% We present additional results, that serve to supplement the ones in the main paper. 
% In \cref{fig:supp_vertical-vs-histogram-64spp} we compare the histogram sampling approach with our \vertical iterative approach on a different set of scenes that the ones in the main paper for the same comparison.
In \cref{fig:supp_staircase-vertical-methods} we showcase tiled error spectra and SCIELAB images for several methods for the Wooden Staircase scene. SCIELAB acts very similarly to the pointwise squared error, as confirmed by its preference for methods that minimize the pointwise error. Thus it does not make for a very good metric for quantifying the quality of the distribution of the noise unlike VDP or the tiled error spectra.
% We also compare our \vertical methods to \citeauthor{Heitz2019}'s permutation and histogram approaches in \cref{fig:supp_vertical-vs-permutation-heitz} and \cref{fig:supp_vertical-methods}. This supplements a similar comparison in the main paper on additional scenes and including our \vertical dithering method.
\fi

\ifdefined\arXiv
\else
\begin{figure*}[t!]
    \centering
    \input{figures/supp_cornellbox-textured/cornellbox-textured}
    \vspace{-2mm}
    \caption{
        The textured Cornell box scene is compared over different samples per pixel using our texture handling approach in \cref{sec:supp_our_demodulation}. The improvements are visible for all spp over~\citeauthor{Heitz2019}'s permutation approach using demodulation. This is especially apparent when comparing the tiled error spectra. Note that the spectra are not normalized to 1.
    }
    \label{fig:supp_cornellbox-textured}
\end{figure*}
\fi
%%%%%%%%%%%%%%%%%%%%%%%%%%%%%%

%%%%%%%%%%%%%%%%%%%%%%%%%%%%%%
\ifdefined\arXiv
\else
\begin{figure*}[t!]
    \centering
    \input{figures/supp_staircase-all-metrics/staircase-all-metrics}
    \vspace{-2mm}
    \caption{
        In the main paper compare all \vertical methods on the \emph{Wooden staircase} scene. All of our methods achieve lower pMSE than the baseline (the averaged image), while the permutation method increases the error both in terms of MSE and pMSE. The tiled error power spectra images confirm the pMSE ranking and provide a visualization of the local pMSE distribution. We also show S-CIELAB error visualizations which suggest that pointwise error is heavily weighted in S-CIELAB, which does not make it a very good predictor for the perceptual quality related to the noise distribution, unlike HDR-VDP-2. %\vassillen{Use permutation instead of histogram sampling?}
    }
    \label{fig:supp_staircase-vertical-methods}
\end{figure*}
\fi

\bibliographystyle{ACM-Reference-Format}
\bibliography{paper}

%%%%%%%%%%%%%%%%%%%%%%%%%%%%%%%%%%%%%%%%%%%%%%%%%%%%%%%%%%%%